\documentclass[A4paper, journal]{IEEEtran}
\usepackage{amsmath,amsfonts}
\usepackage{amssymb}
\usepackage{algorithm}
\usepackage{algorithmic}
\usepackage{array}
\usepackage{textcomp}
\usepackage{stfloats}
\usepackage{url}
\usepackage{verbatim}
\usepackage{graphicx}
\usepackage{subcaption}   
\usepackage{multirow}
\usepackage{makecell}
\usepackage{mathtools}

\usepackage{ulem}
\usepackage{cite}
\def\BibTeX{{\rm B\kern-.05em{\sc i\kern-.025em b}\kern-.08em
    T\kern-.1667em\lower.7ex\hbox{E}\kern-.125emX}}
\usepackage{balance}
\usepackage{orcidlink}
\hypersetup{colorlinks=true, linkcolor=red, citecolor=blue}

\makeatletter
\newif\ifCFGcaseOne
\newif\ifCFGcaseTwo
\@ifclasswith{IEEEtran}{12pt}{%
  \@ifclasswith{IEEEtran}{onecolumn}{%
    \@ifclasswith{IEEEtran}{draftclsnofoot}{\CFGcaseOnetrue}{}%
  }{}%
}{}
\@ifclasswith{IEEEtran}{A4paper}{%
  \@ifclasswith{IEEEtran}{journal}{\CFGcaseTwotrue}{}%
}{}
\makeatother

\begin{document}

\title{Diffusion-aided Task-oriented Semantic Communications with Model Inversion Attack}

\ifCFGcaseTwo
\author{Xuesong Wang$^{\orcidlink{0000-0002-2743-2212}}$, Mo Li$^{\orcidlink{0000-0002-7564-4834}}$, Xingyan Shi$^{\orcidlink{0009-0003-4901-3013}}$, Zhaoqian Liu$^{\orcidlink{0009-0005-8019-4080}}$, and Shenghao Yang$^{\orcidlink{0000-0003-1987-5643}}$, \textit{Member, IEEE}
\thanks{Xuesong Wang is with the College of Electronic Information Engineering, Taiyuan University of Technology, Jinzhong 030600, China. Mo Li, Xingyan Shi, Zhaoqian Liu, and Shenghao Yang are with the School of Science and Engineering, The Chinese University of Hong Kong, Shenzhen, Shenzhen 518172, China.} \thanks{Corresponding author: Shenghao Yang (shyang@cuhk.edu.cn).}
\thanks{This work was supported in part by the National Key R\&D Program of China under Grant 2022YFA1005000, the 1+1+1 CUHK-CUHK(SZ)-GDSTC Joint Collaboration Fund under Grant 2025A0505000082, and the Shenzhen Fundamental Research Program under Grant JCYJ20241202124023031.}
\thanks{
© 2026 IEEE.  Personal use of this material is permitted.  Permission from IEEE must be obtained for all other uses, in any current or future media, including reprinting/republishing this material for advertising or promotional purposes, creating new collective works, for resale or redistribution to servers or lists, or reuse of any copyrighted component of this work in other works.
}
\thanks{
X. Wang, M. Li, X. Shi, Z. Liu, and S. Yang, “Diffusion-aided Task-oriented Semantic Communications with Model Inversion Attack,” IEEE Transactions on Cognitive Communications and Networking, 2026, doi: 10.1109/TCCN.2026.3735677.
}
}
\fi

\markboth{}
{Diffusion-aided Task-oriented Semantic Communications with Model Inversion Attack}

\maketitle

\begin{abstract}

Semantic communication enhances transmission efficiency by conveying semantic information rather than raw input symbol sequences. Task-oriented semantic communication further aims to retain only task-specific information, thereby achieving greater bandwidth savings. However, these neural-network-based communication systems are vulnerable to model inversion attacks, in which adversaries attempt to recover sensitive input information from intercepted semantic features. The key challenge is therefore to preserve privacy while maintaining task accuracy and robustness.
We consider a task-confidential setting in which the adversary attempts to reconstruct the original input from intercepted features without knowing the legitimate receiver's task or model. 
Although PSNR and SSIM are commonly used to assess reconstruction quality, we find that an external classifier can still perform the legitimate receiver’s task with nontrivial accuracy on reconstructions with low PSNR or SSIM, indicating that these reconstructions still contain task-level semantic leakage.
We therefore propose DiffSem, which splits the diffusion process between controlled transmitter-side self-noising and matched receiver-side reverse denoising. Experiments on the MNIST, CIFAR-10, and CelebA datasets show that DiffSem improves the legitimate receiver's task accuracy without increasing either the transmitted feature size or information leakage. 

\end{abstract}



\section{Introduction}

Deep learning based semantic communication systems have emerged as a promising solution to enhance information transmission efficiency in sixth-generation (6G) wireless networks. Unlike traditional paradigms that prioritize the accurate symbol-level transmission, semantic communication shifts the focus to conveying the intended meaning through symbols \cite{weaver1953recent}. In such systems, the transmitter and receiver collaboratively design mappings between meanings and channel symbols, naturally aligning with the principles of joint source-channel coding (JSCC) \cite{shao2024theory, pan2024sc}. This approach accommodates more complex data sources (e.g., images) and diverse distortion measures, such as classification loss \cite{xie2021deep} or task-oriented metrics \cite{Beck2022SemanticCA}. As a result, semantic communication has been studied for various types of sources such as text \cite{yan2022resource, peng2024robust}, visual data \cite{lyu2024semantic, wu2024deep, ma2024direction}, and video \cite{li2024video, wang2022wireless}, and has also been applied in edge computing \cite{cang2023online, yang2022semantic} and autonomous driving \cite{muhammad2022vision, zheng2023distributed}. By adopting a ``comprehend-first, transmit-later'' approach, semantic communication significantly reduces transmitted data volume, minimizes bandwidth consumption, and improves communication efficiency. 

As a variant of semantic communication, task-oriented semantic communication shifts the target from reconstructing the transmitter's raw input to directly enabling the receiver to accomplish a given task, thereby allowing the transmitter to further compress the transmitted content~\cite{zhang2023toward,gunduz2022beyond}. For example, in a typical task-oriented semantic communication scenario targeting the image classification task, the sender sends the semantic representation of the picture, and the receiver uses the semantic information to perform the classification task directly, skipping the process of recovering the picture. 
Studies have demonstrated the efficiency and robustness of task-oriented semantic communication in various scenarios, including image retrieval \cite{jankowski2020deep} and federated learning \cite{sun2025hybrid}, as well as scenarios involving resource-constrained devices \cite{peng2024task} and MIMO channels \cite{cai2025end},
and cloud-edge-device collaborative inference with privacy and QoAIS considerations \cite{jiang2025task}.


Despite its advantages, task-oriented semantic communication, like all neural-based systems, is vulnerable to model inversion attacks, in which an adversary eavesdrops on the transmitter's output and attempts to reconstruct sensitive information of the input \cite{yang2024secure}.
Conventional privacy-preserving techniques, however, are difficult to apply in this context. 
For example, differential privacy perturbs model inputs or outputs \cite{ha2019differential}, but such perturbations can significantly degrade semantic communication performance \cite{pan2024differential}. 
Similarly, federated learning avoids direct data sharing by transmitting parameters, yet remains vulnerable to gradient leakage attacks \cite{zhao2023data}.

Existing privacy-preserving approaches in related settings can be broadly grouped into two lines. One line learns privacy-preserving representations, often through adversarial training or bottleneck-inspired objectives, so that private information becomes harder to reconstruct from intermediate or transmitted features; in many such works, the privacy effect is mainly reflected by degraded image-level reconstruction quality at the adversary, often measured by metrics such as PSNR or SSIM \cite{xiao2020adversarial, li2020tiprdc, wang2024privacy, wang2023privacy}. The other line improves privacy by actively redirecting the eavesdropper’s observation away from the original private source and toward misleading or alternative content \cite{wei2025optimizing, chen2025physical, han2025semantic, tang2024secure}.


This paper focuses on the first line of privacy-preserving approaches. We consider a task-confidential setting~\cite{erak2024contrastive} in which the adversary performs model inversion without knowledge of the receiver's downstream task, and thus aims to reconstruct the original input rather than directly optimizing task performance.
To this end, following the notion of semantic fidelity in~\cite{zhang2023toward, shi2021new}, we evaluate whether the adversary's reconstruction still preserves task-relevant information. Specifically, we introduce an external strong classifier \cite{zhang2023toward}, which is trained for the same downstream task but is not accessible to the adversary. The reconstructed sample is fed into this classifier to perform the same task as the legitimate receiver. Our experiments show that reconstructions with low PSNR and SSIM can still be accurately classified by the strong classifier, indicating that they retain the information required for the task. The retention of such task-relevant information constitutes information leakage.

To mitigate such leakage, we propose DiffSem, a diffusion-aided task-oriented semantic communication system that enhances task performance while limiting adversarial recovery. Several diffusion-aided semantic communication studies provide direct technical foundations for receiver-side recovery under wireless distortion. CDDM introduces a post-equalization diffusion module to remove channel noise from received semantic signals \cite{wu2024cddm}, while DIFFSC employs diffusion probabilistic models to enhance denoising in receiver-side image reconstruction \cite{jiang2024diffsc}. SP-Latent-Diff EDNSC further formulates channel-adaptive equalization and denoising as an inverse problem in the semantic domain and performs latent-diffusion posterior sampling with semantic priors \cite{xu2025semantic}. A structurally related framework models wireless transmission as the forward diffusion process and reconstructs the source at threceiver under bandwidth constraints \cite{guo2025diffusion}.
These studies establish diffusion-based recovery as an effective means of handling wireless distortion. The distinctive use of diffusion in DiffSem lies in assigning its forward and reverse processes to different functions across the communication link. Before transmission, the transmitter halts the forward diffusion process at a controllable intermediate step, making the diffusion level a variable that regulates the recoverability of the exposed representation. After transmission, the receiver performs the matched reverse process to recover task-relevant features for inference. The channel-aware noise-margin design supports this paired process by coordinating the transmitter-side self-noising with the physical channel noise. The receiver can be further strengthened by self-referential label embeddings without requiring external guidance. Conceptually, this design exploits the semantic-information mismatch by reducing task-agnostic recoverability from the transmitted features without depriving a task-aware receiver of the semantics needed for the downstream task.

In general, our contributions can be summarized as follows.
\begin{itemize}
    \item We reveal and quantify a mismatch between pixel-level reconstruction quality and task-level semantic leakage under reconstruction-based model inversion attacks. Our experiments show that adversarial reconstructions with low PSNR or SSIM may still retain sufficient task-relevant information to support nontrivial classification accuracy. Based on this observation, we introduce a semantic leakage evaluation method that uses an external strong classifier to measure the classification accuracy on the reconstructed samples.
    \item We propose DiffSem, a privacy-preserving diffusion-aided framework for task-oriented semantic communications under the considered model inversion threat model. Its core technical innovation lies in splitting the diffusion process between the transmitter and the receiver, with controlled forward diffusion for transmitter-side self-noising and matched reverse denoising for receiver-side task recovery. The self-noising level controls the transmitted information, while the channel-aware noise-margin design aligns the intended diffusion level with the physical channel noise. We further develop E-DiffSem, which uses self-referential soft label embeddings generated by the receiver to guide the denoising process without requiring external guidance or adversarial training.
    \item Experiments on MNIST, CIFAR-10, and CelebA demonstrate the effectiveness of DiffSem and E-DiffSem under different self-noising levels and channel conditions. The proposed methods improve the legitimate receiver's task accuracy without increasing the transmitted feature size, while maintaining a favorable gap between legitimate task utility and the task-level semantic leakage retained in adversarial reconstructions. The results further show that the self-noising level provides a controllable utility--leakage tradeoff, that the channel-aware noise-margin design enables the diffusion process to operate under wireless noise, and that the self-referential guidance in E-DiffSem further improves receiver-side task recovery.
\end{itemize}

This paper is organized as follows. Section II describes the task-oriented semantic architecture as well as the model inversion attack, and general evaluations of the system. Section III introduces the modules and the training algorithms in our system in detail. Section IV includes the experiment results and discussion. Section V concludes the paper.

\section{System architecture}

This section describes the system architecture, followed by the model inversion attack and the evaluation framework for privacy.

\subsection{Diffusion based Task-oriented Semantic System}

As shown in Fig. \ref{fig:whole-system}(a), assume there is a transmitter-receiver pair that performs a specific task using the input raw data $x$ to get target output $y$. First, $x \in X$ is sent into a feature extractor module and a channel coding module to get neural-coded features $f \in F$. In our system, these two modules are both implemented by neural networks, and they can be regarded as a whole module named encoder $\Phi_\phi$ with parameter set $\phi$, like any other neural-network based joint-source-channel-coding (JSCC) system in related literature. Next, a self-noising module $\Psi$ is utilized to add a certain level of noise, based on system requirements, to get $z$ that is ready to transmit. These steps can be expressed as
\begin{equation}
\label{eq:transmitter}
z = \mathcal{T}_{(\phi)}(x) = \Psi (\Phi_{\phi} (x)) = \Psi (f)
\end{equation}
where $\mathcal{T}$ means the transmitter. 
Note that the self-noising module $\Psi$ does not have any learnable parameters and is implemented by a controlled partial forward-diffusion operation, which is introduced in the next section in detail.

\begin{figure*}[t]
    \centering
    \includegraphics[width=\linewidth]{systemfigs/system-1.eps}
    \caption{
System models. 
(a) Task-oriented transmission system, where the transmitter encodes the input into semantic features, applies self-noising before transmission, and the receiver performs diffusion-based denoising followed by task inference. (b) Model inversion attack, where the adversary eavesdrops on a separate channel and attempts to reconstruct the source input from the intercepted transmitted features.
}
    \label{fig:whole-system}
\end{figure*}

Assuming additive white Gaussian noise (AWGN) channel is used\footnote{Our model also covers block fading channel\cite{Schilcher2011} with receiver's channel state information (CSI). Consider $x$ being transmitted within a block, a one-tap coherent equalizer maps $\hat z = h z + \epsilon$ to $z + \tilde\epsilon$ where the noise variance of $\tilde\epsilon$ is rescaled by $1/|h|^{2}$ compared to $\epsilon$ \cite{lyu2024semantic, tse2005fundamentals, goldsmith2005wireless}. We leave a detailed treatment of fading channels to our future work.}. After transmitting over the air, the receiver first observes the complex-baseband signal
\begin{equation}
\label{eq:channel}
\hat{z} = z + \epsilon,
\end{equation}
where $\epsilon\!\sim\!\mathcal{N}(0,\sigma^2 \mathbf{I})$ is the noise in the wireless channel $\mathcal{C}$. Then, $\mathcal{R}$ performs its task in two steps. First, it uses a diffusion module $\Delta_\delta$ to get the estimated feature $\hat{f}$ through denoising; second, $\hat{f}$ is channel decoded and is used to perform a specific task. Decoding and task execution are also combined into a single module, referred to as the classifier $\Theta_{\theta}$, analogous to the encoder module introduced above. Hence the receiver is defined as
\begin{equation}
\label{eq:receiver}
\hat{y} = \mathcal{R}_{(\delta, \theta)}(\hat{z}) = \Theta_{\theta} (\Delta_{\delta} (\hat{z})) = \Theta_{\theta}(\hat{f}),
\end{equation}
where $\delta$ and $\theta$ are the parameter sets in denoising module $\Delta_\delta$ and classifier module $\Theta_{\theta}$, respectively, and $\hat{y}$ is the task output. 
Throughout the system description, hatted symbols denote received, estimated, or inferred quantities corresponding to their unhatted counterparts.
Note that in wireless transmissions, $z, \hat{z} \in \mathbb{C}^{\kappa}$, where $\kappa$ means complex channel use. Since neural networks only process real numbers, without further notice, let $k=2\kappa$, and the length of $z$ ($f$ as well) is $k$ in our system's implementation.

\noindent\textit{Remarks.} We assume that there is a trustworthy third party that can define the specific task and arrange models in each node of the system. Thus, we define that $\Theta_{\theta}$ is reachable only in $\mathcal{R}$ when models are trained and deployed by the third-party, which is a common prerequisite in task-oriented communications \cite{gu2019reaching}.

\begin{figure*}[t]
    \centering
    \includegraphics[width=0.95\linewidth]{systemfigs/system-2.eps}
    \caption{
    Adversary-side data preparation and evaluation. The adversary constructs surrogate inputs \(\dot{x}\), repeatedly queries the transmitter to obtain post-channel observations \(\hat{\dot{z}}\), and trains the inversion model \(A_a\) to reconstruct \(\hat{\dot{x}}\). The strong classifier \(\bar{\Theta}_{\bar{\theta}}\) is not part of the adversary; it is a separate third-party evaluator used only to assess whether the reconstructed images still preserve task-relevant semantics. Green dashed lines indicate the datasets used to train the strong classifier.
}
    \label{fig:adversary-system}
\end{figure*}

\subsection{Model Inversion Attack}

Model inversion attack, originally introduced by Fredrikson et al.~\cite{fredrikson2015model}, considers an adversary that attempts to reconstruct sensitive input information from model-exposed representations or outputs. In semantic communication systems, the privacy risk of such inversion attacks has also been explicitly discussed in recent studies~\cite{wang2024privacy}. In this paper, we consider a model inversion attack in which an eavesdropper attempts to reconstruct an approximation of the source input from the intercepted transmitted feature, as illustrated in Fig.~\ref{fig:whole-system}(b). The attack consists of two phases: a setup phase and an attack phase.
\\\indent
In the setup phase, the adversary $\mathcal{A}_a$ with parameter set $a$ is assumed to have black-box query access to the transmitter $\mathcal{T}_{(\phi)}$, following the common privacy-oriented threat-modeling practice in semantic communication systems~\cite{wang2024privacy,fang2024privacy}. That is, it does not know the parameters of the legitimate transmitter--receiver pair, cannot access internal gradients, and has no access to the receiver-side denoising or task module in $\mathcal{R}_{(\delta,\theta)}$. And we consider a chosen-plaintext adversary, who may possess an auxiliary dataset $\dot{X}$ drawn from the same distribution as the legitimate source dataset $X$. Using samples $\dot{x}\in\dot{X}$, the adversary can repeatedly query the transmitter and collect the corresponding observations on the wiretap branch, as shown in Fig.~\ref{fig:adversary-system}.
\\\indent
The adversary then trains its own inversion model offline to reconstruct an input-like sample from the intercepted wiretap signal. Consistent with the wiretap observation setting considered in wireless semantic communication studies~\cite{wang2022wireless}, we model the adversarial observation as
\begin{equation}
\label{eq:attacker}
    \hat{x} = \mathcal{A}_a(\hat{\bar{z}})
            = \mathcal{A}_a(\bar{z}+\dot{\epsilon}),
\end{equation}
where $\bar{z}$ denotes the transmitted feature on the adversarial observation branch, and $\dot{\epsilon}\sim\mathcal{N}(0,\dot{\sigma}^2\mathbf{I})$ is the corresponding wiretap-channel noise.
\\\indent
The inversion model $\mathcal{A}_a$ is trained by minimizing the pixel-space mean squared error (MSE) loss, as commonly adopted in prior inversion-based privacy studies~\cite{wang2024privacy}:
\begin{equation}
\label{eq:loss_mse}
    \mathcal{L}_{\mathrm{MSE}}(x,\hat{x})
    = \frac{1}{M}\sum_{i=1}^{M}\|x_i-\hat{x}_i\|^2,
\end{equation}
where $M$ is the number of pixels (or, more generally, the input dimension). When applied to the surrogate samples in $\dot{X}$, this loss corresponds to reconstructing $\dot{x}$ from its associated wiretap observation. This objective makes clear that the adversary is solving a reconstruction problem, rather than directly performing downstream task inference.
\\\indent
It is important to emphasize that the adversary is not assumed to possess the legitimate receiver's task-aware recovery objective, the receiver parameters, or the receiver-side denoising/reconstruction mechanism. Moreover, the external strong classifier introduced below is not part of the adversary and is never used in the training or inference of $\mathcal{A}_a$.
\\\indent
After the setup phase, the adversary enters the attack phase. In this online phase, it becomes purely passive: it can only observe the wiretap signal and must perform inversion using the pretrained model obtained in the setup phase, namely by Eq.~(\ref{eq:attacker}).

\subsection{System Performance Evaluation}

The proposed system is evaluated from two aspects: whether the legitimate receiver can reliably accomplish the intended downstream task, and whether the adversary's reconstructions retain information useful for that task. We measure these two aspects using the receiver-side task accuracy ${\rm Acc}_{\rm rx}$ and the reconstruction leakage accuracy ${\rm Acc}_{\rm adv}$, respectively. 
\\\indent
 Let
\begin{equation}
\mathcal{S}_{\rm test}
=
\{(x_1,y_1),\ldots,(x_n,y_n)\}
\end{equation}
be the test set, where $x_i$ is the source input and $y_i$ is the corresponding ground-truth task label. For an input $x_i$, the legitimate receiver uses its task-oriented processing pipeline in Eq.~(\ref{eq:receiver}) to produce the task output $\hat{y}_i$. Let $s(\hat{y},y)\in[0,1]$ denote the task success score. 
The receiver-side task accuracy is naturally given by
\begin{equation}
{\rm Acc}_{\rm rx}
=
\frac{1}{n}
\sum_{i=1}^{n}
s\!\left(
    \hat{y}_i,
    y_i
\right).
\label{eq:acc_rx}
\end{equation}
\\\indent
For the adversarial branch, our interest is whether the reconstructed samples retain task-relevant information about the original inputs, which we refer to as semantic leakage. Let $\hat{x}_i$ denote the input-like reconstruction obtained by the adversary using Eq.~(\ref{eq:attacker}).
These reconstructed samples induce the leakage-evaluation set
\begin{equation}
\widehat{\mathcal{S}}_{\rm adv}
=
\{(\hat{x}_1,y_1),\ldots,(\hat{x}_n,y_n)\}.
\end{equation}
Since each reconstructed sample is paired with the original task label, the performance on $\widehat{\mathcal{S}}_{\rm adv}$ indicates whether the adversarial reconstructions still carry task-oriented information associated with the original samples. 
\\\indent
We employ an external strong classifier $\bar{\Theta}_{\bar{\theta}}$ to evaluate $\widehat{\mathcal{S}}_{\rm adv}$. The classifier is trained separately for the same downstream task and is used only during evaluation, as mentioned in Sec. \ref{Sec:System Training Procedure}. It is not part of the adversary and is never involved in the training or inference of the legitimate transmitter-receiver system in Fig.~\ref{fig:whole-system}.
Based on the $\bar{\Theta}_{\bar{\theta}}$, we define the reconstruction leakage accuracy as
\begin{equation}
{\rm Acc}_{\rm adv}
=
\frac{1}{n}
\sum_{i=1}^{n}
s\!\left(
    \bar{\Theta}_{\bar{\theta}}(\hat{x}_i),
    y_i
\right).
\label{eq:acc_leak}
\end{equation}
The operational meaning of ${\rm Acc}_{\rm adv}$ is as follows. If the external strong classifier performs well on the clean test set $\mathcal{S}_{\rm test}$ but performs close to a no-observation baseline on the leakage-evaluation set $\widehat{\mathcal{S}}_{\rm adv}$, then the adversary's reconstructions are not task-valid with respect to their original labels in the protected task. In other words, the model inversion attack does not recover enough task-relevant semantics for an independent task evaluator to infer the original labels. Conversely, a high ${\rm Acc}_{\rm adv}$ indicates that the adversarial reconstructions still preserve usable task semantics.
\\\indent
Notably, our experiments (in Sec.~\ref{sec: Experiments and Results}) reveal that pixel-level reconstruction quality and task-level semantic leakage can diverge. Reconstructions with low PSNR or SSIM may still yield a high ${\rm Acc}_{\rm adv}$, indicating that task-relevant information can remain available even when pixel-level fidelity is poor. This mismatch is quantified in Fig.~\ref{fig:mnist-3} and further illustrated by the adversarial reconstructions in Fig.~\ref{fig:mnist_images}. This observation motivates us to introduce the transmitter-side self-noising mechanism in DiffSem that directly regulates the transmitted task-relevant information, rather than treating degraded pixel-level reconstruction quality as evidence of reduced leakage.

\section{Modules and Training}


In this section, we first introduce the system modules and the corresponding data flow, and then describe the training procedure based on the overall system design. This design is motivated by the difference between the two recovery objectives: the legitimate receiver only needs to recover the prescribed task output, whereas the inversion model attempts to recover the complete source input. Since complete source recovery generally requires richer information than task inference, transmitter-side self-noising may create an intermediate regime in which the intended task remains recoverable while source inversion is substantially hindered. A detailed asymptotic source--channel interpretation of this mechanism is provided in Sec.~\ref{subsec:mechanism}.

\subsection{Encoder, Classifier and Adversary}

In our system, encoder module $\Phi_{\phi}$ extracts neural-coded feature $f$ whose shape is $\sqrt{k}\times\sqrt{k}$ from source images $x^{CHW}$ with channel $C$, height $H$ and width $W$, and the compression rate is $k/(CHW)$. Feature $f$ is self-distorted to $z$ and is transmitted over the air, then the receiver gets channel-interfered signal $\hat{z}$. Receiver uses denoising module $\Delta_\delta$ and classifier module $\Theta_{\theta}$ to generate $\hat{y}$.

As mentioned before, the self-noising module \(\Psi\) has no learnable parameters. On the transmitter side, it performs a controlled partial forward-diffusion operation on the encoded feature before transmission. Starting from the undistorted feature, the transmitter selects a self-noising step and injects Gaussian noise according to the diffusion schedule, so that only partially noised semantic feature is radiated. When wireless channel compensation is taken into account, the transmitted feature is further adjusted according to the channel condition. On the receiver side, the diffusion module first processes the received signal to recover a task-relevant feature, and the classifier then maps it to the task output. The detailed implementations of the diffusion process and its wireless adaptation are introduced in the following two subsections.

In related work \cite{wang2024privacy}, an auxiliary decoder $\mathcal{D}$ is commonly introduced to simulate an attacker during training, particularly in the context of adversarial learning frameworks. However, our approach does not adopt such attacker-aware training strategies. Instead, we define $\mathcal{A}_a$ as a model trained via model inversion attacks, solely for the purpose of evaluating the privacy risks after the transmission model is established. Importantly, $\mathcal{A}_a$ is not allowed to participate in the training or inference of the transmitter $\mathcal{T}$ and receiver $\mathcal{R}$. Likewise, if $\mathcal{D}$ is used in other methods, it should be restricted to the training phase and excluded from the evaluation phase. In our framework, since no adversarial training is involved, the auxiliary decoder $\mathcal{D}$ is entirely omitted. 
The strong classifier $\bar{\Theta}_{\bar{\theta}}$, as mentioned above, is also external to $\mathcal{A}_a$, and is employed solely for evaluation of $\mathcal{A}_a$.

\subsection{Diffusion Module}

We use denoising diffusion implicit models (DDIM) \cite{song2020denoising} to train the diffusion module in our system. 
Here we specify the partial forward diffusion used in the transmitter-side self-noising module and the matched reverse DDIM process used at the receiver.

Unlike conventional generative usage, we adopt a paired design in which the transmitter deliberately applies a partial forward diffusion (self-noising) to the encoded feature up to a controllable step $T'\leq T$, while the receiver runs a matching DDIM reverse schedule to denoise. This coupling makes $T'$ an information knob that jointly governs utility and privacy. In practice, $T'$ is chosen once and reused across samples.
For consistency with the diffusion notation used in this subsection, we denote the undistorted encoder output by $f_0$, where $f_0$ is exactly the semantic feature $f \in F$ introduced in Section II before self-noising is applied.
Starting from $f_0 \sim p(f_0)$, the forward diffusion process gradually adds Gaussian noise and produces noisy features $f_1,\ldots,f_T$ through \cite{croitoru2023diffusion}

\begin{equation}
\label{eq:diffuse_forward}
    p(f_t | f_{t-1}) = \mathcal{N} \left( f_t; \sqrt{1 - \alpha_t} f_{t-1}, \alpha_t \mathbf{I} \right), \forall t \in \{1, \dots, T\},
\end{equation}
where $\mathbf{I}$ is the identity matrix that has the same dimensions as $f_0$, and $\mathcal{N}(f_t;\mu,\sigma)$ is the normal distribution of mean $\mu$ and covariance $\sigma$ that generates $f_t$.

This recursive formulation allows $f_t$ to be sampled directly from
\begin{equation}
\label{eq:diffuse_forward_direct}
    p(f_t | f_0) = \mathcal{N}\!\left( f_t;\, \sqrt{\bar{\alpha}_t}\, f_0,\, (1-\bar{\alpha}_t)\,\mathbf{I} \right), \quad t \sim \mathcal{U}\{1,\dots,T\},
\end{equation}
where $\mathcal{U}$ draws a normal distribution, $\bar{\alpha}_t=\prod^{t}_{i=1}\beta_{i}$ and $\beta_t=1-\alpha_t$. Then we have
\begin{equation}
\label{eq:f_t}
f_t = \sqrt{\bar{\alpha}_t} f_0 + \sqrt{1 - \bar{\alpha}_t} \epsilon_t,
\end{equation}
where $\epsilon_t \sim \mathcal{N}(0, \mathbf{I})$, to sample $f_t$ from $p(f_t|f_0)$ in the diffusion module's training step.

\begin{figure}[t]
\centering
\ifCFGcaseOne
\includegraphics[width=0.65\linewidth]{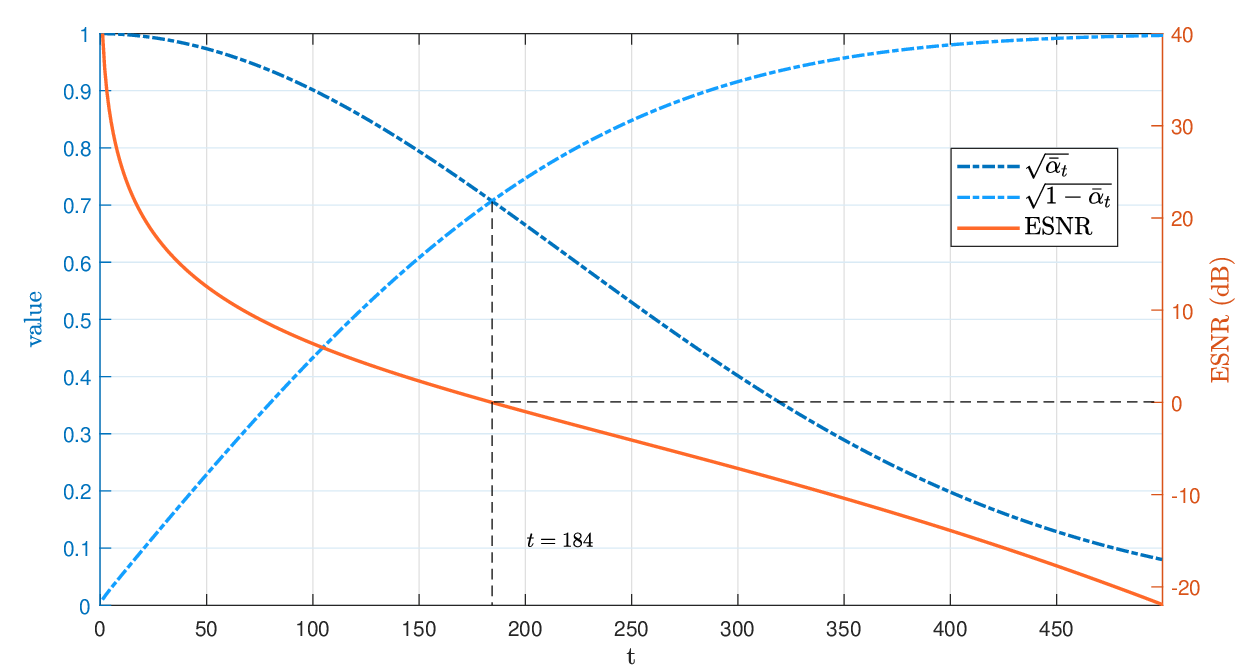}
\fi
\ifCFGcaseTwo
\includegraphics[width=\linewidth]{retultfigs/t.eps}
\fi
\caption{Changes of $\sqrt{\bar{\alpha}_t}$, $\sqrt{1 - \bar{\alpha}_t}$ and $p$ with $t$.}
\label{fig:T}
\end{figure}

To ease the training and using of diffusion module, we choose $(\alpha_t)^T_{t=1}$ to be linear increasing constants, that is
\begin{equation}
\label{eq:alpha_t}
    \alpha_t=\alpha_1+\left(\frac{\alpha_T-\alpha_1}{T-1}\right)(t-1),
\end{equation}
where $\alpha_1=10^{-4}$, $\alpha_T=2\times10^{-2}$. We train the diffusion module using $T=500$, but $T'<T$ is used in the experiment to test the system's robustness. The values of $\sqrt{\bar{\alpha}_t}$ and $\sqrt{1 - \bar{\alpha}_t}$ varying by $t$ are shown in Fig. \ref{fig:T}. We plot the ``ESNR'' line that is defined by 
\begin{equation}
    \mathcal{P} = \frac{\mathbb{E}(\|\sqrt{\bar{\alpha}_t} f_0\|^2)}{\mathbb{E}(\|\sqrt{1 - \bar{\alpha}_t} \epsilon_t\|^2)},
\label{eq:p}
\end{equation}
where $\mathcal{P}$ represents the power ratio between the signal component $\sqrt{\bar{\alpha}_t} f_0$ and noise component $\sqrt{1 - \bar{\alpha}_t} \epsilon_t$ obtained after partially adding noise, and varies with the values of $\alpha_1$, $\alpha_T$ and $T$. In our system, $f_0$ is power-normalized after generated by encoder $\Phi_\phi$. Hence, according to Eq. (\ref{eq:p}), it is equivalent to the signal $f_0$ passing through an AWGN channel with signal-to-noise ratio $\mathcal{P}$, which we refer to as ``equivalent SNR'' (ESNR). As an example, we marked the point $t = 184$ in Fig. \ref{fig:T}, where the energy of the signal component is approximately equal to that of the noise component, and the ESNR $p = 0.01$ dB $\simeq 0$ dB.

The signal $\hat{z}$ needs to be denoised first on the receiver side. Usually, a deep learning network called U-Net can be used to predict $\epsilon_t$ during the sample process. U-Net is a convolutional neural network that can effectively handle multi-scale features and reconstruct high-fidelity output during the reverse process. U-Net learns to estimate $f_0$ from $f_t$ and $t$ by minimizing the difference between the predicted noise
\begin{equation}
\label{eq:epsilon_t_1}
   \epsilon_{f_t} = \Delta_{\delta}(f_t, t)
\end{equation}
and the actual noise $\epsilon_t$, and is trained using the MSE loss of these two noises, denoted by \cite{croitoru2023diffusion}
\begin{equation}
\label{eq:loss_unet}
\mathcal{L}_{\text{U-Net}} = \mathbb{E}_{t \sim [1, T]} \mathbb{E}_{f_0 \sim p(f_0)} \mathbb{E}_{\epsilon_t \sim \mathcal{N}(0, \mathbf{I})} \left\| \epsilon_{f_t} - \epsilon_t \right\|^2.
\end{equation}

With a well-trained U-Net, the goal is to progressively denoise $f_t$ and to recover $f_0$ starting from a latent variable $f_T \sim \mathcal{N}(0, \mathbf{I})$. Specifically, on step $t-1$, given $f_t$ and $\epsilon_{t}$, we have 
\begin{equation}
\label{eq:f_t-1_final}
    f_{t-1} = \sqrt{\bar{\alpha}_{t-1}} \left( \frac{f_t - \sqrt{1 - \bar{\alpha}_t} \epsilon_{f_t}}{\sqrt{\bar{\alpha}_t}} \right) + \sqrt{1 - \bar{\alpha}_{t-1}} \epsilon_{f_t}.
\end{equation}
To generate samples efficiently, DDIM uses a deterministic, stride-based reverse schedule over a subset $\{t_1>\cdots>t_N\}\subseteq\{1,\dots,T\}$ with $N\!\ll\!T$. In our paired design, the receiver’s schedule \textit{snaps} the target step $T'$ to the nearest element in this grid.

A diffusion model learns the distribution $p(f_0)$ rather than an exact per-sample inverse. When initialized from pure noise, it only ensures sampling from the learned distribution and thus cannot guarantee recovery of the specific input instance. However, communication demands recoverability of the transmitted semantics. Our paired design resolves this tension: the transmitter halts the forward diffusion early (self-noising at $T'$) so that sufficient task-relevant structure remains in $f_{T'}$, and the receiver runs the matching reverse schedule to deterministically remove as much of the injected noise as possible.

\textit{{Transmitter-side self-noising.}} Let $0\leq T' \leq T$, and perform
\begin{equation}
\label{eq:diffusion_T_prime}
f_{T'} = \sqrt{\bar{\alpha}_{T'}} f_0 + \sqrt{1 - \bar{\alpha}_{T'}} \epsilon_{T'} ,
\end{equation}
in the self-noising module $\Psi$ before transmitting. In the receiver, $\Delta_\delta$ tries to recover $f_0$ using $T'$ steps' iteration that is the same as Eq. (\ref{eq:f_t-1_final}). This gives a way to control the weight of ``task-relevant information'' $\sqrt{\bar{\alpha}_{T'}} f_0$ in the distorted output. The larger $T'$ is, the less ``task-relevant information'' in $f_{T'}$, and nearly no guidance involved when $T'=T$ (in which diffusion module generates $f_0$ without specific order). Since Eq. (\ref{eq:diffusion_T_prime}) is valid for any $0\leq T' \leq T$, we can train the diffusion module to denoise from $T$ and deploy the well-trained module that is set by $T'$ when transmitting the signal, while receiver recovers the estimation of $f_0$ without other explicit guidance required. In this way, the choice of $T'$ controls the distortion of the transmitter, hence affects the difficulty for adversaries to estimate the original images with task information in them. 

The receiver can operate in two forms. The basic version corresponds to DiffSem, where the recovered feature is directly used for task inference. The enhanced version corresponds to E-DiffSem (``E'' for ``enhance''), where the receiver further reuses its own soft prediction as self-referential guidance to refine the result.
Specifically, the classifier module $\Theta_{\theta}$ generates the output $\hat{y}$ bypassing the diffusion module $\Delta_\delta$ at first. The accuracy of $\hat{y}$ is higher when distortion on $f$ is lower or channel condition is favorable, which means that $\hat{y}$ could be more helpful in such cases. Then $\hat{y}$ can be used to guide the diffusion module $\Delta_\delta$ in the denoising process (Fig. \ref{fig:whole-system}). Different from calculating the accuracy of $\hat{y}$, here we use the soft version of $\hat{y}$ that all the category info in the vector will be utilized as guidance. Next, $\hat{z}$ and $\hat{y}$ are fed into $\Delta_\delta$ to produce $\hat{y}'$, which is further reintroduced into $\Delta_\delta$ for refined denoising. By iterating this feedback loop multiple times, the accuracy of $\mathcal{R}$ is can be further improved. Consistent with guidance-based DDIM implementation, we provide explicit external control to $\Delta_\delta$ by modifying the noise estimation in Eq. (\ref{eq:epsilon_t_1}) to
\begin{equation}
\label{eq:epsilon_t_2}
    \epsilon_{(f_t, \hat{y})} = \Delta_{\delta}(f_t, \hat{y}, t),
\end{equation}
where $\hat{y}$ serves as self-referential guidance. During the training, a classifier-free DDIM is trained with ground-truth labels $y$ by replacing the term $\epsilon_{f_t}$ with $\epsilon_{(f_t, y)}$ in Eq. (\ref{eq:loss_unet}). During sampling, $\hat{y}$ replaces $y$ to adaptively guide denoising. This soft label-embedding method ensures that:
\begin{itemize}
    \item If $\hat{y}$ is more accurate, the denoising effect is enhanced;
    \item If $\hat{y}$ is unreliable (e.g., due to excessive self-noising in $\Psi$ or poor channel conditions), the system performance remains stable without degradation.
\end{itemize}

Although the external guidance from $\hat{y}$ may be suboptimal in extreme cases, it improves robustness in most scenarios. Detailed experimental validation of this iterative denoising framework is provided in Sec. IV.

\subsection{Diffusion Module Under Wireless Channel}

Here we discuss the impact of wireless channel on the diffusion module in our system. To facilitate discussion, let $f_{T'}=z'$, then the receiver gets
\begin{equation}
\begin{split}
    \hat{z}' &= z' + \epsilon_{\text{ch}}= \left(\sqrt{\bar{\alpha}_{T'}} f_0 + \sqrt{1 - \bar{\alpha}_{T'}} \epsilon_{T'}\right) + \epsilon_{\text{ch}},
\end{split}
\end{equation}
where $\epsilon_{\text{ch}} \sim \mathcal{N}(0, \sigma_{\text{ch}}^2 \mathbf{I})$ represents channel noise. Obviously, the channel noise term adds extra distortion to $f_{T'}$ and changes its distribution, thus affects the sampling process at the receiver. To tackle this issue, we introduce a noise margin for possible channel interference during the forward process.

Suppose we need to distort $f_0$ to $f_{T'}$ which will be transmitted at $\mathcal{T}$, given specific timestep $T'$. Let $0\leq T''<T'$ and
\begin{equation}
    f_{T''} = \sqrt{\bar{\alpha}_{T'}} f_0 + \sqrt{1 - \bar{\alpha}_{T''}} \epsilon_{T''},
\end{equation}
where $\epsilon_{T''} \sim \mathcal{N}(0, \mathbf{I})$. Note that the first term of $f_{T''}$ remains the same as that of $f_{T'}$. For simplicity, let $\epsilon_{T'}=\epsilon_{T''}=\epsilon$. Then, $\mathcal{T}$ transmits $f_{T''}$ to the wireless channel, and receiver obtains
\begin{equation}
\begin{split}
    \hat{z}'' &= z'' +  \epsilon_{\text{ch}} = \sqrt{\bar{\alpha}_{T'}} f_0 + \sqrt{1 - \bar{\alpha}_{T''}} \epsilon + \epsilon_{\text{ch}}.
\label{eq:hat_z_primes}
\end{split}
\end{equation}
If 
\begin{equation}
\label{eq:T_double_prime_1}
    \mathbb{E}\left(\|\sqrt{1 - \bar{\alpha}_{T'}} \epsilon\|^2\right) = \mathbb{E}\left(\|\sqrt{1 - \bar{\alpha}_{T''}} \epsilon + \epsilon_{\text{ch}}\|^2\right),
\end{equation}
then $\mathbb{E}\left(\|\hat{z}''\|^2\right)=\mathbb{E}\left(\|z'\|^2\right)$, which is equivalent to that $\mathcal{R}$ receives $f_{T'}$. So we can transmit $z''$ while maintaining the distribution after the wireless transmission, and then the diffusion module can recover the estimation of $f_0$ without further ambiguity.

Now we need to find $T''$. Let $\gamma$ denote channel's signal-to-noise power ratio (SNR), then we have
\begin{equation}
\label{eq:channel_noise_2}
\begin{split}
    \sigma^{2}_{\text{ch}} &= \frac{\mathbb{E}\left(\|z''\|^{2}\right)}{\gamma} = \frac{\bar{\alpha}_{T'}\mathbb{E}\left(\|f_0\|^2\right) + (1-\bar{\alpha}_{T''}) D}{\gamma},
\end{split}
\end{equation}
where $D$ is the length of $\epsilon$ and $\mathbb{E}[\|\epsilon\|^2]=D$.

By considering Eq. (\ref{eq:hat_z_primes}) to (\ref{eq:channel_noise_2}), we can derive that
\begin{equation}
\label{eq:alpha_t_primes}
    \bar{\alpha}_{T''} = \frac{\bar{\alpha}_{T'} \left(\gamma + \mathbb{E}\left(\|f_0\|^2\right)\right) + D}{\gamma + D}.
\end{equation}
Details can be found in Appendix \ref{apx:a}. In Eq. (\ref{eq:alpha_t_primes}), $\bar{\alpha}_t$ can be regarded as a discrete function of $t$ according to Eq. (\ref{eq:alpha_t}), so $T''$ can be approximated given $T'$, $\gamma$ and $f_0$ \footnote{$f_0$ and $\epsilon_{\text{ch}}$ are of the same length.}. In this way, our diffusion module can effectively handle the effects of the wireless channel\footnote{In the implementation, we find the maximum $T''$ that $\alpha_{T''}$ is no larger than the right side of Eq. (\ref{eq:alpha_t_primes}). If not, set $T''=0$; generally, it means that $T'$ is small enough, or $\gamma$ is larger enough. Both situations mean that there will not be significant distortion on $f_0$.}.

\textit{Remark on SNR estimation.}
In implementation, we substitute an estimate $\hat{\gamma}$ for $\gamma$ in Eq.~(\ref{eq:channel_noise_2})--(\ref{eq:alpha_t_primes}). Since the deterministic stride-based DDIM sampler uses a discrete timestep grid, moderate perturbations in $\hat{\gamma}$ typically lead only to a small change in the selected reverse step $T''$, and in many cases do not alter the snapped timestep index. As a result, precise SNR estimation is not critical for our method. We further verify this empirically on CIFAR-10 under multiplicative SNR estimation errors, and provide a more detailed discussion in Appendix~\ref{apx:b}.

\begin{algorithm}[t]
\caption{System Training Preparation}
\label{alg:pretrain}
\begin{algorithmic}[1]
\STATE \textbf{Initialization:} Initialize weights $ W $ and biases $ b $ in $\mathcal{T}$, $\mathcal{R}$, $\mathcal{A}_a$ and $\bar{\Theta}_{\bar{\theta}}$.
\STATE \textbf{Input:} $x \in X$ and $\dot{x} \in \dot{X}$.
\STATE \textbf{Stage 1: Training $\Phi_\phi$ and $\Theta_\theta$.}
\STATE \hspace{0.3cm} $\Phi_{\phi}(x) = z$.
\STATE \hspace{0.3cm} Transmit $z$ over wireless channel and get $\hat{z}$.
\STATE \hspace{0.3cm} $\Theta_{\theta}(z) = \hat{y}$.
\STATE \hspace{0.3cm} Calculate $\mathcal{L}_{\text{CE}}$ by Eq. (\ref{eq:loss_ce}) and update $\phi$, $\theta$.
\STATE \textbf{Stage 2: Training $\Delta_\delta$}
\STATE \hspace{0.3cm} $\Phi_\phi(x) = f$.
\STATE \hspace{0.3cm} Train U-Net by minimizing $\mathcal{L}_{\text{U-Net}}$ using Eq. (\ref{eq:loss_unet}), and update $\delta$.
\STATE \textbf{Stage 3: Training $\mathcal{A}_{a}$ and $\bar{\Theta}_{\bar{\theta}}$}
\STATE \hspace{0.3cm} $\Phi_{\phi}(\dot{x}) = \dot{z}$.
\STATE \hspace{0.3cm} Transmit $\dot{z}$ over wireless channel and get $\hat{\dot{z}}$.
\STATE \hspace{0.3cm} $\mathcal{A}_{a}(\hat{\dot{z}}) = \hat{\dot{x}}$.
\STATE \hspace{0.3cm} Directly distort $x$ to get $\ddot{x}$.
\STATE \hspace{0.3cm} $\bar{\Theta}_{\bar{\theta}}\left([x; \dot{x}; \hat{x}; \ddot{x}]\right) = \hat{\dot{y}}$.
\STATE \hspace{0.3cm} Use $\hat{\dot{x}}$ to calculate MSE loss by Eq. (\ref{eq:loss_mse}); use $\hat{\dot{y}}$ to calculate CE loss by Eq. (\ref{eq:loss_ce}).
\STATE \hspace{0.3cm} Update $a$ and $\bar{\theta}$.
\STATE \textbf{Output:} Pre-trained network $\mathcal{T}$, $\mathcal{R}$, $\mathcal{A}_a$ and $\bar{\Theta}_{\bar{\theta}}$.
\end{algorithmic}
\end{algorithm}

\subsection{System Training Procedure}
\label{Sec:System Training Procedure}

Here we introduce the training procedure used in our experiments. As shown in Fig. \ref{fig:whole-system}, $t$ is used as a global parameter that controls the distortion at the transmitter and helps the denoising at the receiver. All modules are neural-based except for self-noising module and wireless channel modules. The self-noising module can be regarded as a channel compensator, ensuring that the signals received by the diffusion module appear as if they have passed through a Gaussian channel, thereby conforming to the input requirements for denoising. 
We use three datasets, namely MNIST, CIFAR-10, and CelebA, to evaluate the proposed system from relatively simple to more complex visual scenarios. The corresponding model implementations may vary across datasets due to differences in input dimensionality, visual complexity, and task difficulty; however, such variations do not change the overall training logic of the proposed framework. Therefore, this subsection focuses on the unified training procedure, while the dataset-specific module implementations are deferred to the experiment section.

In this paper, we assume that the receiver performs an image classification task, so that $\mathcal{R}$ is trained using the cross-entropy (CE) loss
\begin{equation}
\label{eq:loss_ce}
    \mathcal{L}_{\text{CE}}(\phi,\theta) = -\sum_{i} p(y_i)\, \log q(\hat{y}_i),
\end{equation}
where $p(y)$ and $q(\hat{y})$ denote the true and predicted label distributions, respectively. The classification accuracy is the fraction of correctly classified inputs.
The strong classifier $\bar{\Theta}_{\bar{\theta}}$ also produces task-level classification outputs, so that the receiver-side accuracy and the attacker-side semantic leakage can be evaluated under the same task criterion and are therefore directly comparable.

\begin{algorithm}[t]
\caption{Receiver Fine-tuning}
\label{alg:finetuning}
\begin{algorithmic}[1]
\STATE \textbf{Initialization:} Load pre-trained networks $\mathcal{T}$, $\mathcal{R}$ and $\mathcal{A}$.
\STATE \textbf{Input:} $x \in X$.
\STATE \textbf{Given} $0 \leq T' \leq T$, 
\STATE \hspace{0.3cm} Find $T''$ by Eq. (\ref{eq:alpha_t_primes}).
\STATE \hspace{0.3cm} $\mathcal{T}_{\phi}(x,T',T'') = z$.
\STATE \hspace{0.3cm} Transmit $z$ over wireless channel and get $\hat{z}$.
\STATE \hspace{0.3cm} $\Theta_\theta(\hat{z}, T') = \hat{y}$.
\STATE \hspace{0.3cm} \textbf{if} using E-DiffSem: 
\STATE \hspace{0.6cm} \textbf{while} accuracy of $\hat{y}$ increases:
\STATE \hspace{0.9cm} \textbf{for} $t \in \{t_1=0<t_2< \dots<t_N=T'\}$:
\STATE \hspace{1.2cm} Denoise $\hat{z}$ by Eq.~(\ref{eq:f_t-1_final}) and Eq.~(\ref{eq:epsilon_t_2}) from $f_t$ to $f_{t-1}$, with $\hat{y}$ involved.
\STATE \hspace{0.9cm} \textbf{end for} and get $\hat{f}$. 
\STATE \hspace{0.9cm} $\Theta_\theta(\hat{f},\hat{z}, T') = \hat{y}$.
\STATE \hspace{0.6cm} \textbf{end while}
\STATE \hspace{0.3cm} \textbf{elif} using DiffSem:
\STATE \hspace{0.6cm} \textbf{for} $t \in \{t_1=0<t_2< \dots<t_N=T'\}$:
\STATE \hspace{0.9cm} Denoise $\hat{z}$ by Eq.~(\ref{eq:f_t-1_final}) and Eq.~(\ref{eq:epsilon_t_1}) from $f_t$ to $f_{t-1}$.
\STATE \hspace{0.6cm} \textbf{end for} and get $\hat{f}$. 
\STATE \hspace{0.6cm} $\Theta_\theta(\hat{f},\hat{z}, T') = \hat{y}$.
\STATE \hspace{0.3cm} \textbf{end if}
\STATE \hspace{0.3cm} Calculate $\mathcal{L}_{\text{CE}}(y,\hat{y})$ by Eq. (\ref{eq:loss_ce}) and update $\theta$.
\STATE Change $T'$ and return to Line 3 to repeat.
\STATE \textbf{Output:} Finetuned network $\Theta_\theta$.
\end{algorithmic}
\end{algorithm}

For ease of tracking, we divide the training procedure into two parts. The first part consists of three stages, each designed to prepare a specific module, as outlined in Algorithm \ref{alg:pretrain}. We assume that when an adversary $\mathcal{A}_a$ attempts to ``join'' a transmitter-receiver pair, the transmitter $\mathcal{T}$ and receiver $\mathcal{R}$ have already been fully trained and are successfully deployed on the transmission nodes. Consequently, the training of $\mathcal{A}_a$ must occur after the completion of the $\mathcal{T}$-$\mathcal{R}$ pair. As shown in Fig. \ref{fig:adversary-system}, we use the well-trained transmitter to train the adversary, where receiver is not involved here. Hence, the performance of $\mathcal{A}_a$ will inherently depend on the training outcomes of $\mathcal{T}$. By optimizing the training of the $\mathcal{T}$-$\mathcal{R}$ pair and subsequently training $\mathcal{A}_a$ to its fullest potential, we aim to limit the amount of valid information that the adversary can extract under the given dataset and system. The training of strong classifier $\bar{\Theta}_{\bar{\theta}}$ can be carried out simultaneously with $\mathcal{A}_a$, or after the training of $\mathcal{A}_a$ is finished.

The second part is fine-tuning the receiver given fixed $\Phi_\phi$ and $\Psi$, as described in Algorithm~\ref{alg:finetuning}. We implemented DiffSem, which performs a single-pass pipeline processing on the receiver's input, and an enhanced version named E-DiffSem, where ``E'' stands for ``enhance''. E-DiffSem utilizes soft $\hat{y}$ as self-referential guidance to improve the performance of denoising. Specifically, self-referential $\hat{y}$ is iteratively utilized until the accuracy of $\hat{y}$ reaches stability. For both methods, the diffusion steps $N$ are set to $\min (T', 50)$. Experimental results and discussions will be elaborated in the next section.

\subsection{Design Guidance}
\label{subsec:mechanism}

\begin{figure}[!t]
    \centering
    \includegraphics[width=\linewidth]{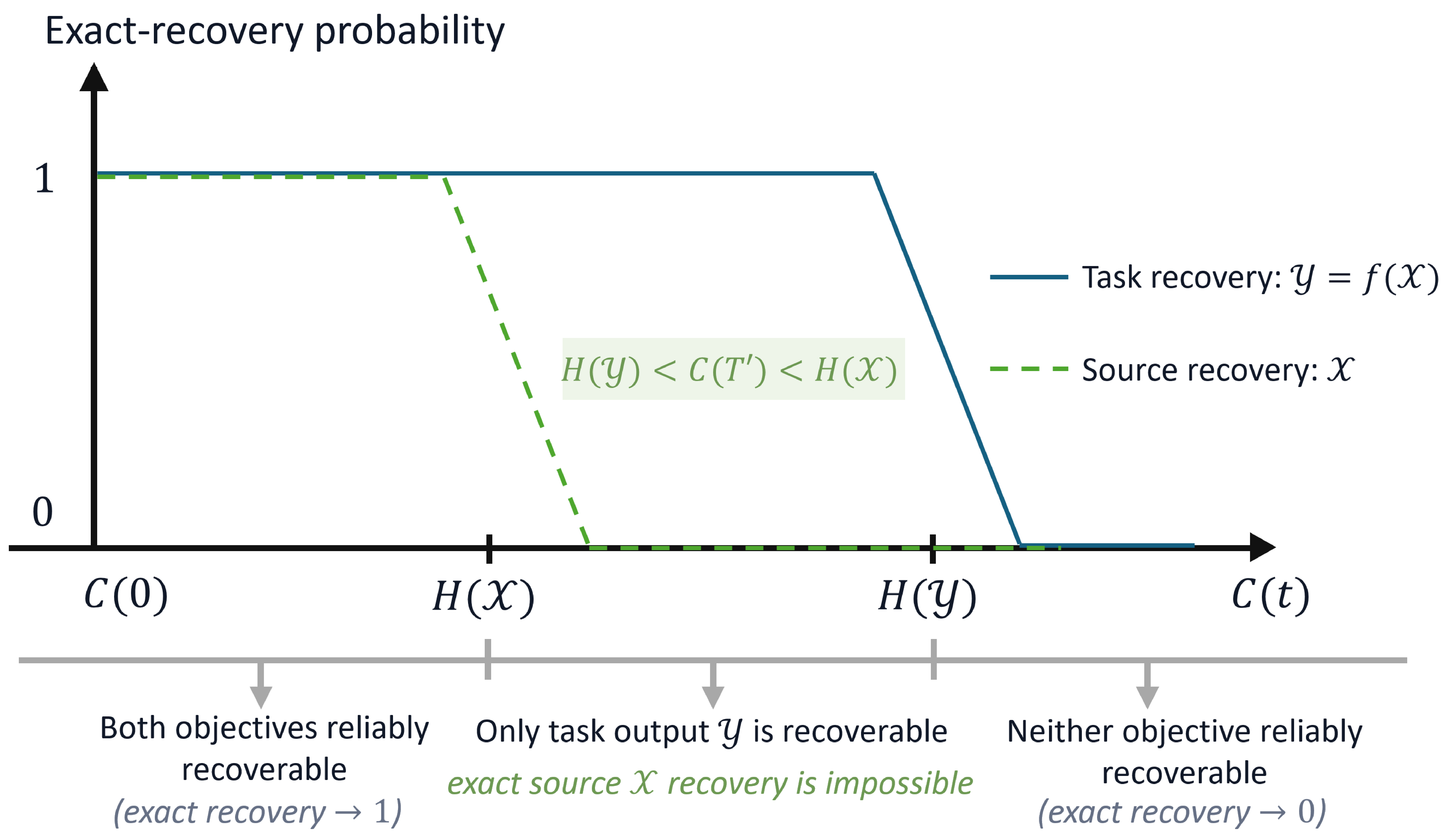}
    \caption{
    Simplified asymptotic separation between task recovery and exact source recovery under transmitter-side self-noising.
    }
    \label{fig:capacity_separation}
\end{figure}

Here we provide an asymptotic source--channel interpretation of why transmitter-side self-noising is effective. 
\\\indent
Consider a discrete source random variable $\mathcal{X}$ with finite support, corresponding to the aforementioned image set $X$, and let $\mathcal{Y}=g(\mathcal{X})$ denote the ground-truth output random variable for a prescribed downstream task $g$.
The transmitter communicates over a self-noising then over an AWGN channel $\mathcal{C}$, the output are observed by two terminals denoted by $\mathcal{R}$ and $\mathcal{A}$. 
At this stage, we do not distinguish between legitimate and adversarial roles; these names indicate only their respective decoding objectives. Terminal $\mathcal{R}$ aims to recover $\mathcal{Y}$, whereas terminal $\mathcal{A}$ aims to recover $\mathcal{X}$. 
\\\indent
The AWGN channel and the transmitter-side self-noising jointly induce a tunable channel, where $C(T')$ denotes its effective capacity per channel use with parameter $T'$. As shown in Fig. \ref{fig:T}, $C(T')$ is non-increasing and continues with $T'$.
\\\indent
For $n$ independent source samples, consider the exact recovery of $\mathcal{X}^n$. According to the source--channel separation theorem \cite{arimoto1973converse, cover2006elements}, if
\begin{equation}
    H(\mathcal{X})<C(T')
\end{equation}
then there exists a sequence of source--channel codes for which the block error probability of recovering $\mathcal{X}^n$ approaches zero as $n\rightarrow\infty$. Conversely, if
\begin{equation}
    H(\mathcal{X})>C(T'),
\end{equation}
then the probability of exact block recovery approaches zero. The same argument applies to the exact recovery of $\mathcal{Y}$.
\\\indent
Suppose that $\mathcal{C}$ is sufficiently capable of supporting source recovery, that is
\begin{equation}
    C(0)>H(\mathcal{X}),
\end{equation}
whereas the maximum considered self-noising level reduces the effective capacity below that required for task recovery,
\begin{equation}
    C(T'_{\max})<H(\mathcal{Y}).
\end{equation}
Since $H(\mathcal{Y}) = H(g(\mathcal{X})) < H(\mathcal{X})$, there then exist an interval of $T'$ over which
\begin{equation}
    H(\mathcal{Y})
    <
    C(T')
    <
    H(\mathcal{X}).
\label{eq:capacity_separation}
\end{equation}
For any $T'$ in this interval, the tunable channel has sufficient capacity to support reliable recovery of $\mathcal{Y}$, but insufficient capacity to support reliable recovery of $\mathcal{X}$, as illustrated in Fig.~\ref{fig:capacity_separation}.
\\\indent
This interpretation clarifies the roles of the two diffusion processes in DiffSem. Transmitter-side partial forward diffusion controls the degradation of the equivalent channel through the self-noising level, while receiver-side reverse diffusion helps the legitimate receiver recover the task-relevant representation from the corrupted observation. DiffSem seeks to operate in the intermediate regime of Eq.~(\ref{eq:capacity_separation}), where sufficient information remains for the prescribed task but not for reliable lossless recovery of the complete source.

\section{Experiments and Results}
\label{sec: Experiments and Results}

In this section, we will first explain the module design in our system and introduce the details of experiment initialization. Next, we present the results and discussions based on our experimental findings.

\subsection{Module Implementation and baselines}

\ifCFGcaseOne
\begin{table*}[!t]
\small
\begin{center}
\caption{Summary of Non-diffusion Modules}
\label{table:codec_modules_three_datasets}
\resizebox{\linewidth}{!}{%
\begin{tabular}{c|c|c|c|c|c}
\textbf{Dataset} & \textbf{\makecell{Image\\$\rightarrow$ Latent}} & \textbf{Encoder $\Phi$} & \textbf{\makecell{Receiver\\Classifier $\Theta$}} & \textbf{\makecell{Third-party\\Classifier $\bar{\Theta}$}} & \textbf{Adversary $\mathcal{A}$} \\
\hline
MNIST &
\makecell{$1\times 28\times 28$\\$\rightarrow 1\times 8\times 8$} &
\makecell{\makecell{Conv2d$(1,16)$, \\[-0.66em]stride 2 + LReLU}\\\makecell{Conv2d$(16,32)$, \\[-0.66em]stride 2 + LReLU}\\Conv2d$(32,1)$ + LayerNorm} &
\makecell{Concat$(\hat{z},\hat{f},t)$\\Linear + ReLU\\Linear ($10$ classes)} &
\makecell{Linear + Tanh\\Linear + Tanh\\Linear + Sigmoid} &
\makecell{Flatten\\Linear + Tanh\\Linear\\Reshape to $1\times 28\times 28$} \\
\hline
CIFAR-10 &
\makecell{$3\times 32\times 32$\\$\rightarrow 3\times 8\times 8$} &
\makecell{Conv2d$(3,16)$ + LReLU\\\makecell{Conv2d$(16,32)$, \\[-0.66em]stride 2 + LReLU}\\Residual Block\\\makecell{Conv2d$(32,3)$, \\[-0.66em]stride 2 + LReLU + LayerNorm}} &
\makecell{Concat$(\hat{z},\hat{f},t)$\\Conv + Tanh\\Strided Conv + Tanh\\Residual Block\\\makecell{Global Pool + \\[-0.66em]Linear ($10$ classes)}} &
\makecell{Strided Conv + ReLU\\Strided Conv + ReLU\\Residual Block\\Global Pool\\Linear + ReLU\\Linear ($10$ classes)} &
\makecell{Conv + Tanh\\Deconv + Tanh\\Residual Block\\Deconv + Tanh\\Residual Block\\Conv + Tanh} \\
\hline
CelebA &
\makecell{$3\times 128\times 128$\\$\rightarrow 3\times 32\times 32$} &
\makecell{Conv2d$(3,32)$ + BN + ReLU\\Conv2d$(32,32)$ + BN + ReLU\\\makecell{Conv2d$(32,64)$, stride 2 + \\[-0.66em]BN + ReLU}\\\makecell{Conv2d$(64,96)$, stride 2 + \\[-0.66em]BN + ReLU}\\Conv2d$(96,64)$ + BN + ReLU\\Conv2d$(64,3)$ + LayerNorm} &
\makecell{MobileNetV2\\\makecell{Input \\[-0.66em]$3\times 32\times 32$ latent}\\Linear head ($3$ logits)} &
\makecell{ResNet-18\\\makecell{Input \\[-0.66em]$3\times 128\times 128$ image}\\Linear head ($3$ logits)} &
\makecell{Conv2d$(3,128)$ + BN + ReLU\\Residual Block\\\makecell{Upsample +\\[-0.66em]Conv2d$(128,192)$ + BN + ReLU}\\Residual Block\\\makecell{Upsample + \\[-0.66em]Conv2d$(192,160)$ + BN + ReLU}\\Conv2d$(160,96)$ + ReLU\\Conv2d$(96,3)$ + Tanh} \\
\end{tabular}%
}
\end{center}
\end{table*}

\begin{table*}[!t]
\small
\begin{center}
\caption{Summary of Dataset-specific U-Nets}
\label{table:unet_modules_three_datasets}
\resizebox{\linewidth}{!}{%
\begin{tabular}{c|c|c|c}
\textbf{Component} & \textbf{MNIST} & \textbf{CIFAR-10} & \textbf{CelebA} \\
\hline
\textbf{Latent Domain}
& $1\times 8\times 8$
& $3\times 8\times 8$
& $3\times 32\times 32$ \\
\hline
\textbf{DiffSem U-Net}
& \makecell{Time EmbedFC ($2\times$)\\ResidualConvBlock $(128)$\\UnetDown $(128)$\\UnetDown $(256)$\\AvgPool2d + GELU\\\makecell{ConvTranspose2d +\\[-0.66em] GN + ReLU $(256)$}\\UnetUp $(128)$\\UnetUp $(128)$\\\makecell{Conv2d + GroupNorm +\\[-0.66em] ReLU + Conv2d}}
& \makecell{Time EmbedFC ($2\times$)\\ResidualConvBlock $(3,128)$\\UnetDown $(128)$\\UnetDown $(256)$\\AvgPool2d + GELU\\\makecell{ConvTranspose2d +\\[-0.66em] GN + ReLU $(256)$}\\UnetUp $(128)$\\UnetUp $(128)$\\\makecell{Conv2d + GroupNorm +\\[-0.66em] ReLU + Conv2d$(128,3)$}}
& \makecell{Time embedding (sinusoidal + MLP)\\Conv2d$(3,96)$\\ResBlock$\times 2$\\Downsample$(96,128)$ + ResBlock$\times 2$\\Downsample$(128,192)$ + ResBlock$\times 2$\\Mid ResBlock + Self-Attention + ResBlock\\Upsample$(192,128)$ + ResBlock$\times 2$\\Upsample$(128,96)$ + ResBlock$\times 2$\\GroupNorm + SiLU + Conv2d$(96,3)$} \\
\hline
\textbf{E-DiffSem U-Net}
& \makecell{Same U-Net backbone as DiffSem\\Time EmbedFC ($2\times$)\\Context EmbedFC ($2\times$)}
& \makecell{Same U-Net backbone as DiffSem\\Time EmbedFC ($2\times$)\\Context EmbedFC ($2\times$)}
& \makecell{Same U-Net backbone as DiffSem\\Time embedding (sinusoidal + MLP)\\Attribute embedding MLP ($3\rightarrow64\rightarrow256$)\\Time and condition fused in every ResBlock} \\
\end{tabular}%
}
\end{center}
\end{table*}
\fi

\ifCFGcaseTwo
\begin{table*}[!t]
\small
\renewcommand\arraystretch{1.4}
\renewcommand\cellset{\renewcommand\arraystretch{1.4}}
\begin{center}
\caption{Summary of Non-diffusion Modules}
\label{table:codec_modules_three_datasets}
\resizebox{\linewidth}{!}{%
\begin{tabular}{c|c|c|c|c|c}
\textbf{Dataset} & \textbf{\makecell{Image\\$\rightarrow$ Latent}} & \textbf{Encoder $\Phi$} & \textbf{\makecell{Receiver\\Classifier $\Theta$}} & \textbf{\makecell{Third-party\\Classifier $\bar{\Theta}$}} & \textbf{Adversary $\mathcal{A}$} \\
\hline
MNIST &
\makecell{$1\times 28\times 28$\\$\rightarrow 1\times 8\times 8$} &
\makecell{\makecell{Conv2d$(1,16)$, \\[-0.44em]stride 2 + LReLU}\\\makecell{Conv2d$(16,32)$, \\[-0.44em]stride 2 + LReLU}\\Conv2d$(32,1)$ + LayerNorm} &
\makecell{Concat$(\hat{z},\hat{f},t)$\\Linear + ReLU\\Linear ($10$ classes)} &
\makecell{Linear + Tanh\\Linear + Tanh\\Linear + Sigmoid} &
\makecell{Flatten\\Linear + Tanh\\Linear\\Reshape to $1\times 28\times 28$} \\
\hline
CIFAR-10 &
\makecell{$3\times 32\times 32$\\$\rightarrow 3\times 8\times 8$} &
\makecell{Conv2d$(3,16)$ + LReLU\\\makecell{Conv2d$(16,32)$, \\[-0.44em]stride 2 + LReLU}\\Residual Block\\\makecell{Conv2d$(32,3)$, \\[-0.44em]stride 2 + LReLU + LayerNorm}} &
\makecell{Concat$(\hat{z},\hat{f},t)$\\Conv + Tanh\\Strided Conv + Tanh\\Residual Block\\\makecell{Global Pool + \\[-0.44em]Linear ($10$ classes)}} &
\makecell{Strided Conv + ReLU\\Strided Conv + ReLU\\Residual Block\\Global Pool\\Linear + ReLU\\Linear ($10$ classes)} &
\makecell{Conv + Tanh\\Deconv + Tanh\\Residual Block\\Deconv + Tanh\\Residual Block\\Conv + Tanh} \\
\hline
CelebA &
\makecell{$3\times 128\times 128$\\$\rightarrow 3\times 32\times 32$} &
\makecell{Conv2d$(3,32)$ + BN + ReLU\\Conv2d$(32,32)$ + BN + ReLU\\\makecell{Conv2d$(32,64)$, stride 2 + \\[-0.44em]BN + ReLU}\\\makecell{Conv2d$(64,96)$, stride 2 + \\[-0.44em]BN + ReLU}\\Conv2d$(96,64)$ + BN + ReLU\\Conv2d$(64,3)$ + LayerNorm} &
\makecell{MobileNetV2\\\makecell{Input \\[-0.44em]$3\times 32\times 32$ latent}\\Linear head ($3$ logits)} &
\makecell{ResNet-18\\\makecell{Input \\[-0.44em]$3\times 128\times 128$ image}\\Linear head ($3$ logits)} &
\makecell{Conv2d$(3,128)$ + BN + ReLU\\Residual Block\\\makecell{Upsample +\\[-0.44em]Conv2d$(128,192)$ + BN + ReLU}\\Residual Block\\\makecell{Upsample + \\[-0.44em]Conv2d$(192,160)$ + BN + ReLU}\\Conv2d$(160,96)$ + ReLU\\Conv2d$(96,3)$ + Tanh} \\
\end{tabular}%
}
\end{center}
\end{table*}

\begin{table*}[!t]
\small
\renewcommand\arraystretch{1.35}
\renewcommand\cellset{\renewcommand\arraystretch{1.35}}
\begin{center}
\caption{Summary of Dataset-specific U-Nets}
\label{table:unet_modules_three_datasets}
\resizebox{\linewidth}{!}{%
\begin{tabular}{c|c|c|c}
\textbf{Component} & \textbf{MNIST} & \textbf{CIFAR-10} & \textbf{CelebA} \\
\hline
\textbf{Latent Domain}
& $1\times 8\times 8$
& $3\times 8\times 8$
& $3\times 32\times 32$ \\
\hline
\textbf{DiffSem U-Net}
& \makecell{Time EmbedFC ($2\times$)\\ResidualConvBlock $(128)$\\UnetDown $(128)$\\UnetDown $(256)$\\AvgPool2d + GELU\\\makecell{ConvTranspose2d +\\[-0.44em] GN + ReLU $(256)$}\\UnetUp $(128)$\\UnetUp $(128)$\\\makecell{Conv2d + GroupNorm +\\[-0.44em] ReLU + Conv2d}}
& \makecell{Time EmbedFC ($2\times$)\\ResidualConvBlock $(3,128)$\\UnetDown $(128)$\\UnetDown $(256)$\\AvgPool2d + GELU\\\makecell{ConvTranspose2d +\\[-0.44em] GN + ReLU $(256)$}\\UnetUp $(128)$\\UnetUp $(128)$\\\makecell{Conv2d + GroupNorm +\\[-0.44em] ReLU + Conv2d$(128,3)$}}
& \makecell{Time embedding (sinusoidal + MLP)\\Conv2d$(3,96)$\\ResBlock$\times 2$\\Downsample$(96,128)$ + ResBlock$\times 2$\\Downsample$(128,192)$ + ResBlock$\times 2$\\Mid ResBlock + Self-Attention + ResBlock\\Upsample$(192,128)$ + ResBlock$\times 2$\\Upsample$(128,96)$ + ResBlock$\times 2$\\GroupNorm + SiLU + Conv2d$(96,3)$} \\
\hline
\textbf{E-DiffSem U-Net}
& \makecell{Same U-Net backbone as DiffSem\\Time EmbedFC ($2\times$)\\Context EmbedFC ($2\times$)}
& \makecell{Same U-Net backbone as DiffSem\\Time EmbedFC ($2\times$)\\Context EmbedFC ($2\times$)}
& \makecell{Same U-Net backbone as DiffSem\\Time embedding (sinusoidal + MLP)\\Attribute embedding MLP ($3\rightarrow64\rightarrow256$)\\Time and condition fused in every ResBlock} \\
\end{tabular}%
}
\end{center}
\end{table*}
\fi

We summarize the non-diffusion modules in Table \ref{table:codec_modules_three_datasets} and the dataset-specific diffusion backbones in Table \ref{table:unet_modules_three_datasets}. Specifically, MNIST compresses a $1\times 28\times 28$ image to a $1\times 8\times 8$ latent, CIFAR-10 compresses a $3\times 32\times 32$ image to a $3\times 8\times 8$ latent, and CelebA first resizes the image to $3\times 128\times 128$ and then maps it to a $3\times 32\times 32$ latent. Accordingly, the compression rates are $8.16\%$ for MNIST and $6.25\%$ for both CIFAR-10 and CelebA. In all three cases, the encoder output is power-normalized before entering the self-noising/channel module, so that the transmitted semantic feature satisfies the power constraint and no extra normalization is required afterwards.
\\\indent We have two classifiers in each system, $\Theta_{\theta}$ and $\bar{\Theta}_{\bar{\theta}}$. The receiver classifier performs direct semantic recognition in the classifier-only branch and also provides the semantic hypothesis used by E-DiffSem. $\bar{\Theta}_{\bar{\theta}}$ learns jointly from the clean image $x$, the adversarial reconstruction $\hat{x}$, the direct self-noised image $\dot{x}$, and the image that is directly distorted according to the channel condition, denoted as $\ddot{x}$. The resulting stacks are reflected by the module descriptions in Table \ref{table:codec_modules_three_datasets}. For CelebA, both classifiers predict the three binary attributes \textit{Male}, \textit{Smiling}, and \textit{Black\_Hair}; for MNIST and CIFAR-10 they output the $10$ semantic classes.
\\\indent The diffusion part is described separately in Table \ref{table:unet_modules_three_datasets}. For all three datasets, DiffSem and E-DiffSem are implemented as two independently trained U-Nets rather than one shared network. Their common design principle is that DiffSem uses only time conditioning, whereas E-DiffSem additionally injects the semantic condition through a lightweight embedding branch. Once trained, the diffusion module remains fixed during subsequent classifier fine-tuning and inference. This avoids joint end-to-end retraining of the whole system and therefore helps reduce both computational overhead and energy consumption \cite{grassucci2023generative}. We use a maximum diffusion timestep of $T=500$ for MNIST and CIFAR-10, and $T=200$ for CelebA. The learning rate is chosen in the range from $1\times 10^{-3}$ to $1\times 10^{-4}$, Adam is used for optimization, and the training data are partitioned so that $60\%$ is allocated to the $\mathcal{T}$-$\mathcal{R}$ pair and the remaining $40\%$ to $\mathcal{A}_a$.

We reproduced several baseline models to provide a clear comparison and better evaluate the performance of our system.
\begin{itemize}
    \item Classifier-only mode (named ``Classifier''), where $\hat{z}$ is directly sent into the classifier $\Theta_{\theta}$ and bypasses the diffusion module.
    \item Information bottleneck theory and adversarial learning (named ``IBAL'') \cite{wang2024privacy}, that could extract and transmit task-related features while enhancing privacy by significantly decreasing the SSIM and PSNR values of estimated images that the adversary could generate;
    \item IBAL under dynamic channel conditions (named ``IBAL-D'') \cite{wang2024privacy}, that can adjust the weights of the classifier's decision term and the attacker's recovery term in the loss function based on the channel conditions.
    \item Direct mode (named ``Direct''), where images $x\in X$ are directly distorted according to the channel conditions to get $\ddot{x}$, and then are classified by the strong classifier.
    \item Permutation mode (named ``Permutation'')~\cite{chen2023model}, where the order of elements of images are rearranged according to a specific permutation rule or key, without altering the actual content of the elements. Then permuted images are classified by the strong classifier.
\end{itemize}

For the main semantic-transmission methods compared in this paper, including classifier-only, DiffSem, E-DiffSem, IBAL, and IBAL-D, the communication setting is kept consistent within each dataset. Specifically, these methods share the same encoder-to-channel pipeline, transmitted feature dimension, compression ratio, average power normalization, and channel setting. For the diagnostic references including Direct and Permutation modes, the signal transmission is also implemented under the corresponding power normalization and noise injection setting, so that all compared branches are evaluated under properly defined channel corruption. However, Direct and Permutation modes are not used as bandwidth-matched communication baselines; they are included only to illustrate the possible mismatch between pixel-level image quality metrics and task-level semantic fidelity.

\subsection{Results on MNIST dataset}

\begin{figure}[t]
    \centering
    \ifCFGcaseOne
    \includegraphics[width=0.65\linewidth]{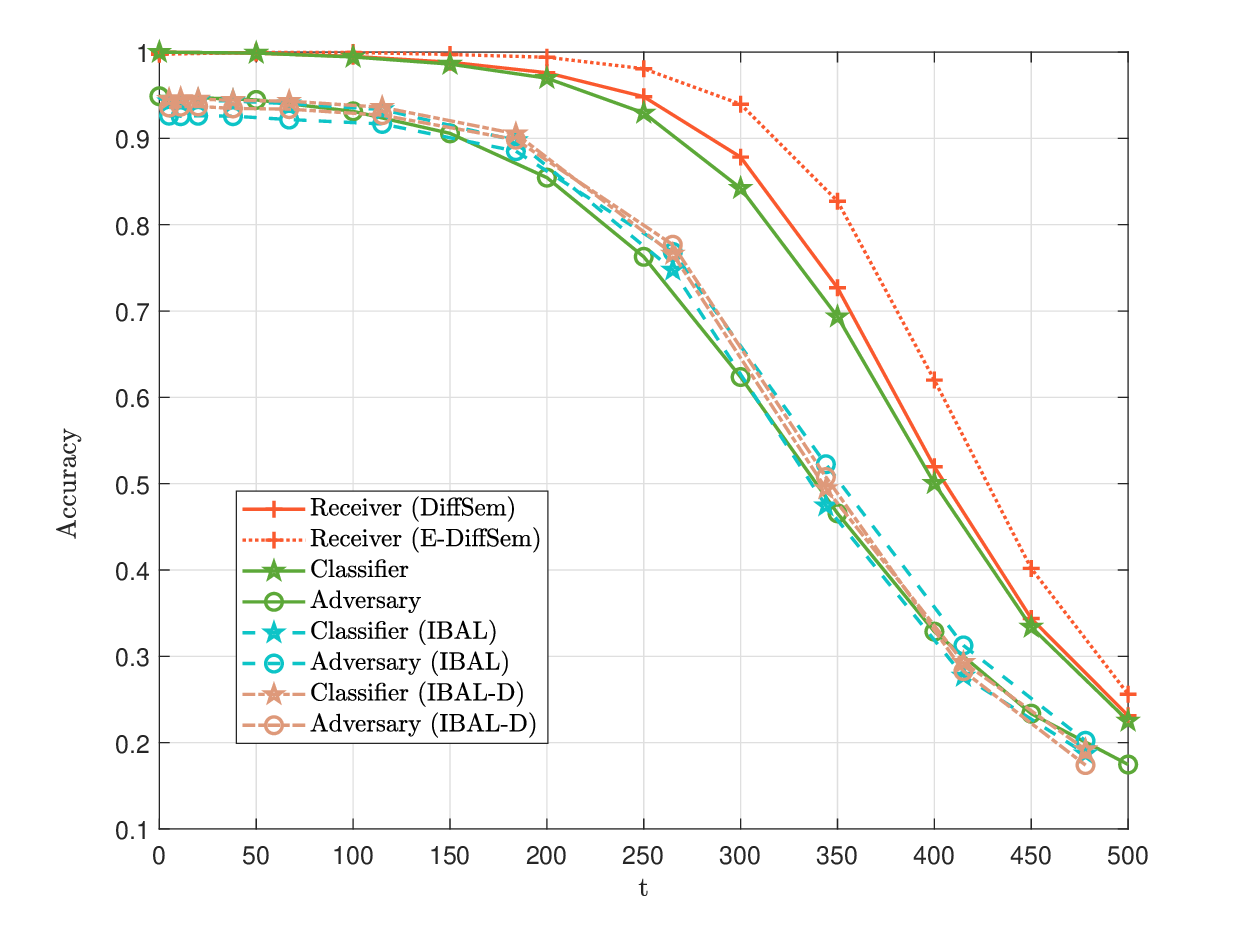}
    \fi
    \ifCFGcaseTwo
    \includegraphics[width=\linewidth]{retultfigs/mnist-1-20250917.eps}
    \fi
    \caption{System performances and baselines using MNIST dataset.}
    \label{fig:mnist-1}
\end{figure}

\begin{figure}[t]
    \centering
    \ifCFGcaseOne
    \includegraphics[width=0.65\linewidth]{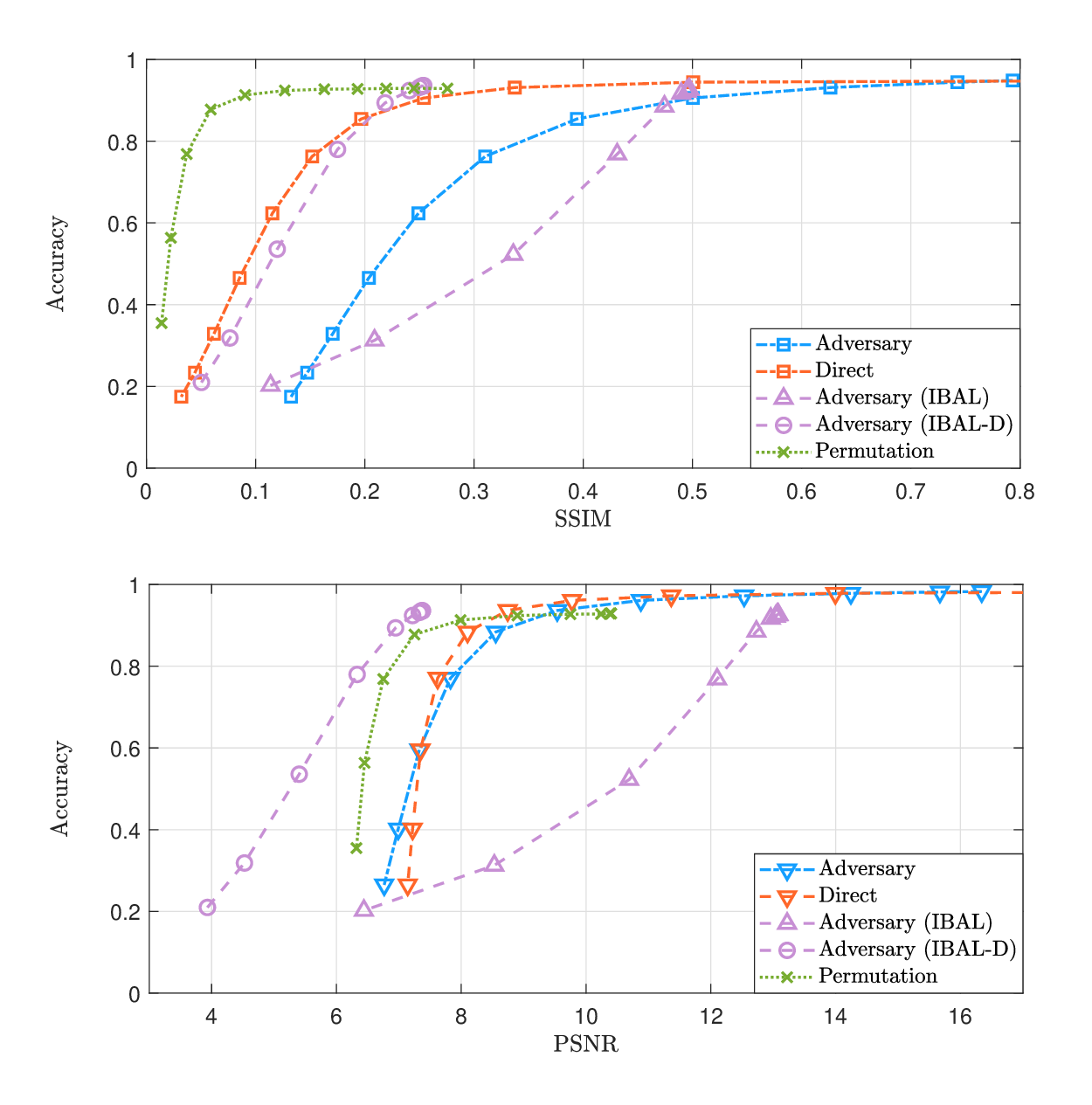}
    \fi
    \ifCFGcaseTwo
    \includegraphics[width=\linewidth]{retultfigs/mnist-3-20250917.eps}
    \fi
    \caption{Plot of accuracy versus SSIM and PSNR using MNIST dataset. }
    \label{fig:mnist-3}
\end{figure}

Fig. \ref{fig:mnist-1} provides a comprehensive performance comparison of various semantic communication systems for the MNIST dataset with a particular focus on accuracy, under channel SNR 15 dB. The green solid lines with notations stars and circles represent the classification accuracy of $\hat{z}$ and $\hat{x}$, without other modules involved. They can be simply regarded as a classification task and an image restoration task under different channel conditions that are determined by $t$, and they reveal a fundamental gap that reflects the intrinsic difference between recovering task-specific features (\textit{task-oriented semantic information}) and pixel-level details (\textit{syntactic information}) given the specific task and dataset \cite{lyu2024semantic}, confirming that task-oriented recovery is inherently more efficient. For example, at $t=300$, the classifier achieves $84\%$ accuracy, while the adversary's reconstructed images yield only $62\%$ accuracy for the same task. Since semantic information is a highly condensed version of syntax and requires less feature to describe, reconstructing syntax is inherently more demanding than recovering semantics.

\ifCFGcaseTwo
\begin{figure}[!t]
\centering
\setlength{\tabcolsep}{2pt}
\hfill
\begin{subfigure}[t]{1.8cm}
    \includegraphics[width=1.8cm]{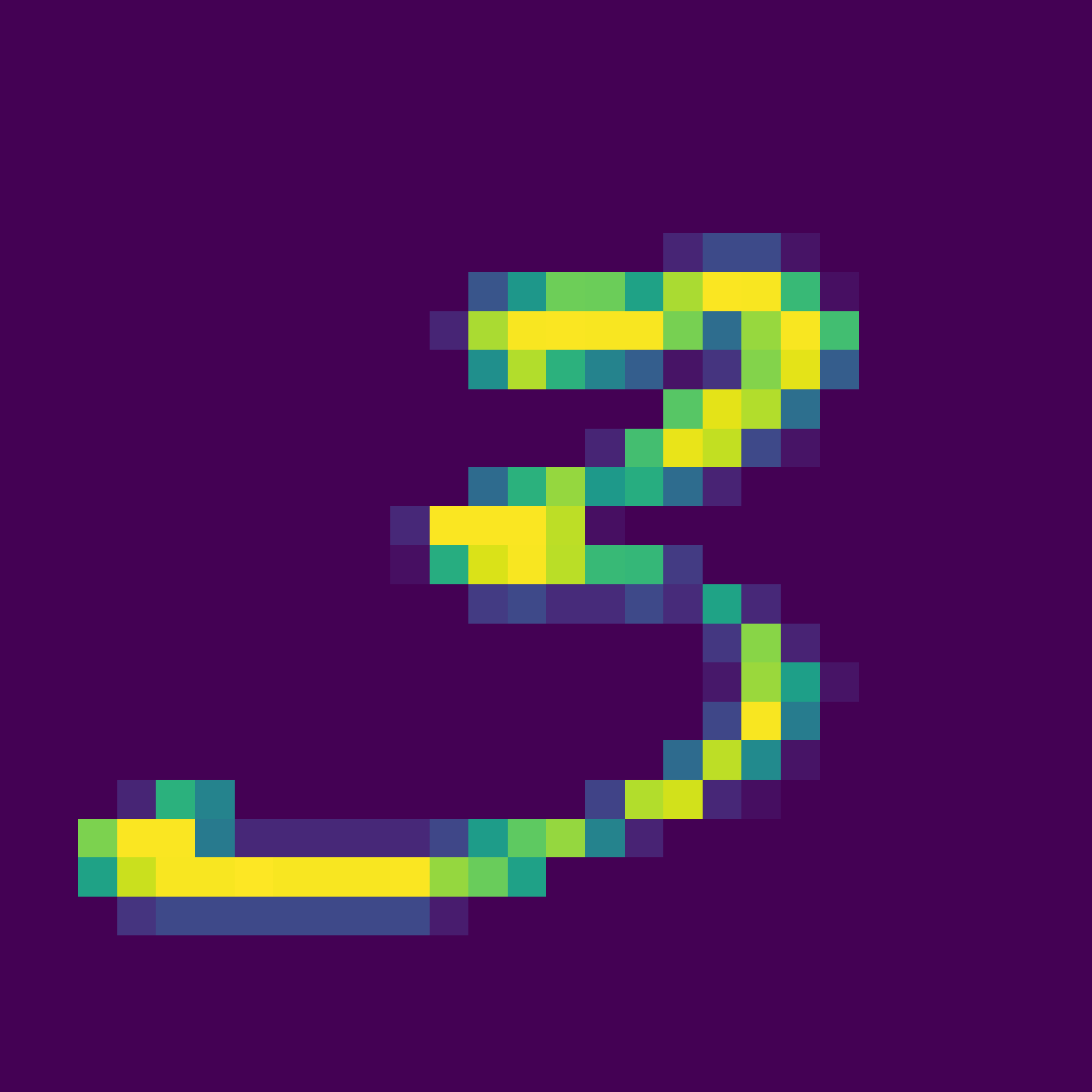}
    \caption{}
\end{subfigure}
\hfill
\begin{subfigure}[t]{1.8cm}
    \includegraphics[width=1.8cm]{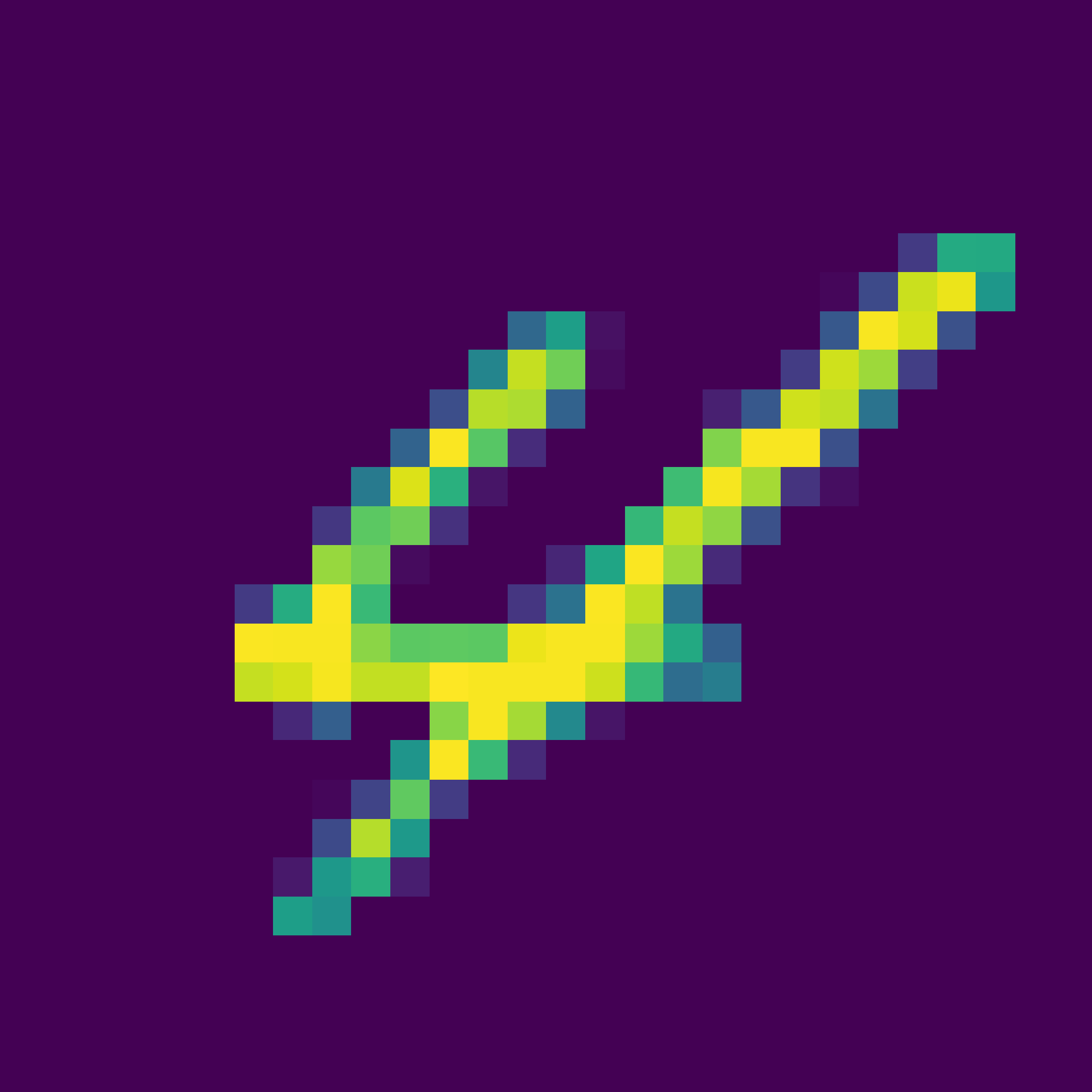}
    \caption{}
\end{subfigure}
\hfill
\begin{subfigure}[t]{1.8cm}
    \includegraphics[width=1.8cm]{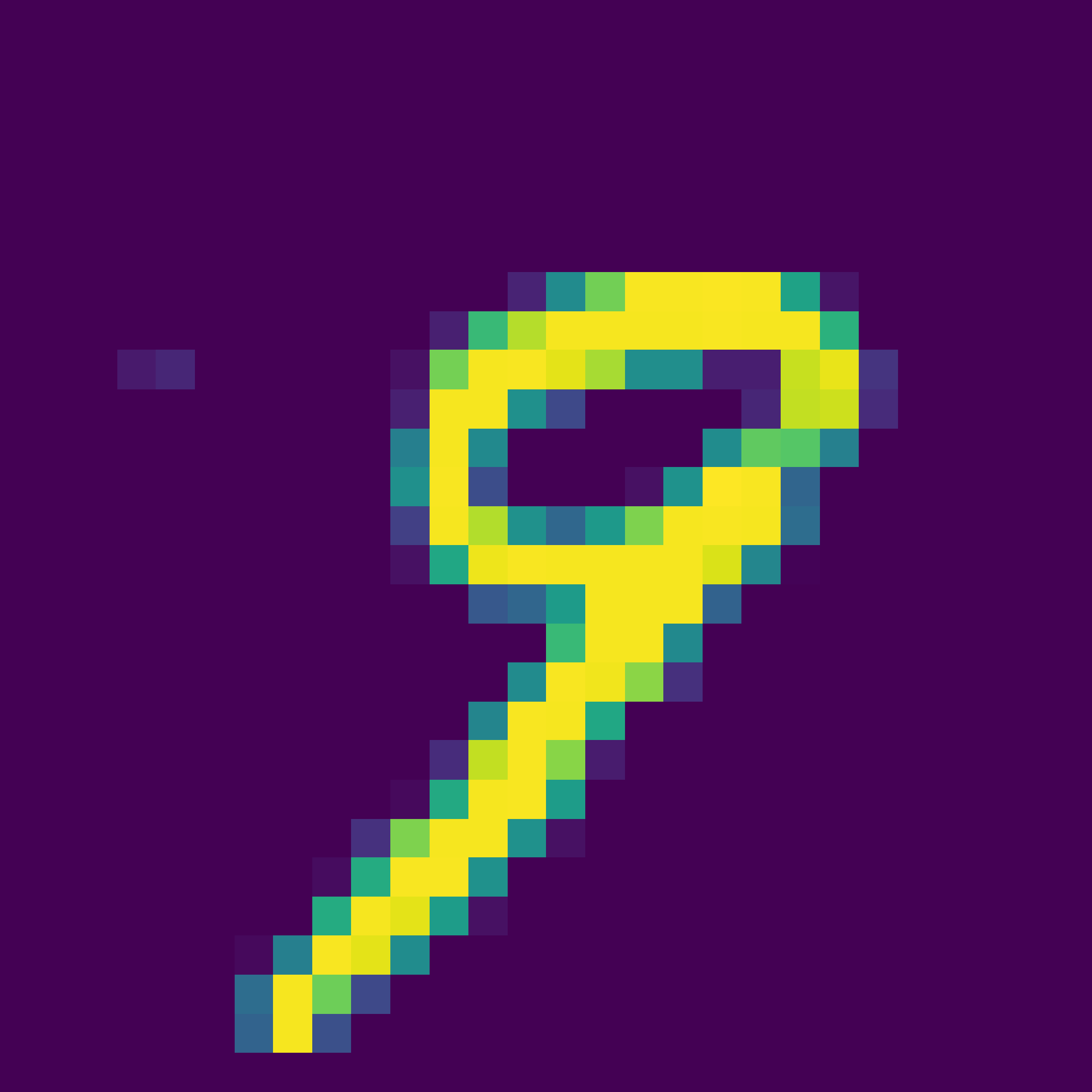}
    \caption{}
\end{subfigure}
\hfill\hfill

\vspace{2mm}

\hfill
\begin{subfigure}[t]{1.8cm}
    \includegraphics[width=1.8cm]{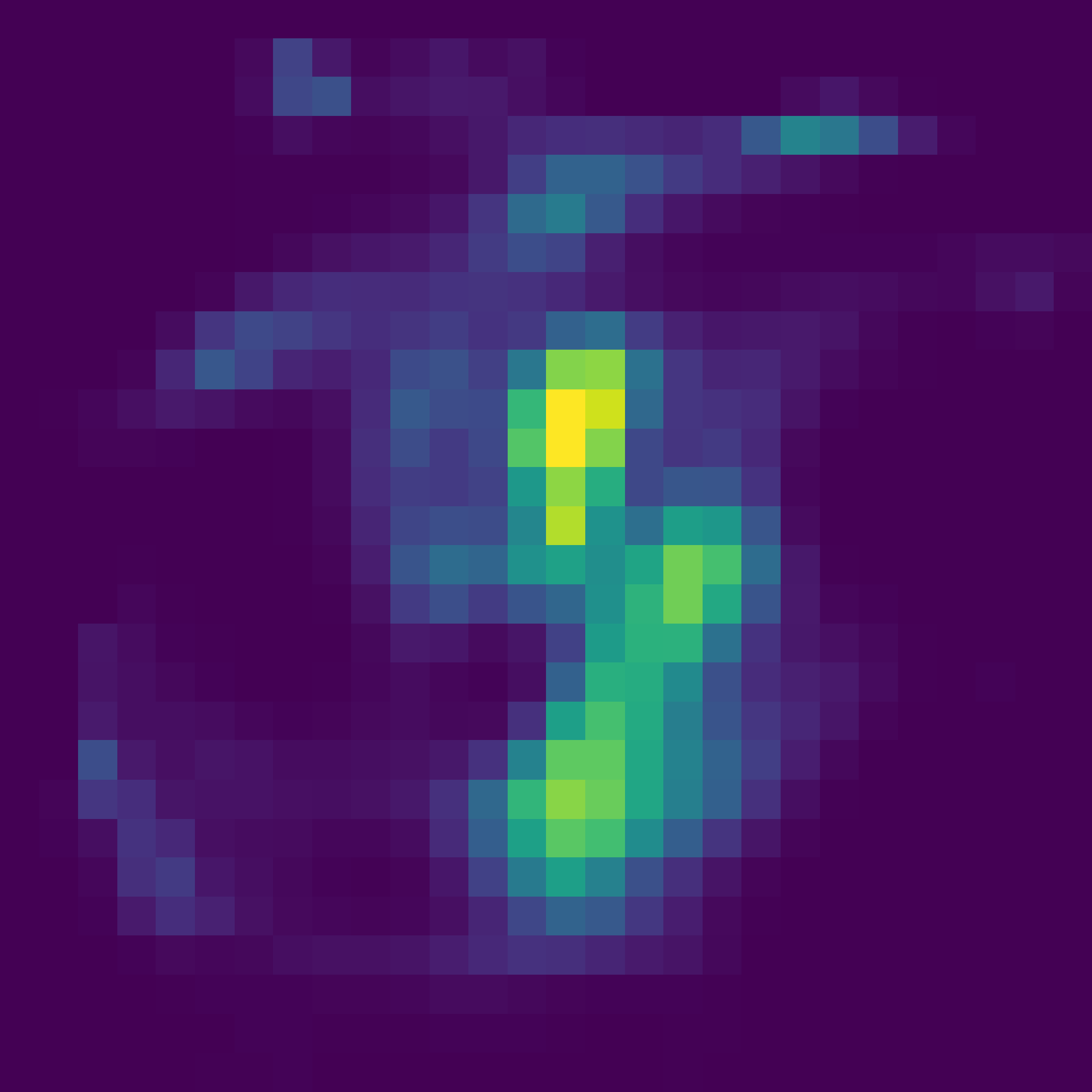}
    \caption{}
\end{subfigure}
\hfill
\begin{subfigure}[t]{1.8cm}
    \includegraphics[width=1.8cm]{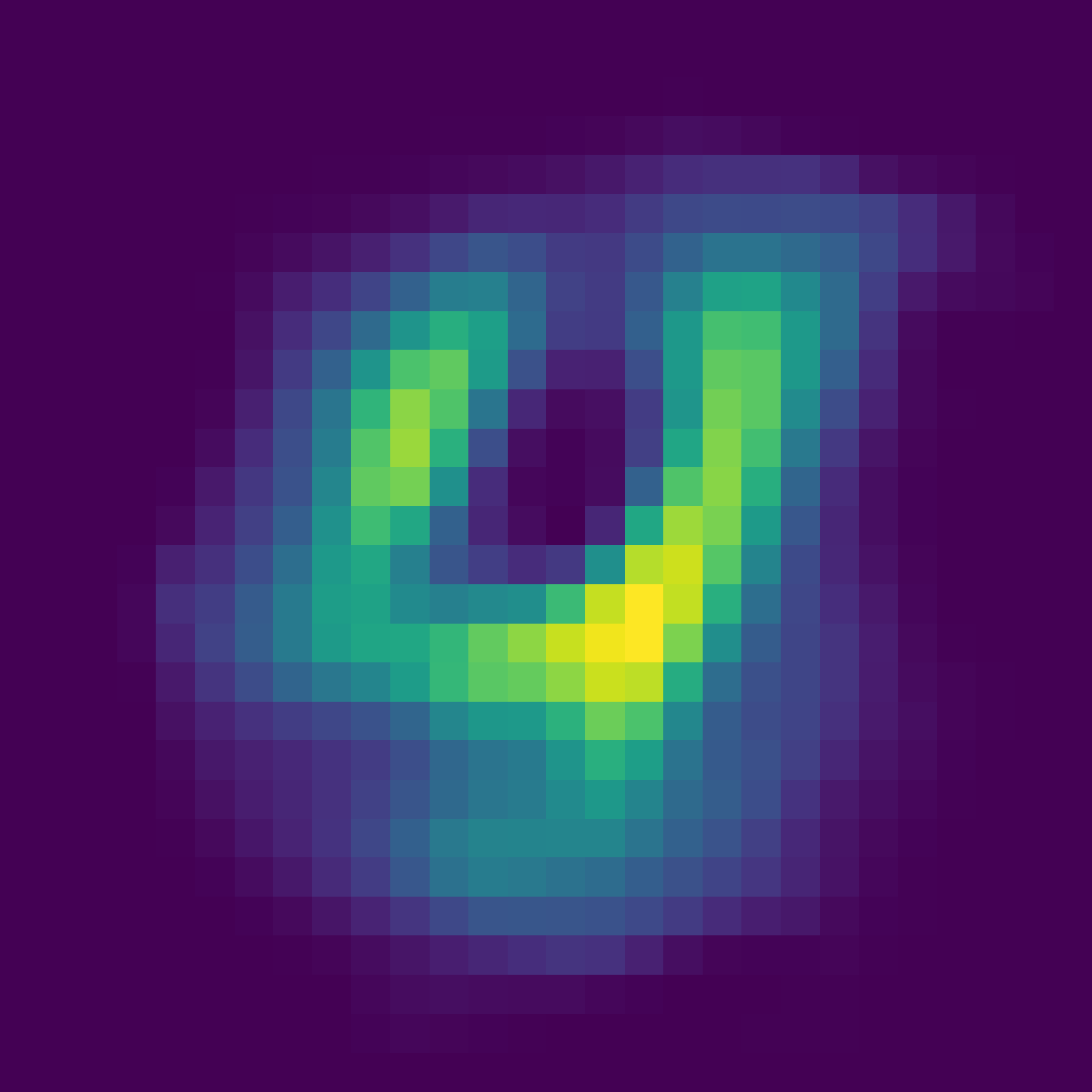}
    \caption{}
\end{subfigure}
\hfill
\begin{subfigure}[t]{1.8cm}
    \includegraphics[width=1.8cm]{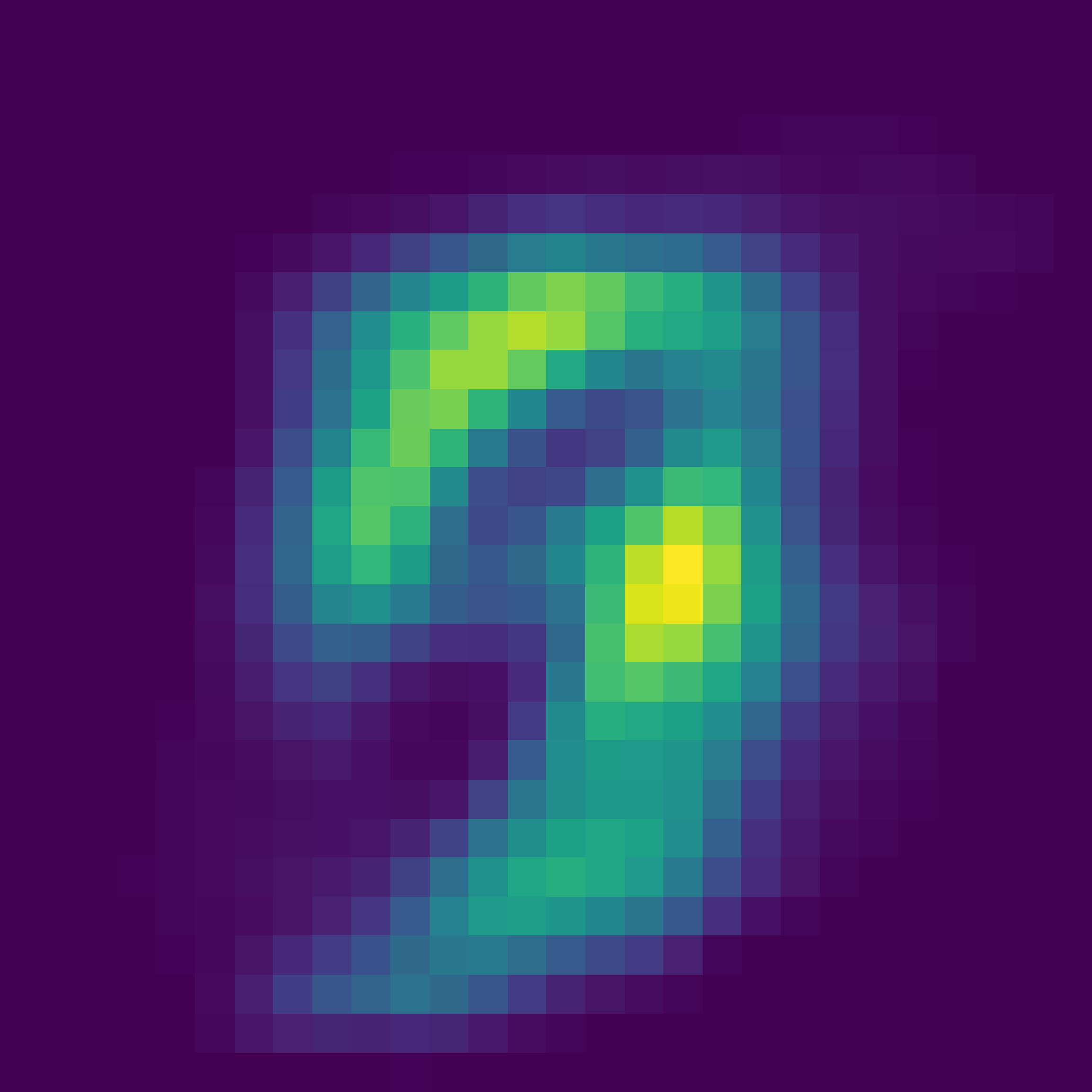}
    \caption{}
\end{subfigure}
\hfill\hfill

\vspace{2mm}

\hfill
\begin{subfigure}[t]{1.8cm}
    \includegraphics[width=1.8cm]{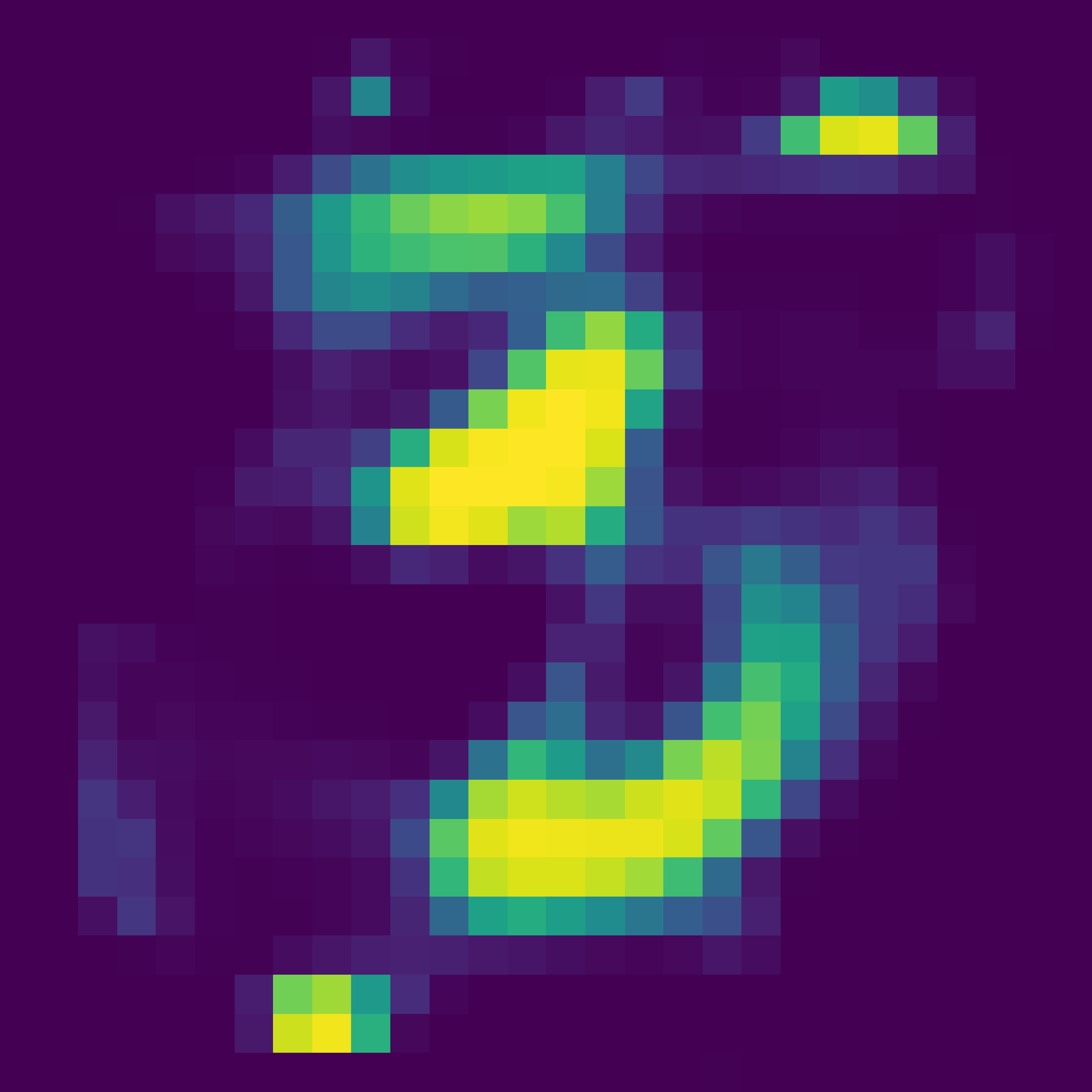}
    \caption{}
\end{subfigure}
\hfill
\begin{subfigure}[t]{1.8cm}
    \includegraphics[width=1.8cm]{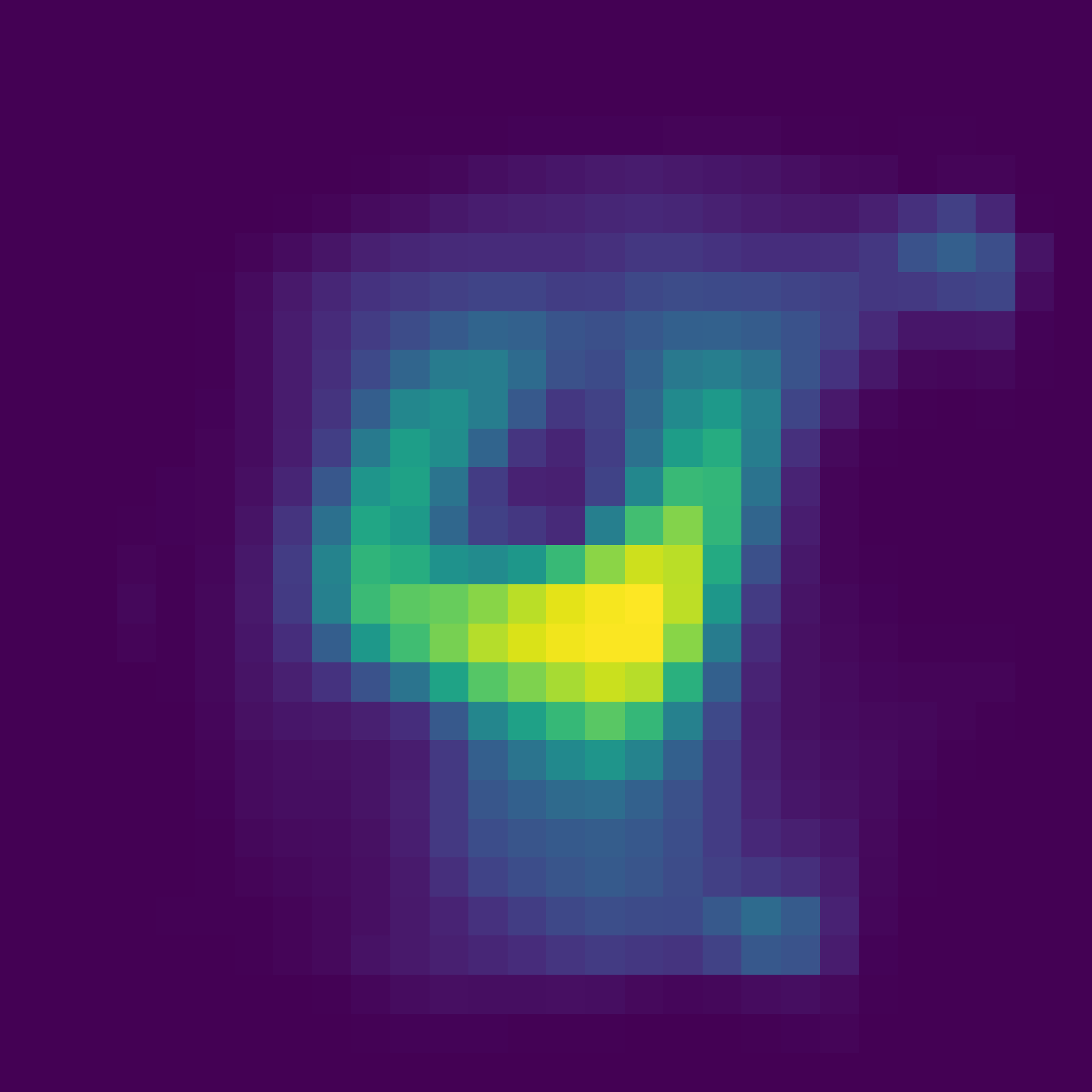}
    \caption{}
\end{subfigure}
\hfill
\begin{subfigure}[t]{1.8cm}
    \includegraphics[width=1.8cm]{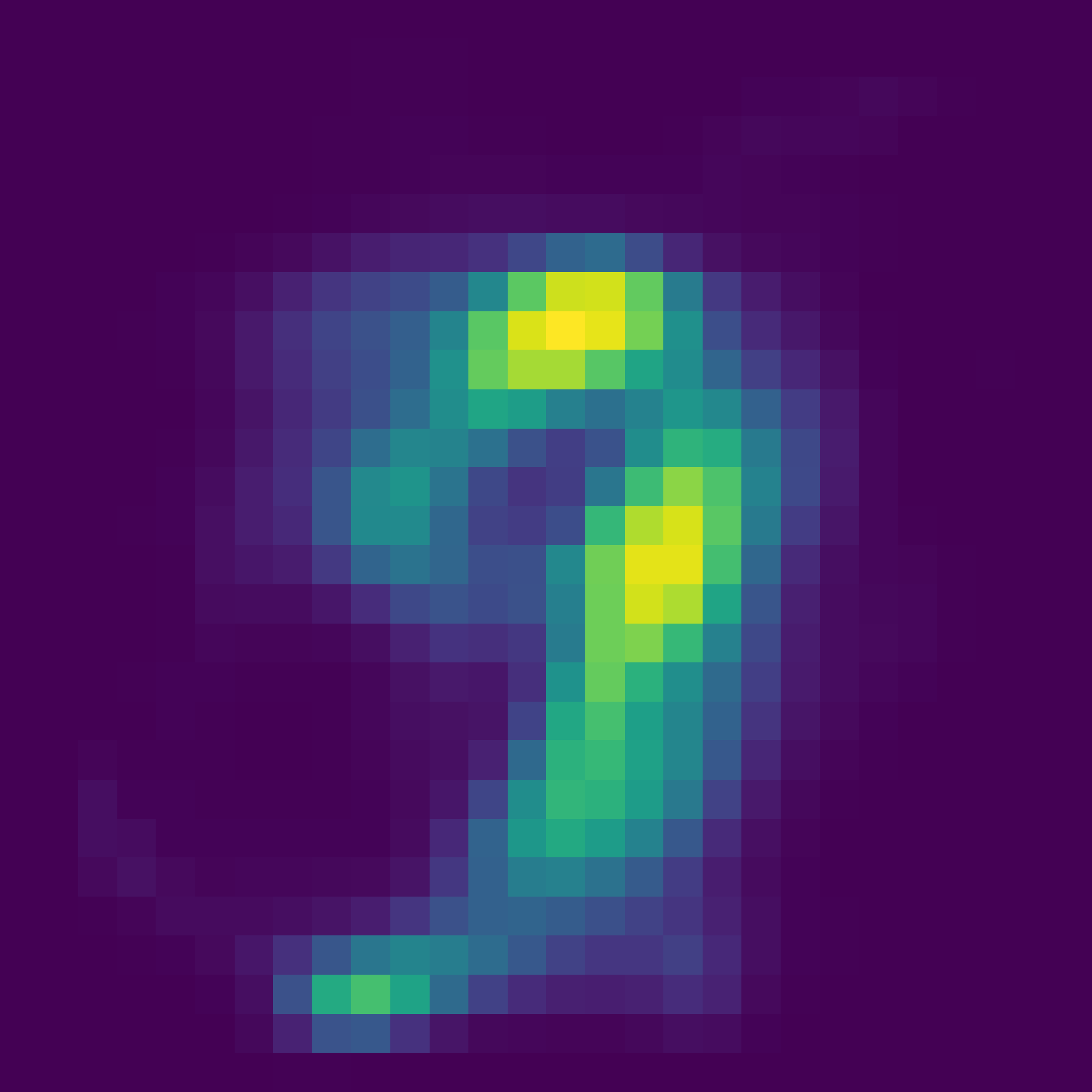}
    \caption{}
\end{subfigure}
\hfill\hfill

\vspace{2mm}

\hfill
\begin{subfigure}[t]{1.8cm}
    \includegraphics[width=1.8cm]{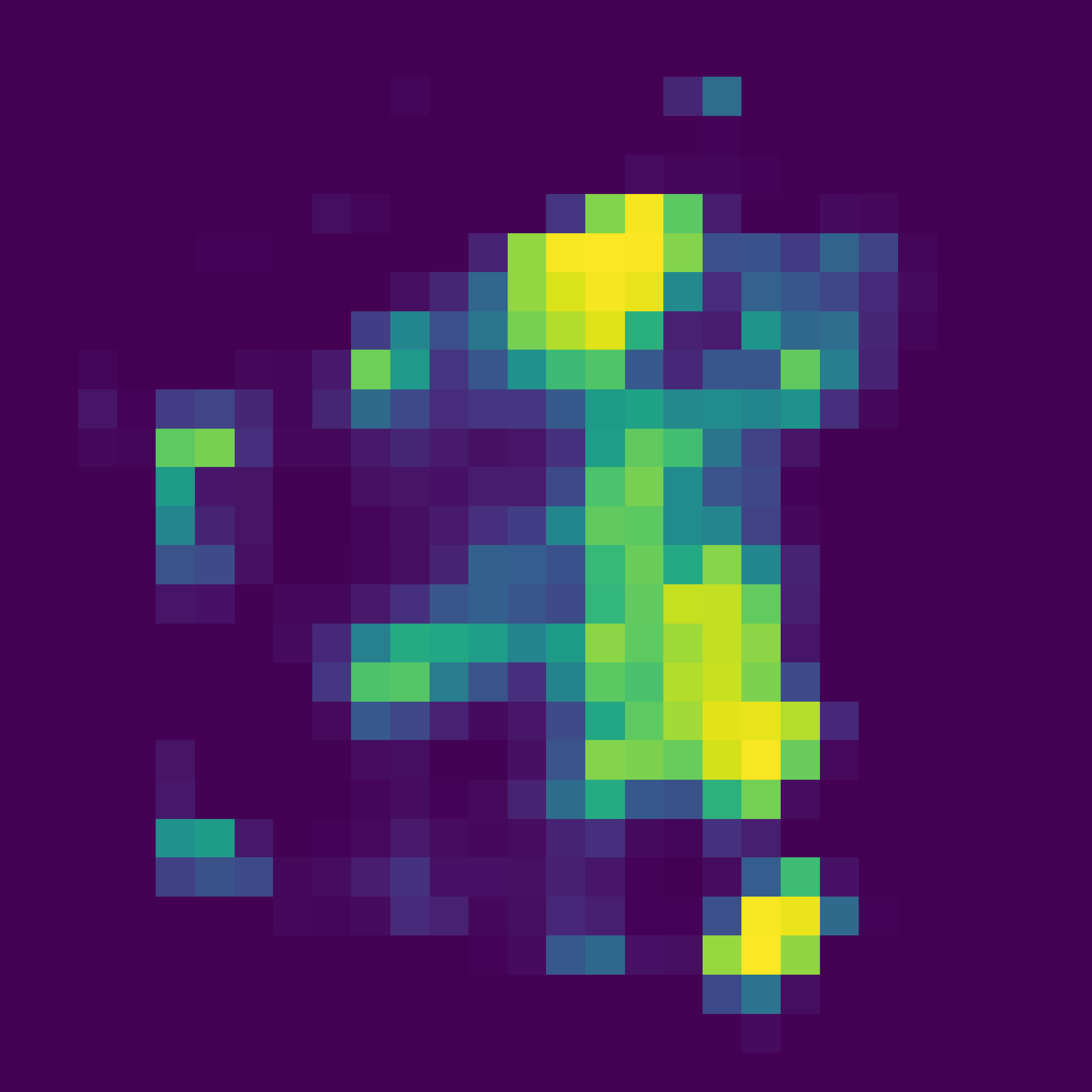}
    \caption{}
\end{subfigure}
\hfill
\begin{subfigure}[t]{1.8cm}
    \includegraphics[width=1.8cm]{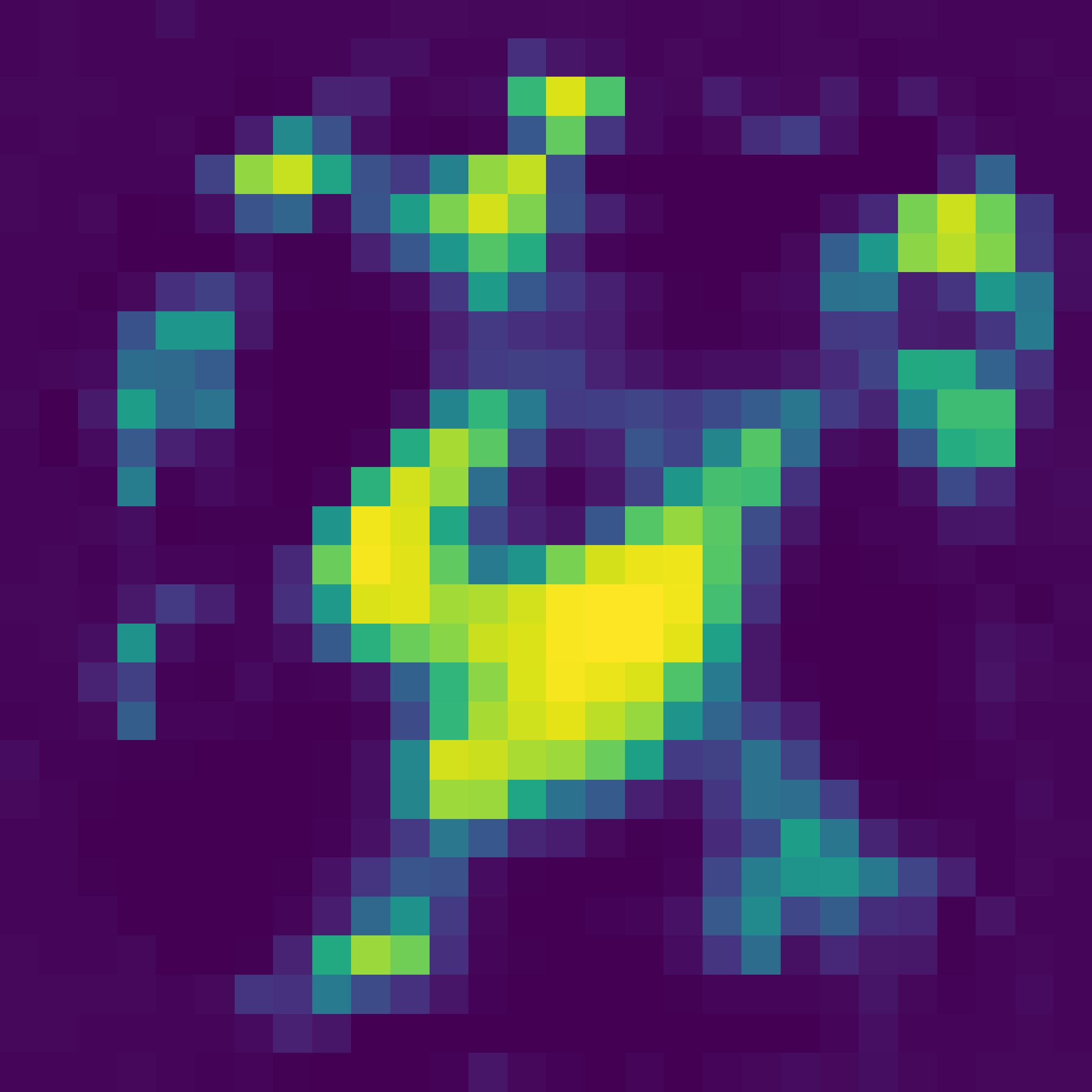}
    \caption{}
\end{subfigure}
\hfill
\begin{subfigure}[t]{1.8cm}
    \includegraphics[width=1.8cm]{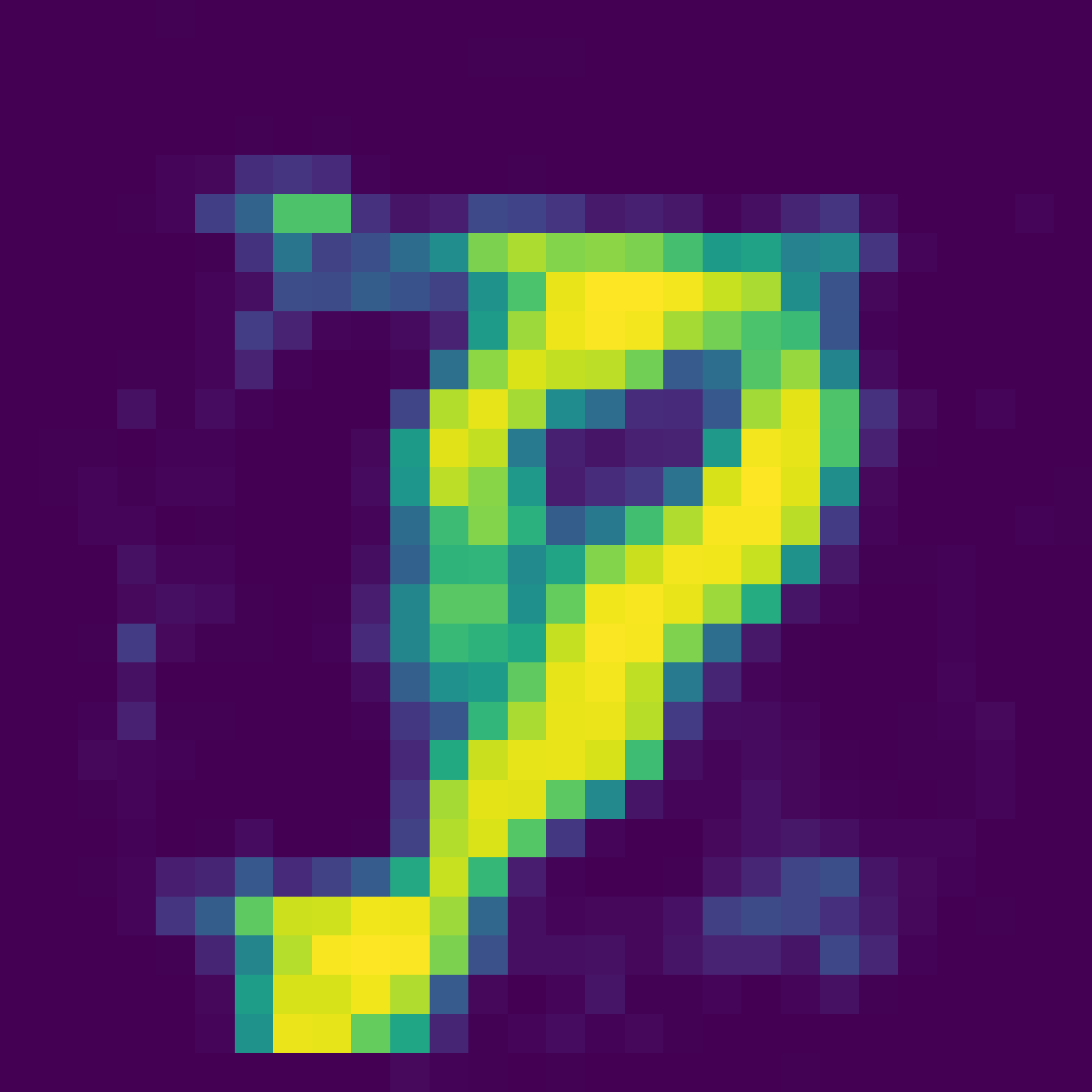}
    \caption{}
\end{subfigure}
\hfill\hfill

\caption{Adversary's output images with different SNR or $t$ under MNIST dataset. 1st row: $-10$ dB; $t=344$. 2nd row: $0$ dB; $t=184$. 3rd row: $10$ dB; $t=67$. (a)(b)(c) Original images. (d)(e)(f) IBAL system. (g)(h)(i) IBAL-D system. (j)(k)(l) Our system.} 
\label{fig:mnist_images} 
\end{figure}
\fi
\ifCFGcaseOne
\begin{figure}[!t]
\centering
\hfill
\begin{subfigure}{1.8cm}
    \renewcommand{\thesubfigure}{a}
    \includegraphics[height=1.8cm]{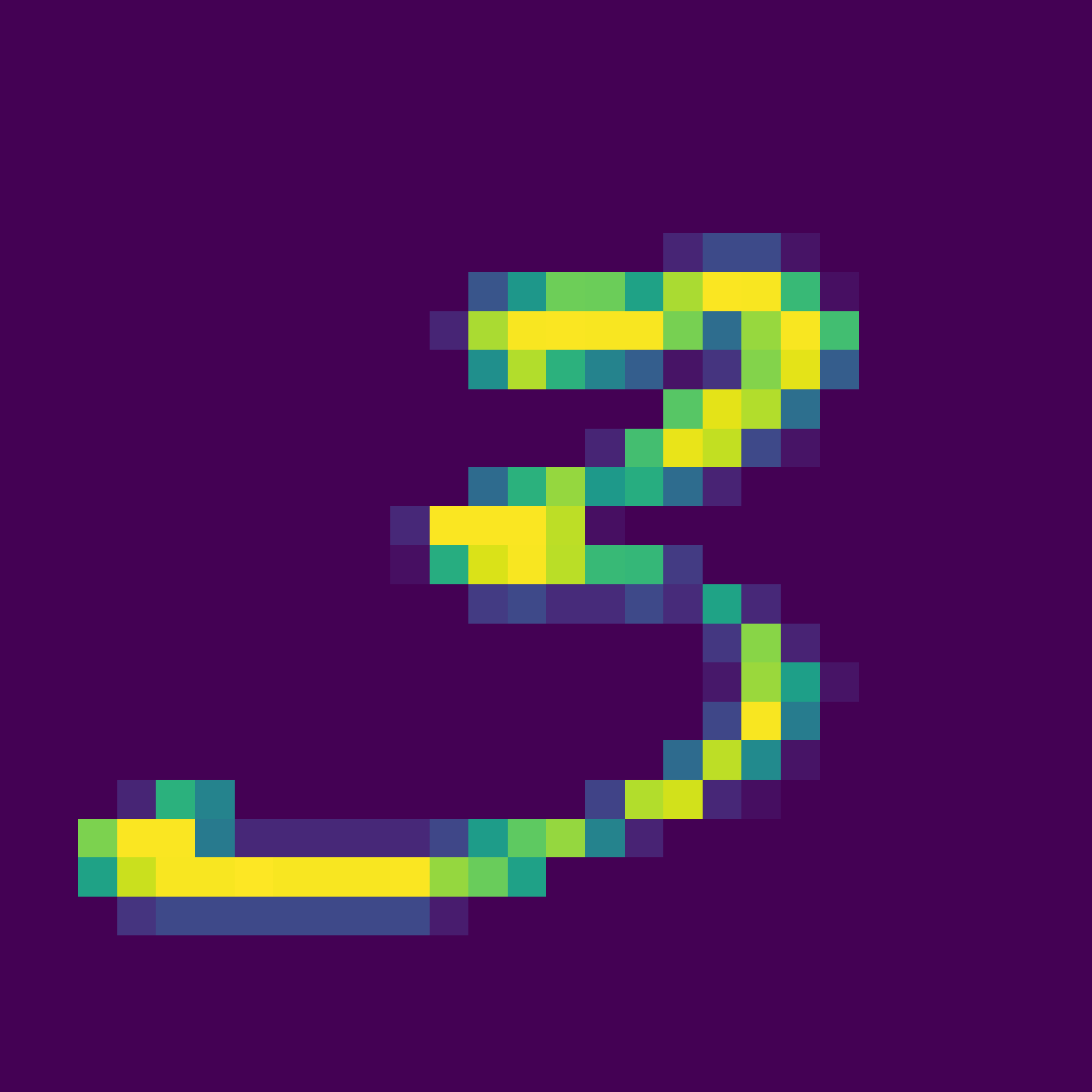}
    \caption{}
\end{subfigure}
\hfill
\begin{subfigure}{1.8cm}
    \renewcommand{\thesubfigure}{d}
    \includegraphics[height=1.8cm]{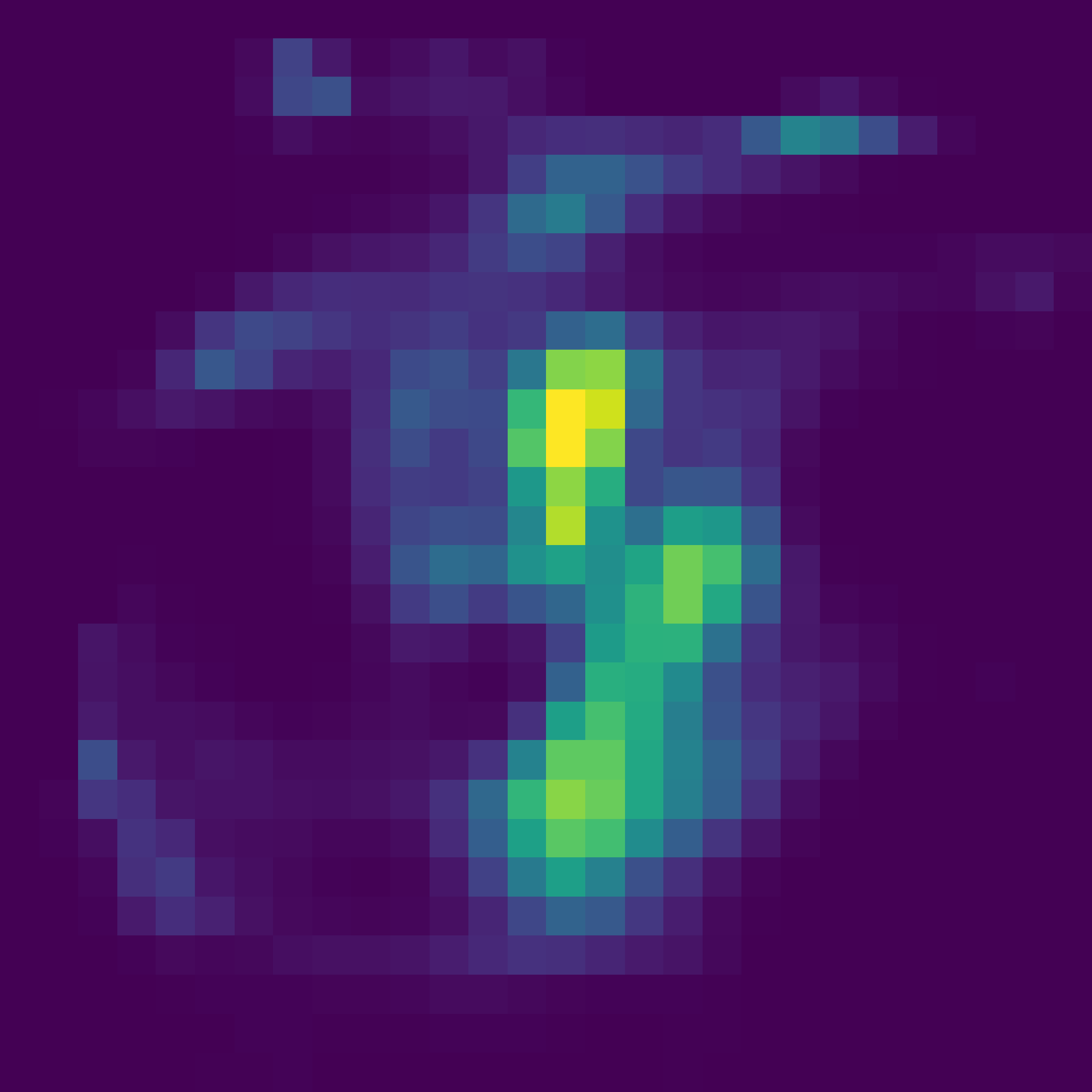}
    \caption{}
\end{subfigure}
\hfill
\begin{subfigure}{1.8cm}
    \renewcommand{\thesubfigure}{g}
    \includegraphics[height=1.8cm]{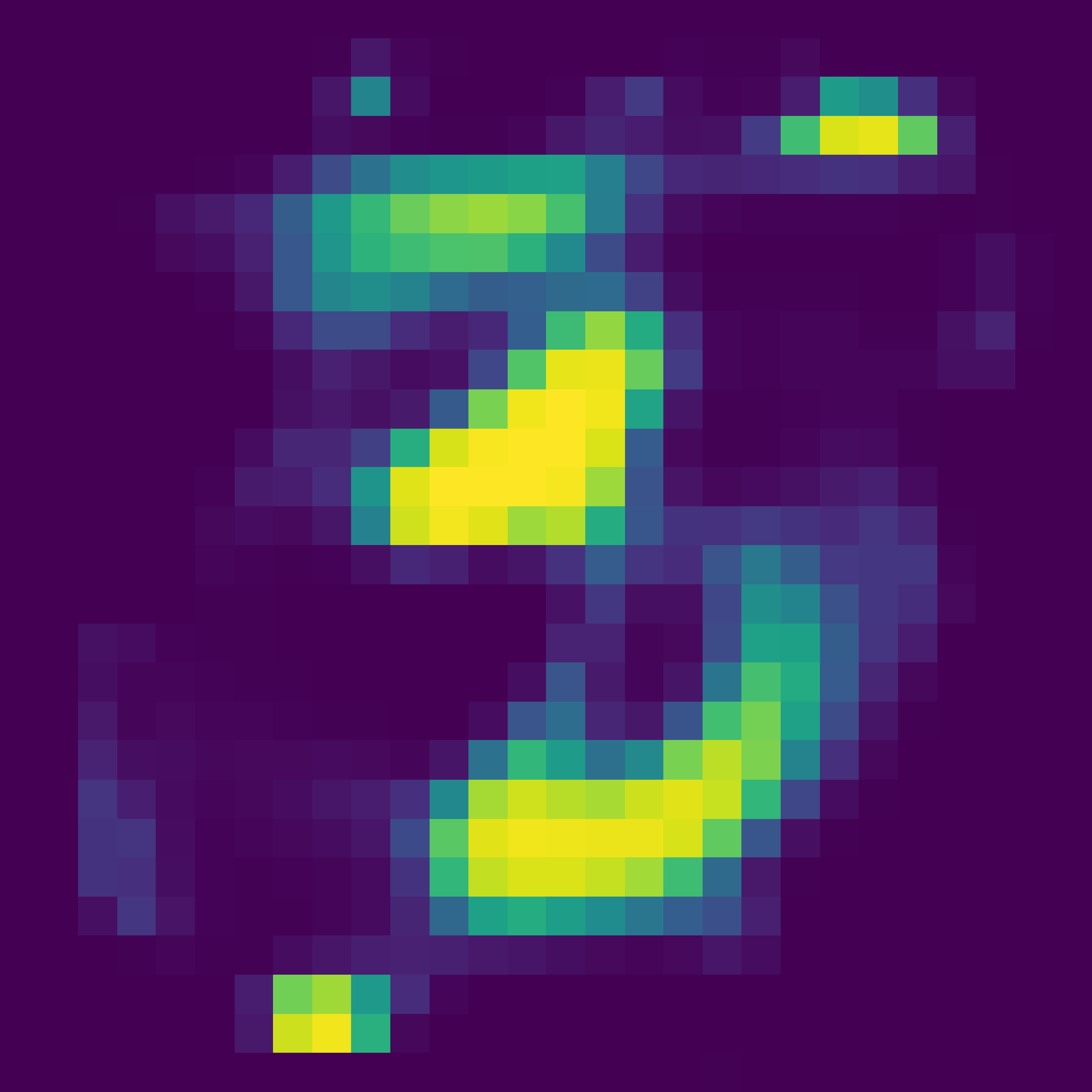}
    \caption{}
\end{subfigure}
\hfill
\begin{subfigure}{1.8cm}
    \renewcommand{\thesubfigure}{j}
    \includegraphics[height=1.8cm]{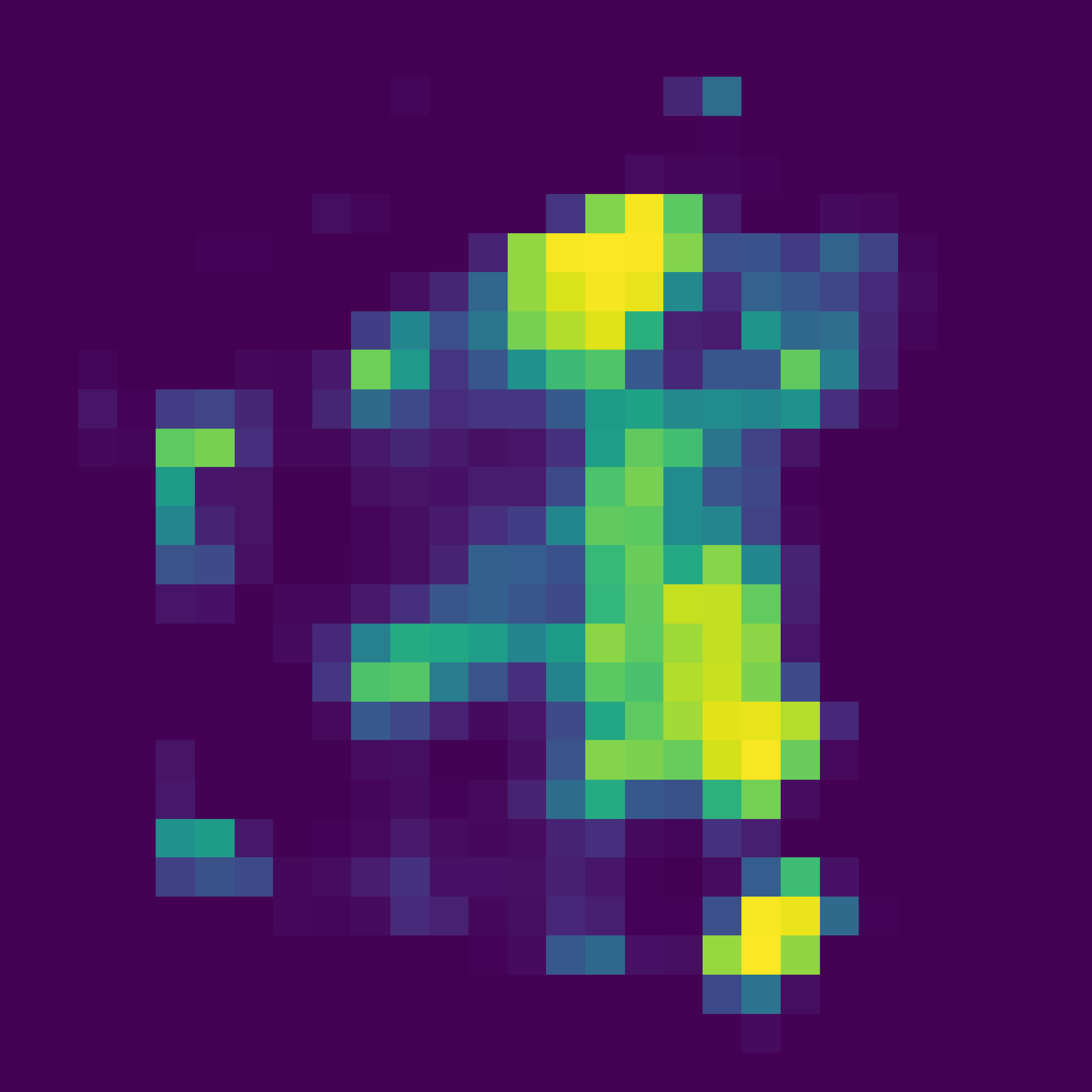}
    \caption{}
\end{subfigure}
\hfill\hfill

\vspace{2mm}
\hfill
\begin{subfigure}{1.8cm}
    \renewcommand{\thesubfigure}{b}
    \includegraphics[height=1.8cm]{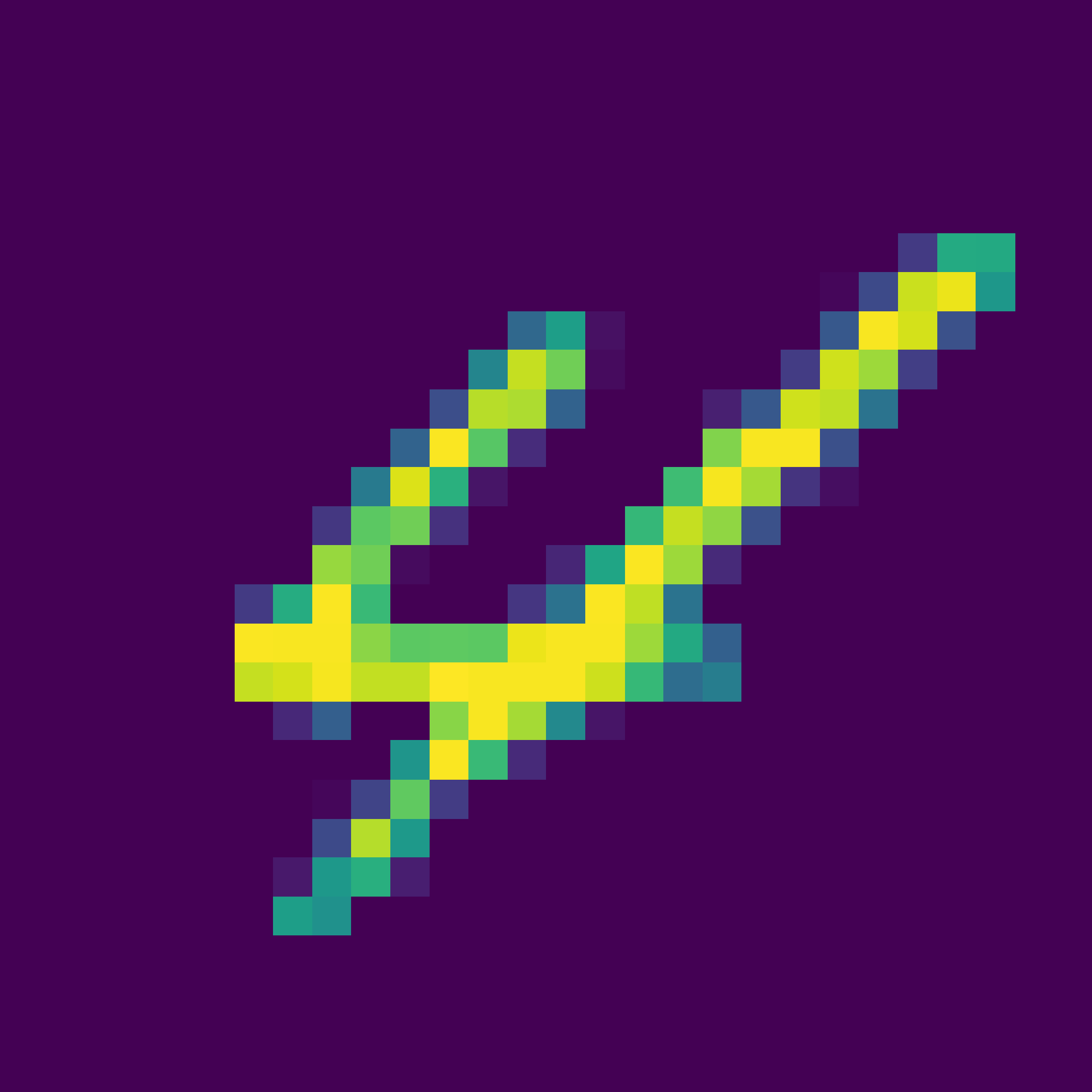}
    \caption{}
\end{subfigure}
\hfill
\begin{subfigure}{1.8cm}
    \renewcommand{\thesubfigure}{e}
    \includegraphics[height=1.8cm]{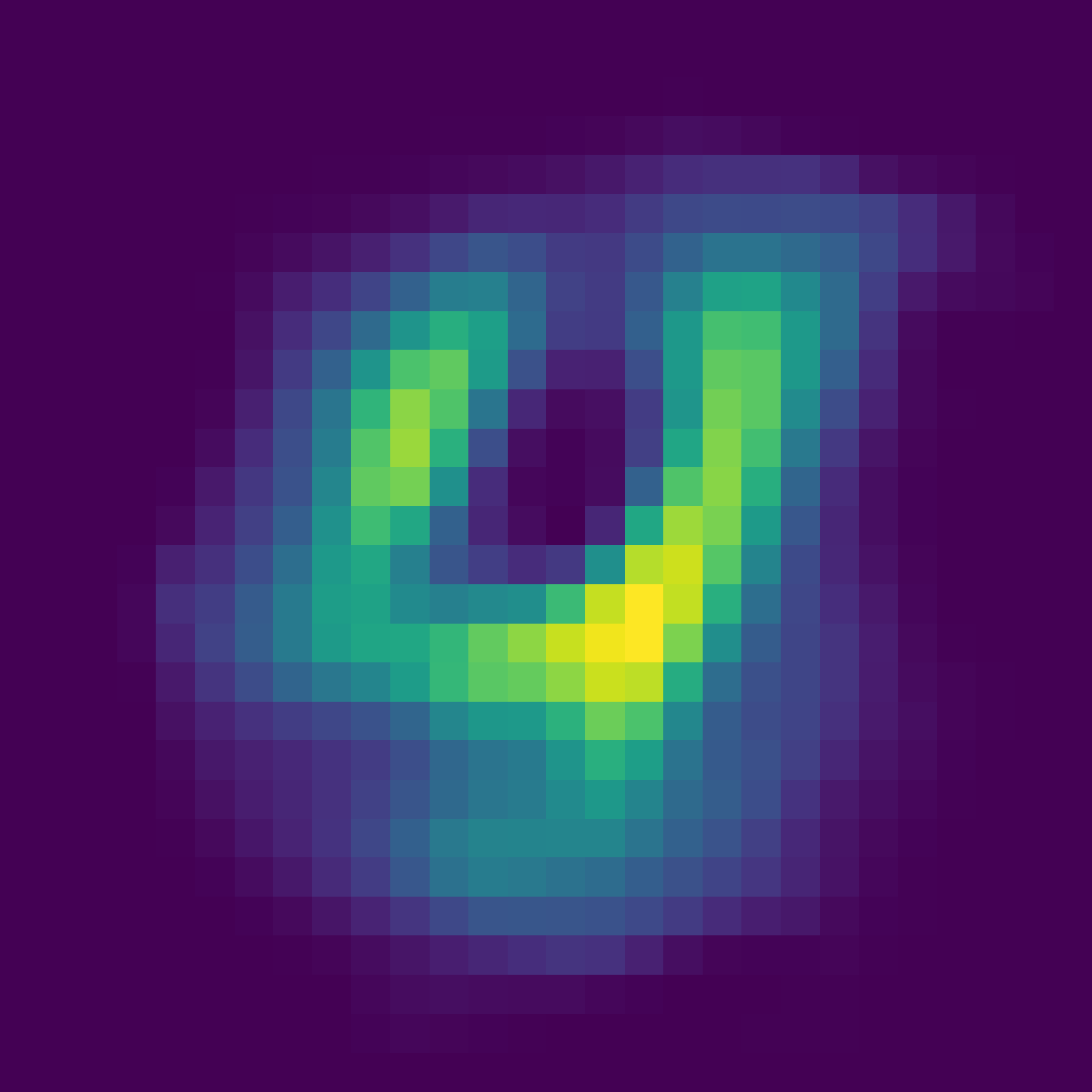}
    \caption{}
\end{subfigure}
\hfill
\begin{subfigure}{1.8cm}
    \renewcommand{\thesubfigure}{h}
    \includegraphics[height=1.8cm]{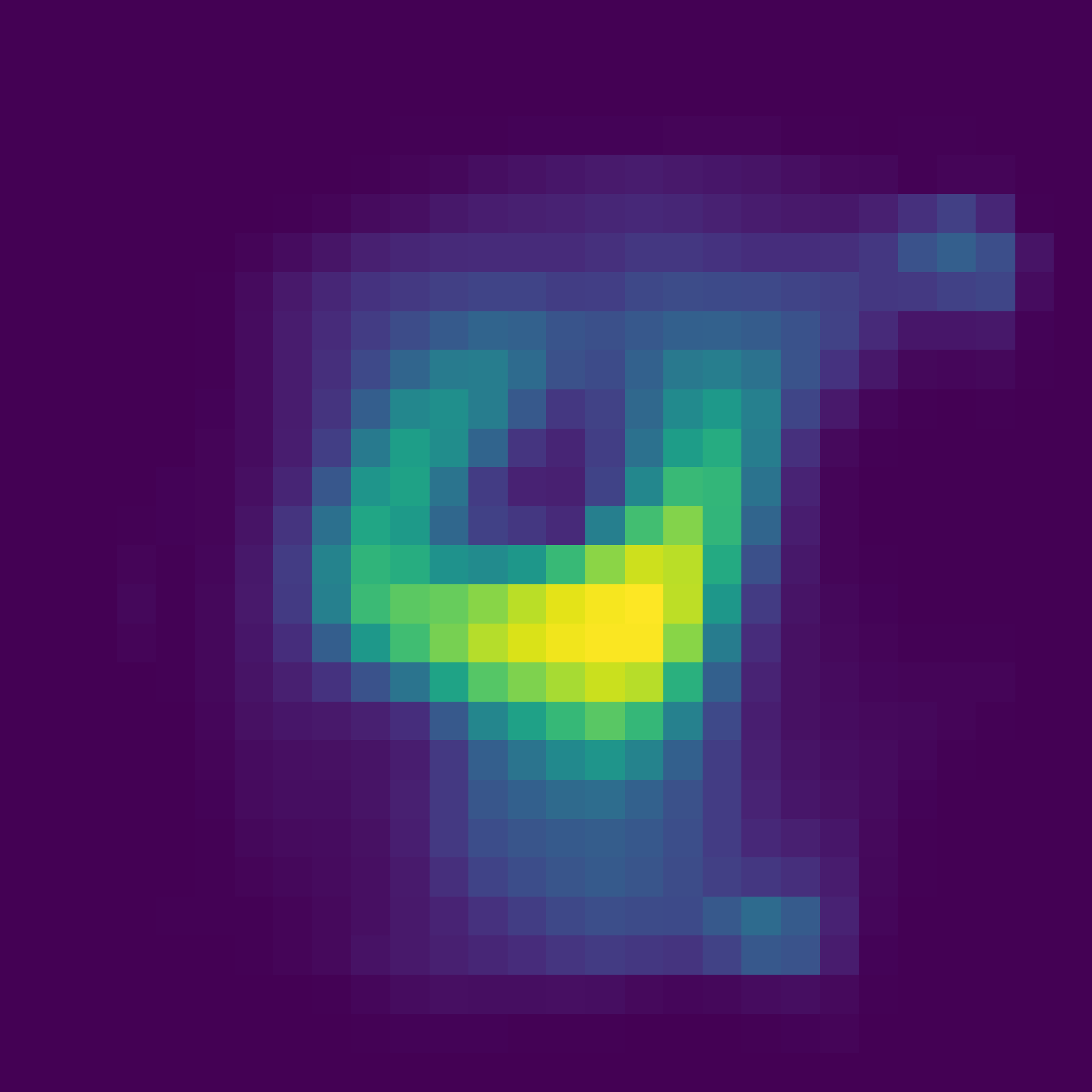}
    \caption{}
\end{subfigure}
\hfill
\begin{subfigure}{1.8cm}
    \renewcommand{\thesubfigure}{k}
    \includegraphics[height=1.8cm]{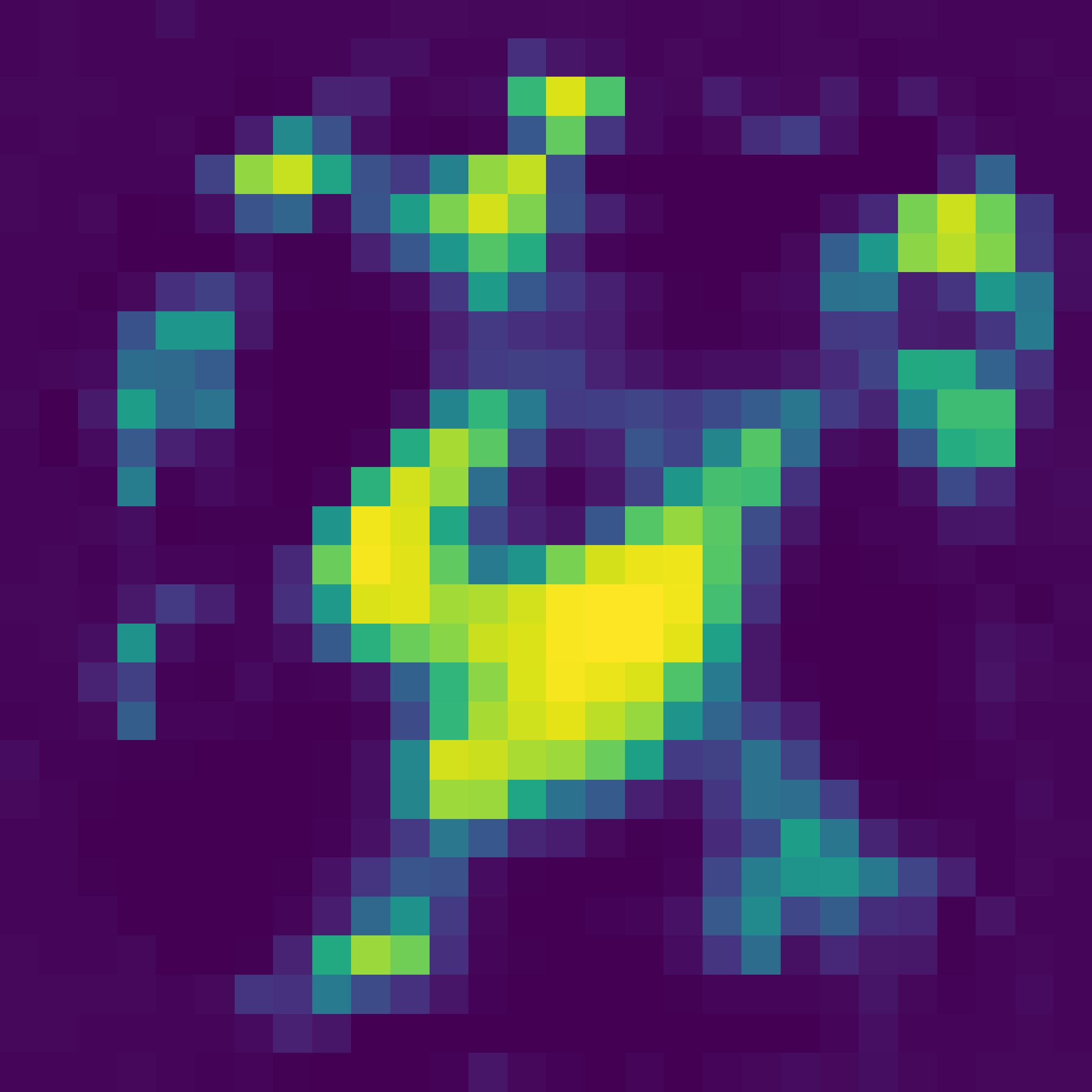}
    \caption{}
\end{subfigure}
\hfill\hfill

\vspace{2mm}
\hfill
\begin{subfigure}{1.8cm}
    \renewcommand{\thesubfigure}{c}
    \includegraphics[height=1.8cm]{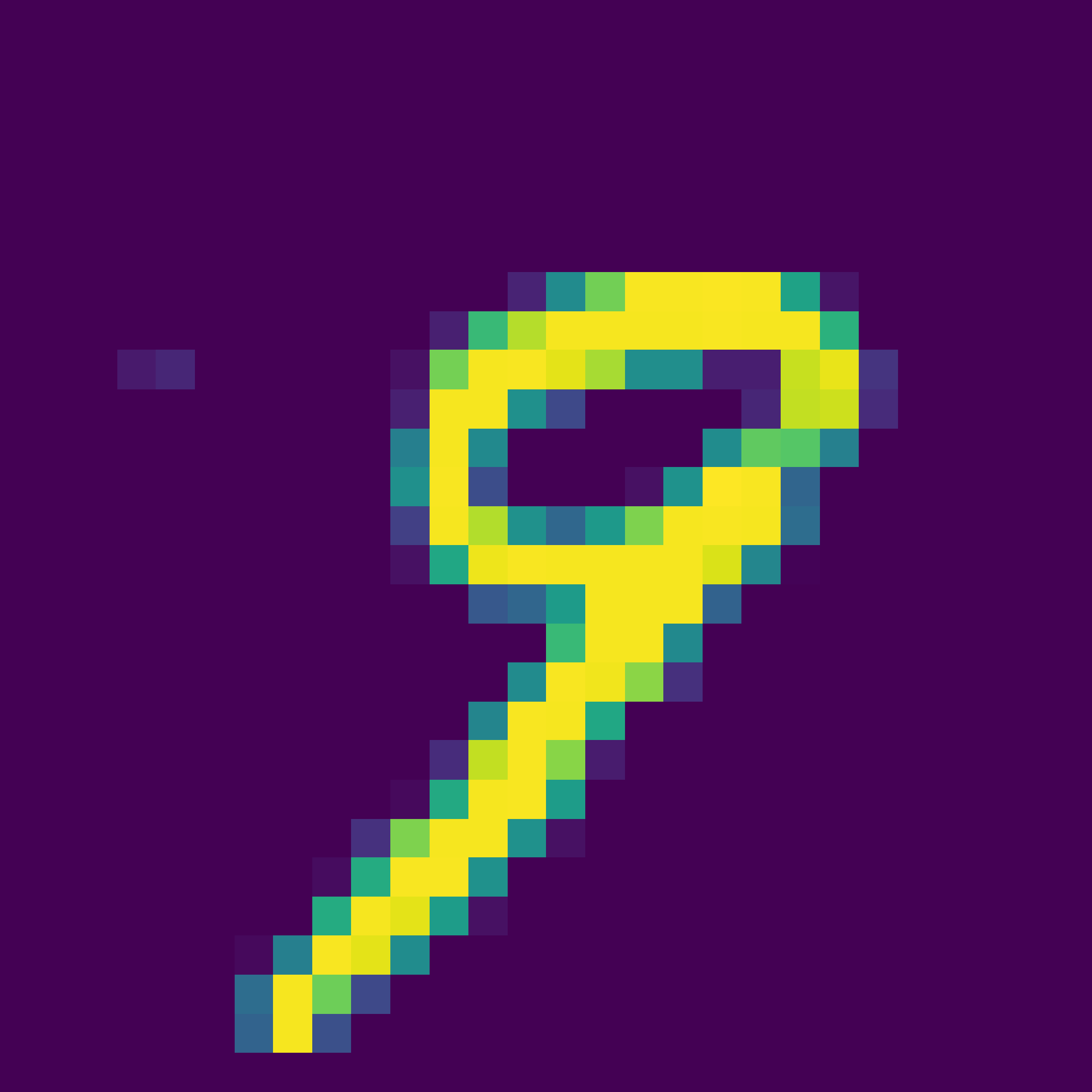}
    \caption{}
\end{subfigure}
\hfill
\begin{subfigure}{1.8cm}
    \renewcommand{\thesubfigure}{f}
    \includegraphics[height=1.8cm]{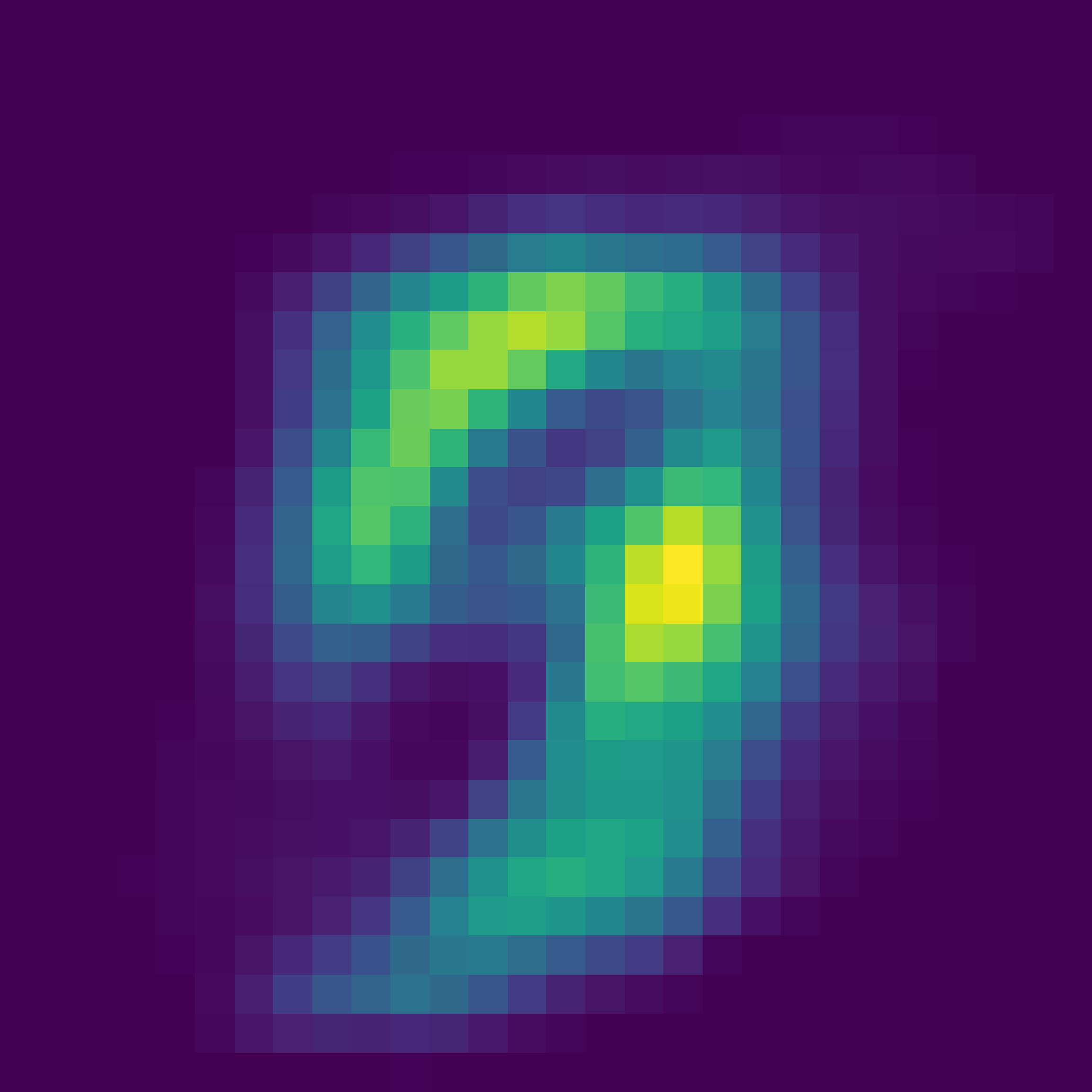}
    \caption{}
\end{subfigure}
\hfill
\begin{subfigure}{1.8cm}
    \renewcommand{\thesubfigure}{i}
    \includegraphics[height=1.8cm]{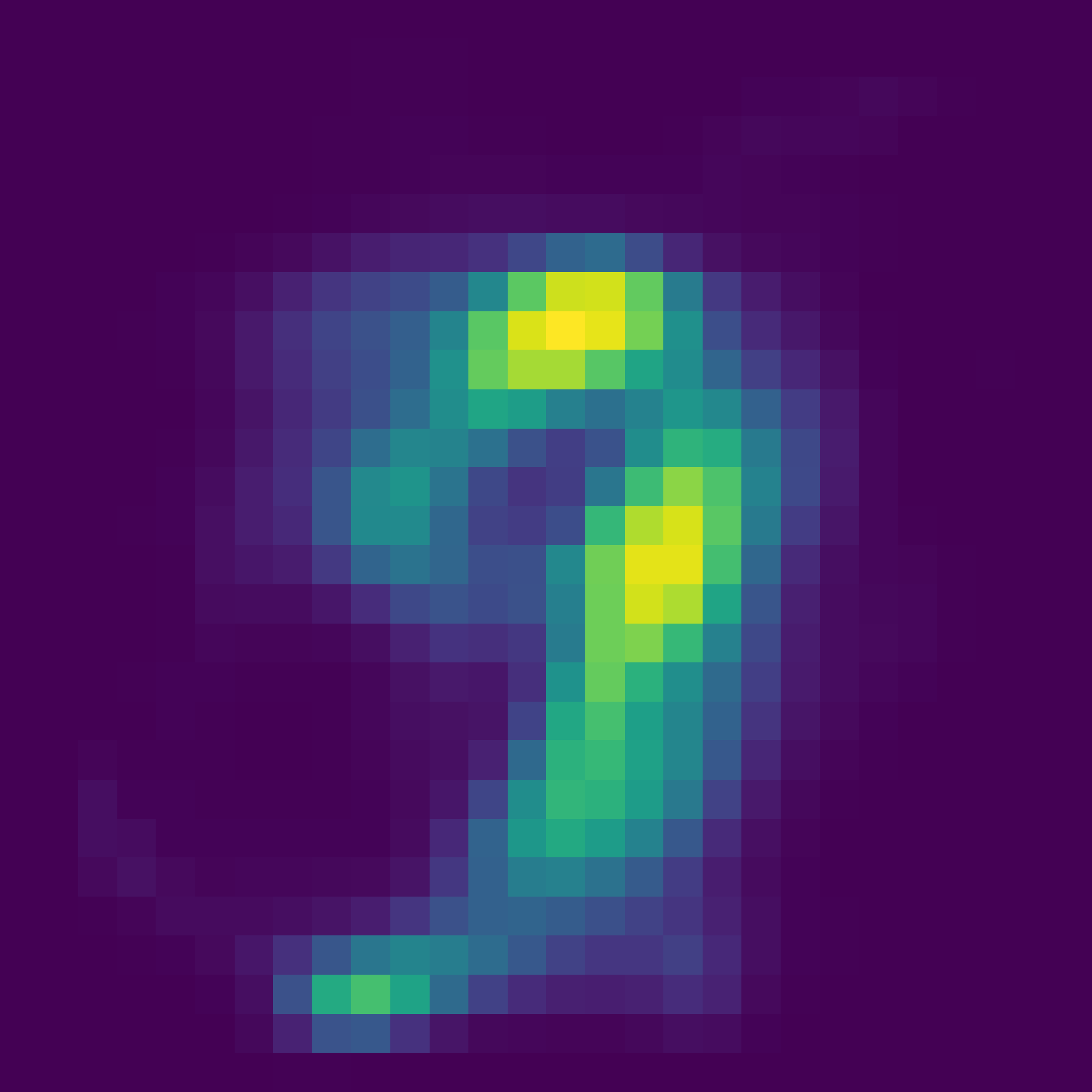}
    \caption{}
\end{subfigure}
\hfill
\begin{subfigure}{1.8cm}
    \renewcommand{\thesubfigure}{l}
    \includegraphics[height=1.8cm]{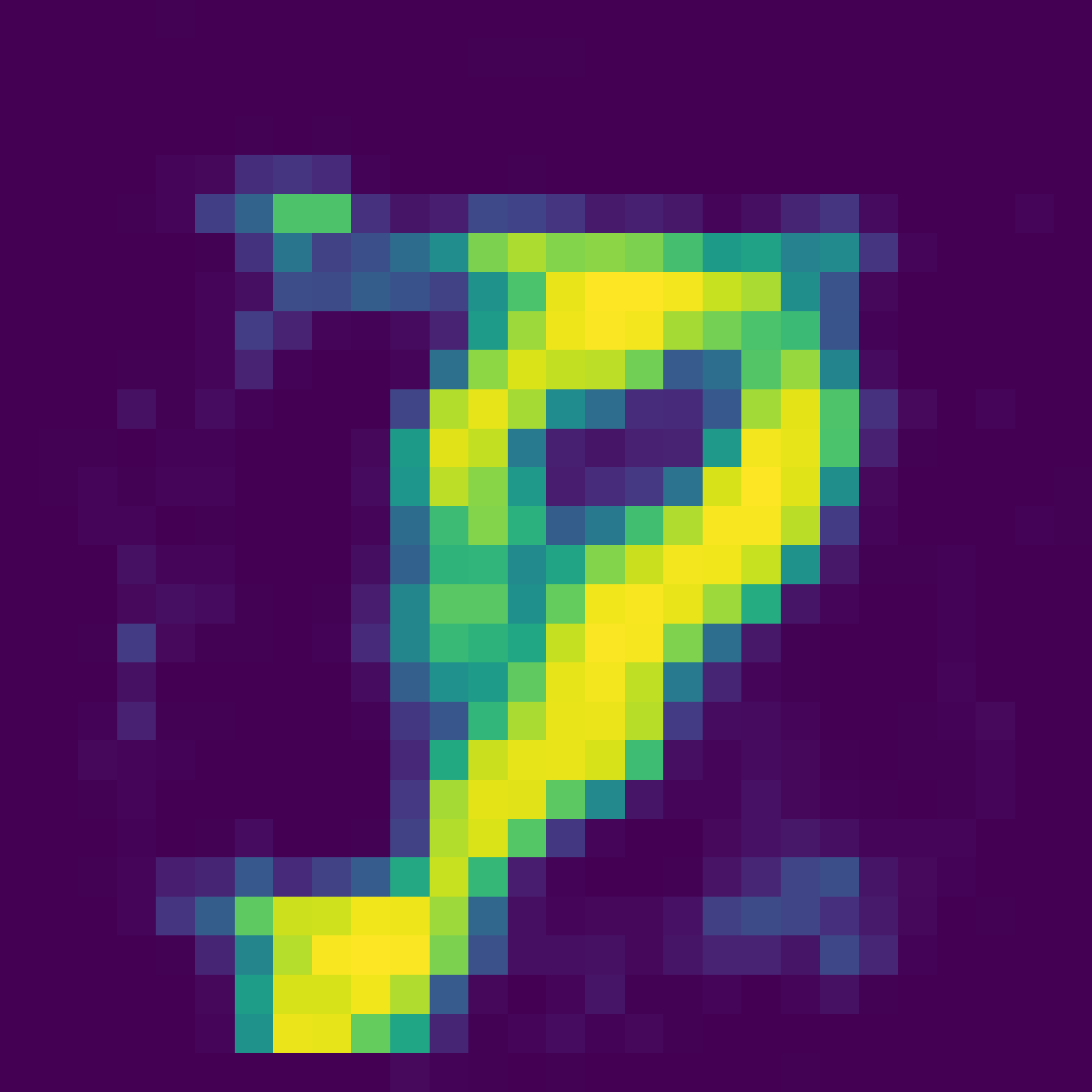}
    \caption{}
\end{subfigure}
\hfill\hfill

\caption{Adversary's output images with different SNR or $t$ under MNIST dataset. 1st row: $-10$ dB; $t=344$. 2nd row: $0$ dB; $t=184$. 3rd row: $10$ dB; $t=67$. (a)(b)(c) Original images. (d)(e)(f) IBAL system. (g)(h)(i) IBAL-D system. (j)(k)(l) Our system.} 
\label{fig:mnist_images} 
\end{figure}
\fi

The two red lines at the top of the figure represent the classification accuracy of the receiver after incorporating the diffusion module. It can be observed that using DiffSem provides a slight improvement in the receiver's recovery, while the label-embedded E-DiffSem significantly enhances system performance. This aligns with the earlier discussion that \(\sqrt{\bar{\alpha}_t}f_0\) can be regarded as an intrinsic guide, and the receiver's preliminary recovery \(\hat{y}\) can serve as guidance to effectively boost diffusion. Most notably, when \(t = 300\), DiffSem and E-DiffSem improve classification accuracy by 3.35\% and 10.03\% compared to the system without diffusion, respectively. 
One possible explanation for the superior performance of diffusion-aided methods lies in their ability to model complex distributions and generate high-quality reconstructions. Diffusion models iteratively refine the output by design, which aligns well with the requirements of semantic communication, where accurate reconstruction of semantic information is crucial. E-DiffSem further optimizes the diffusion process by effectively leveraging the label estimated by the classifier, leading to higher accuracy.

The IBAL and IBAL-D baselines provide an interesting contrast to the proposed methods. Note that the curves in Fig. \ref{fig:mnist-1} use SNR as the horizontal axis during training; here they have been recalculated based on Eq. (\ref{eq:p}) and presented with \(t\) as the horizontal axis. We employ the same strong classifier structure for all adversaries including the one in our system, ensuring a consistent ability to assess the amount of task-relevant information in reconstructed images.
Since adversarial training, which is grounded in the information bottleneck principle, demands the careful navigation of a dynamic trade-off between the transmission of task-relevant information and the mitigation of adversarial threats, this inevitable compromise inherently limits the amount of information that can be transmitted during standard communication, ultimately causing a pronounced decline in the receiver's accuracy.
This also explains why several IBAL/IBAL-D-related curves remain relatively close under the task-accuracy metric: these baselines reduce pixel-level reconstruction quality, but do not create a comparably large separation in task-relevant semantics.
In stark contrast, our diffusion-aided receiver demonstrates remarkable superiority, achieving significantly higher performance compared to both IBAL systems.


\begin{figure}[t]
\centering

\hfill
\begin{subfigure}{1.8cm}
  \includegraphics[height=1.8cm]{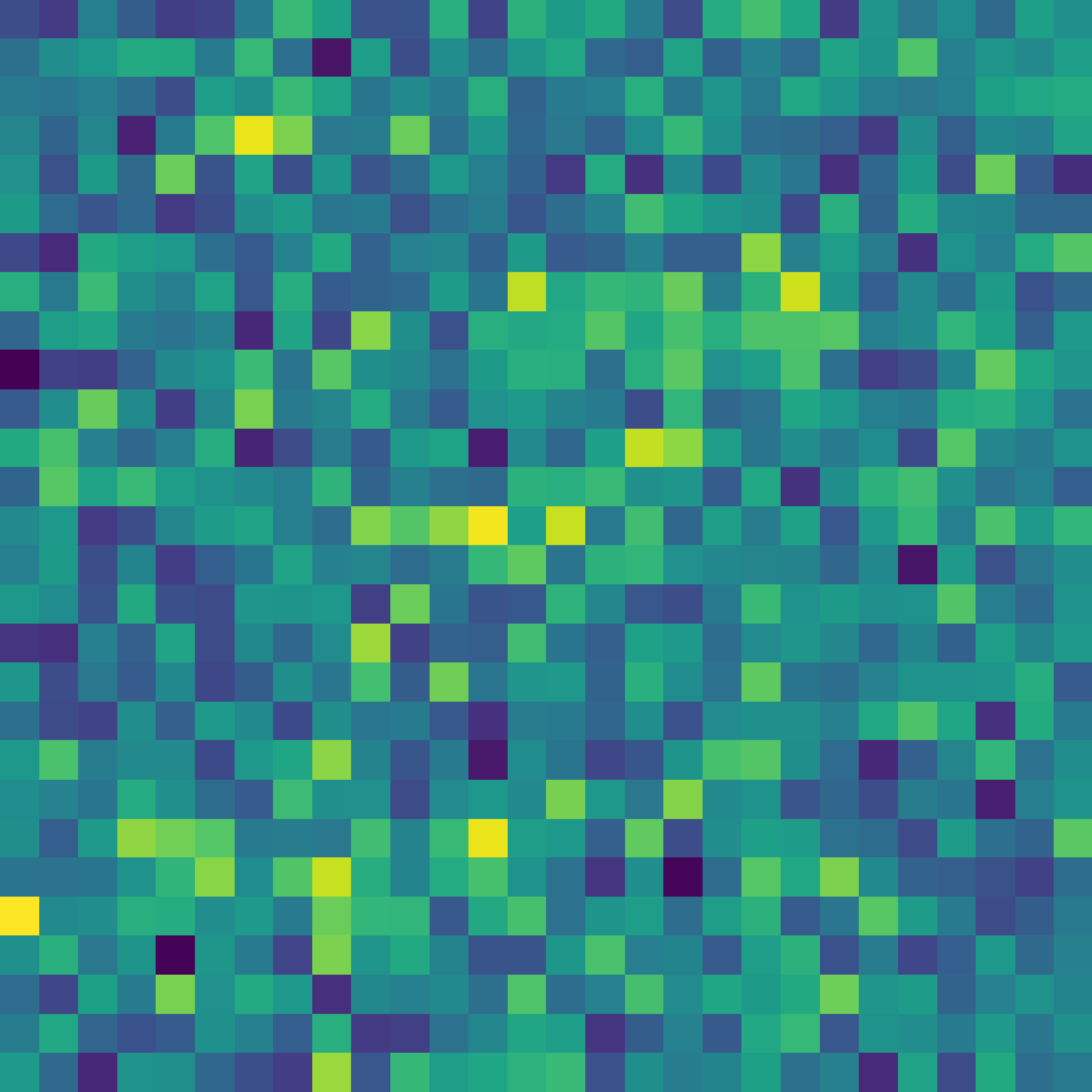}
  \caption{}
\end{subfigure}
\hfill
\begin{subfigure}{1.8cm}
  \includegraphics[height=1.8cm]{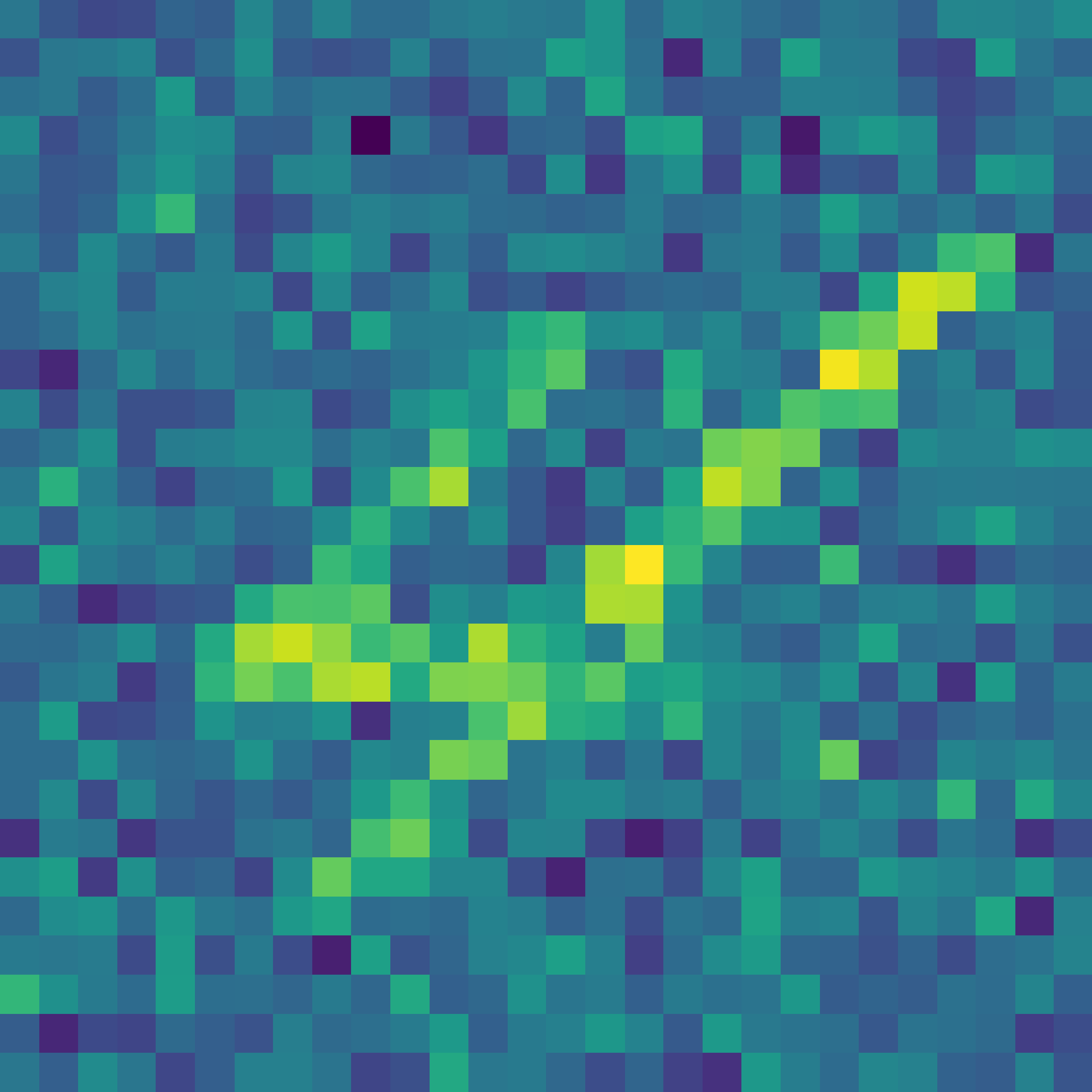}
  \caption{}
\end{subfigure}
\hfill
\begin{subfigure}{1.8cm}
  \includegraphics[height=1.8cm]{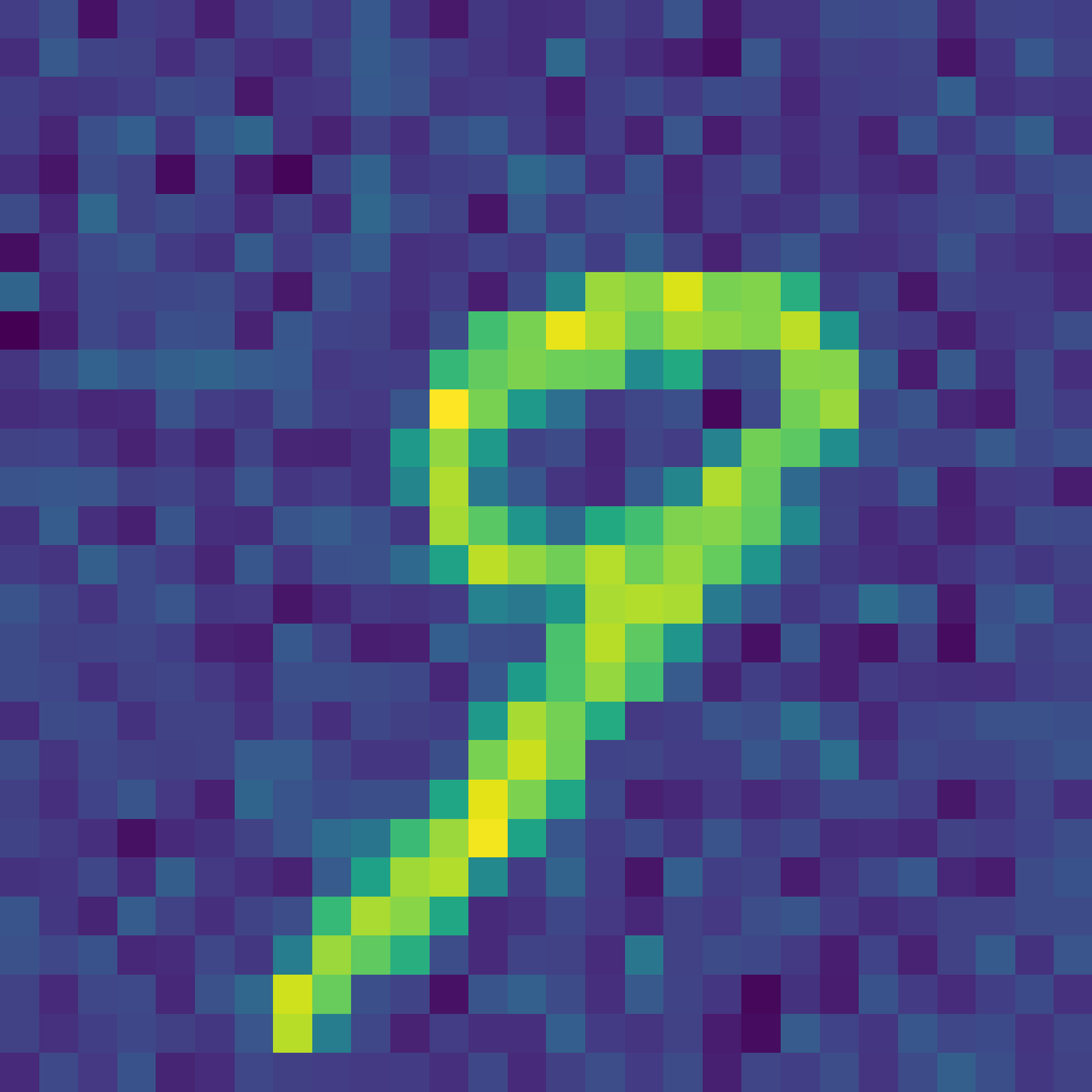}
  \caption{}
\end{subfigure}
\hfill\hfill

\vspace{2mm} 

\hfill
\begin{subfigure}{1.8cm}
  \includegraphics[height=1.8cm]{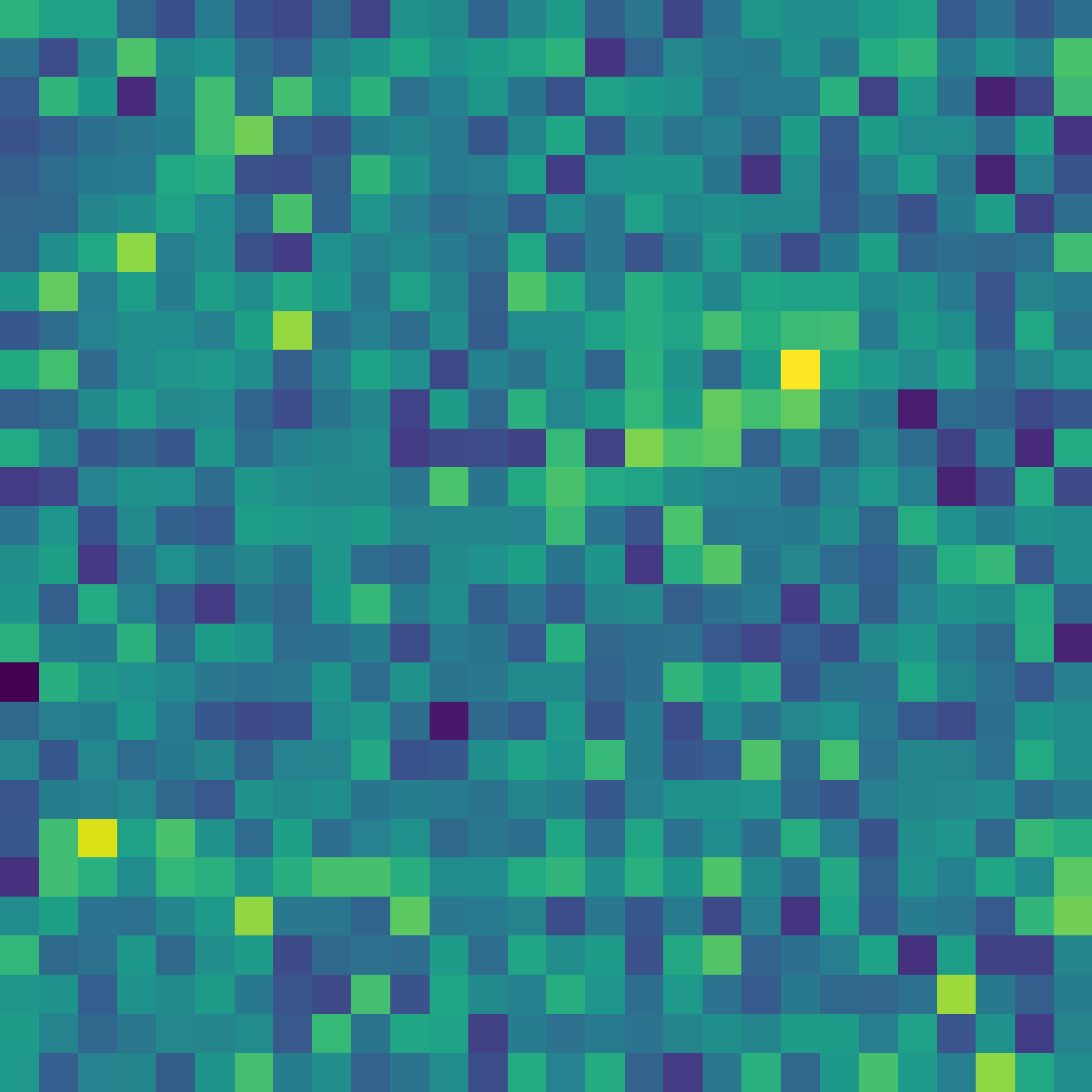}
  \caption{}
\end{subfigure}
\hfill
\begin{subfigure}{1.8cm}
  \includegraphics[height=1.8cm]{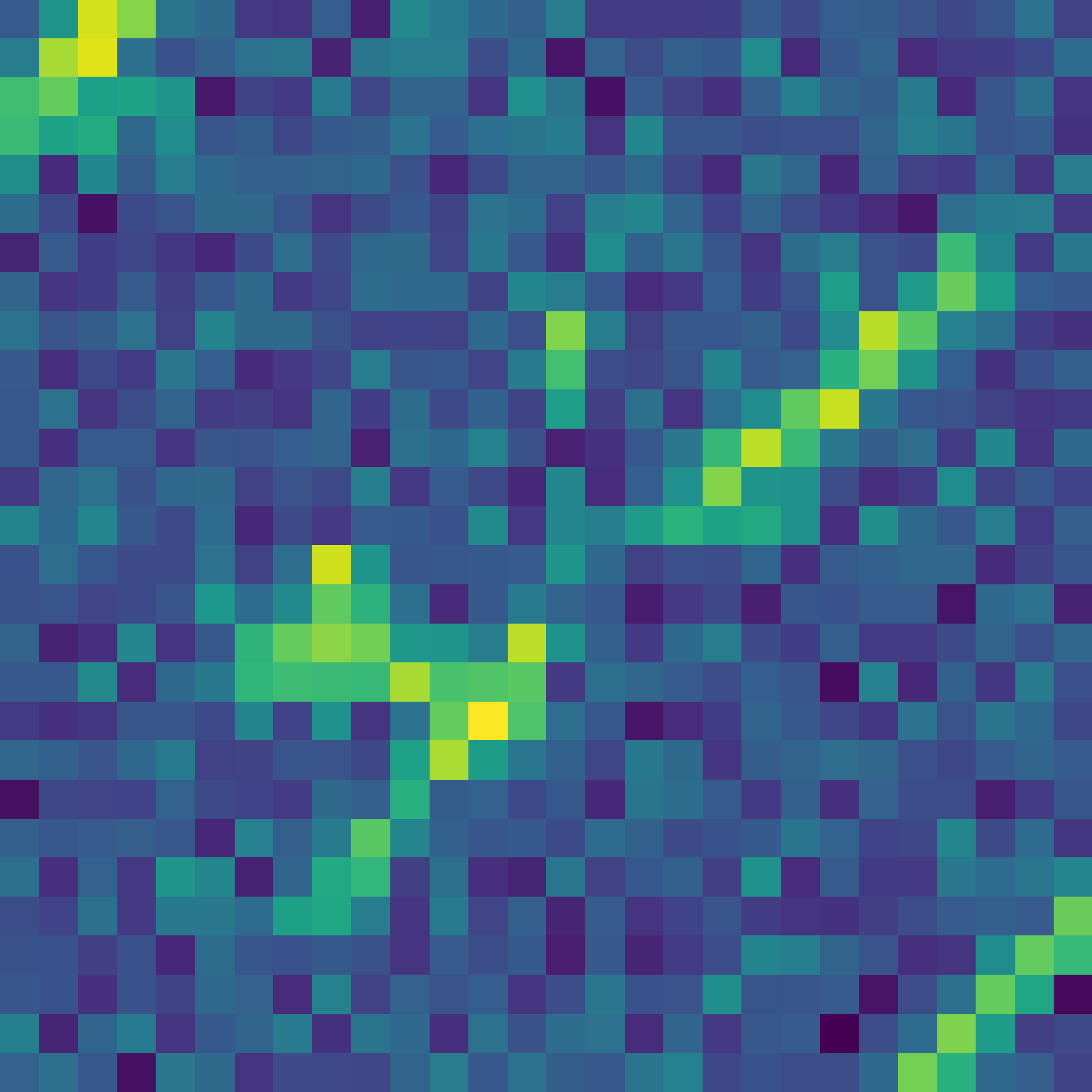}
  \caption{}
\end{subfigure}
\hfill
\begin{subfigure}{1.8cm}
  \includegraphics[height=1.8cm]{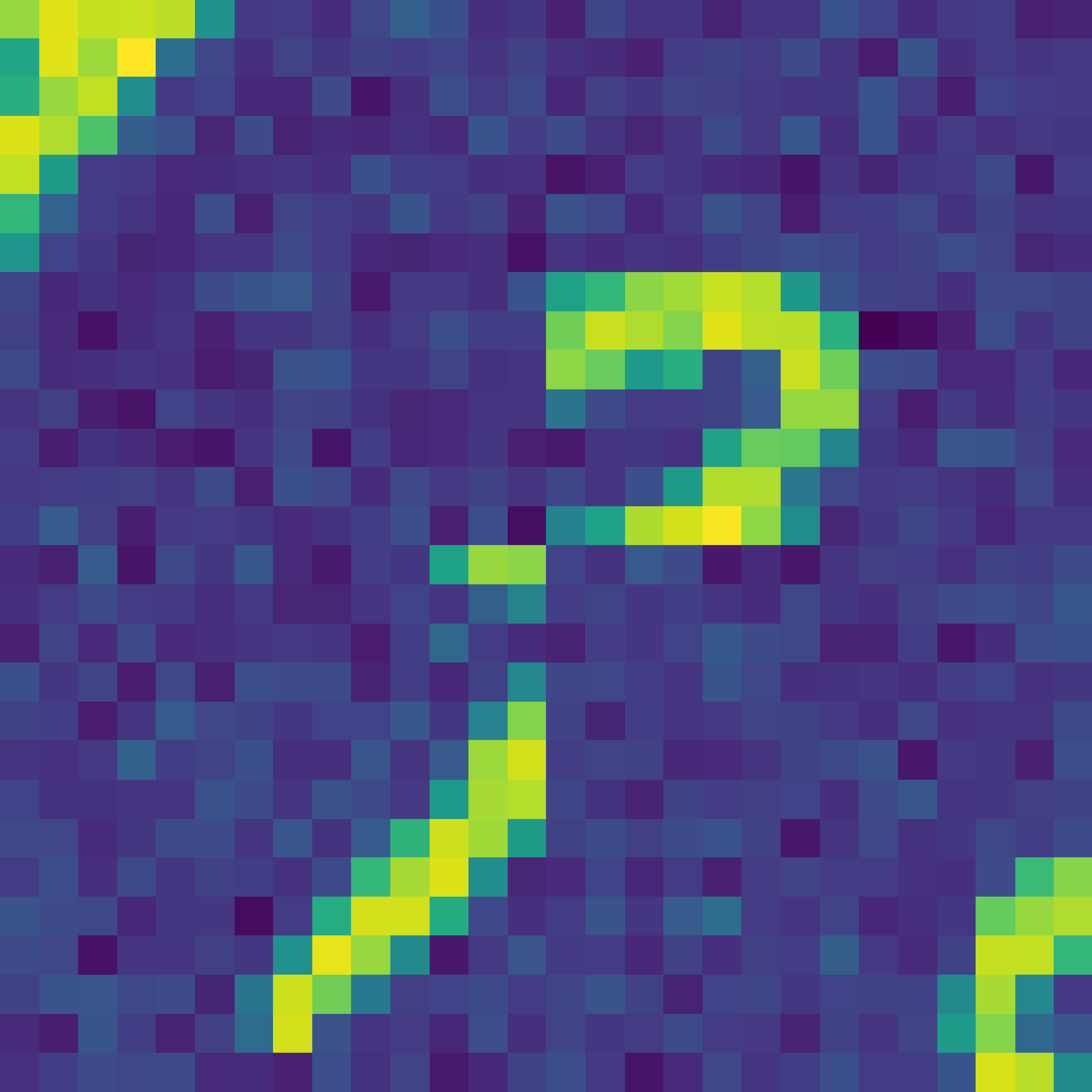}
  \caption{}
\end{subfigure}
\hfill\hfill

\caption{(a)(b)(c) Direct mode and (d)(e)(f) permutation mode, with channel noise $-10$ dB, $0$ dB, $10$ dB. Original images are as (a)(b)(c) in Fig.~\ref{fig:mnist_images}.}
\label{fig:permutation}
\end{figure}

Fig. \ref{fig:mnist-3} shows how accuracy varies with SSIM and PSNR on the MNIST dataset. Here, ``Direct'' and ``Permutation'' mean the accuracy of $\bar{\Theta}_{\bar{\theta}}$'s classification to the original images and permuted images that are directly noised by a certain level of Gaussian white noise \footnote{Here we just use those two baselines as comparisons to show that SSIM or PSNR may be disconnected to the semantic fidelity in some cases, and the application of permutation is beyond our scope of this paper. Future work may utilize permutation to further enhance the system performance against model inversion attack.} (as shown in Fig. \ref{fig:permutation}), based on Eq. (\ref{eq:f_t}); we use a simple permutation method that divides the image into four parts and swaps the position of up-left corner and down-right corner parts. Essentially, our system's adversary $\mathcal{A}_a$, as well as the adversaries in the two IBAL systems, can partially achieve the recovery of the input $x$, as evidenced by their SSIM values being higher than those of the direct-noising scenario in most cases with the same accuracy. This observation is consistent with neural-based systems that possess denoising capabilities \cite{wang2023implementation}. Compared to the adversary in our system, despite the adversaries in both information-bottleneck-based systems effectively reduce the estimated SSIM and PSNR of the original input obtained by the attacker, we observe that there are cases that the adversary's classification accuracy can remain relatively high. For example, when the SSIM is 0.2, the attacker in our system (blue dashed line) achieves a accuracy of 0.47, whereas IBAL-D (purple dashed line with circle markers) reaches 0.78 at SSIM $=0.175$. This indicates that adversarial training merely distorts syntactic features without suppressing semantic information, since the key features remain intact or largely unchanged from the perspective of neural networks. It can be evidenced in Fig. \ref{fig:mnist_images} that the adversary's images in IBAL still have enough semantic information that can be used to categorize, even though their SSIM or PSNR values are lower. The lower plot in Fig. \ref{fig:mnist-3}, showing the complicated correspondence between PSNR and accuracy, further supports the discussion above, since PSNR ignores perceptual quality and structural information that is important to a classification task \cite{wang2004image}. In task-oriented semantic communication, the primary goal is to accurately reconstruct and understand the transmitted task-related information, rather than merely preserving pixel-level fidelity. Therefore, metrics like accuracy, which directly measures the system's ability to correctly interpret the semantic content, are more appropriate for evaluating the effectiveness of adversarial attacks.

The training strategy of our system prioritizes the recovery of semantic information. The adversary is only tasked with reconstructing images after the $\mathcal{T}-\mathcal{R}$ pair is fully trained to recover task-relevant semantic information, and the amount of information in $\mathcal{T}$'s output can be considered constant. Since we further process the adversary's output through a strong classifier that effectively converting the image reconstruction task into a classification task, it allows us to evaluate how much of the ``task-relevant'' information is contained within the reconstructed images, which is directly tied to the quality of the image reconstruction. However, due to the limited information provided by $\mathcal{T}$ and the fact that the $\mathcal{T}$'s output is not primarily designed for image reconstruction, the adversary's ability to recover syntactic information is inherently biased, resulting in less ``task-relevant'' information embedded in the reconstructed images. If the $\mathcal{T}-\mathcal{A}$ pair is trained first, followed by the receiver $\mathcal{R}$, the training process would inevitably prioritize a ``high-information task'' (image reconstruction) before transitioning into a ``low-information task'' (classification). This approach would likely lead to the encoder producing excessive information, which would benefit the adversary by providing more access to task-relevant semantic information, even if the recovered images have relatively low SSIM (or PSNR) values. Consequently, this strategy would compromise the privacy protection intended for the classifier.

\begin{figure}[t]
    \centering
    \ifCFGcaseOne
    \includegraphics[width=0.65\linewidth]{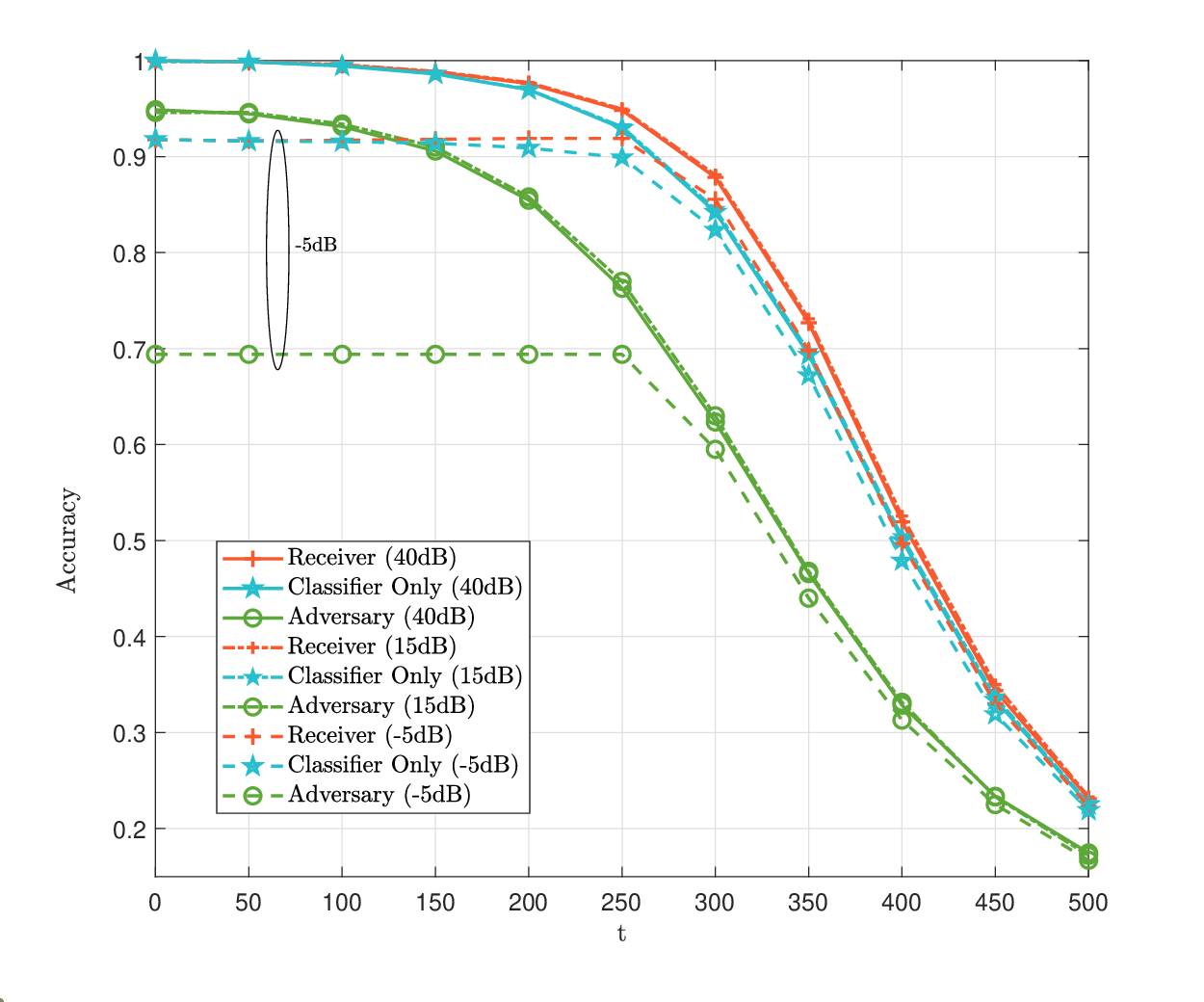}
    \fi
    \ifCFGcaseTwo
    \includegraphics[width=\linewidth]{retultfigs/mnist-4-20250917.eps}
    \fi
    \caption{System performances on MNIST dataset under AWGN channel with different channel SNRs.}
    \label{fig:mnist-4}
\end{figure}


Fig. \ref{fig:mnist-4} illustrates the robustness of our DiffSem system across diverse channel conditions. As outlined in Sec. III Part C, the transmitter dynamically adjusts the level of self-noising pre-added noise based on the estimated channel condition. When facing high SNR regimes, the self-noising module becomes the primary driver of signal distortion, meaning that variations in $t$ more directly influence the performance of both the receiver and the attacker, allowing precise control via $t$. Conversely, in poor channel conditions, the wireless channel itself dominates, allowing the self-noising module to flexibly control the quality of the transmitted signal while preventing over-compression to exert minimal impact on the signal state at the receiver. For instance, at a channel SNR of \(-5\) dB, the dominance of channel noise results in a nearly constant information gap between the classifier and the adversary. We observe that for \( t > 250 \), the performance of both the receiver and the adversary slightly deteriorates compared to scenarios with higher channel SNR. This discrepancy may stem from the discrete nature of \(\bar{\alpha}_t\), causing deviations between practical estimates and theoretical calculations. Overall, our self-noising module effectively modulates the degree of signal distortion in response to wireless channel conditions, maintaining an appropriate noise margin and ensuring reliable operation under both high and low SNR environments.

\subsection{Results on CIFAR-10 dataset}

\begin{figure}[!t]
    \centering
    \ifCFGcaseOne
    \includegraphics[width=0.65\linewidth]{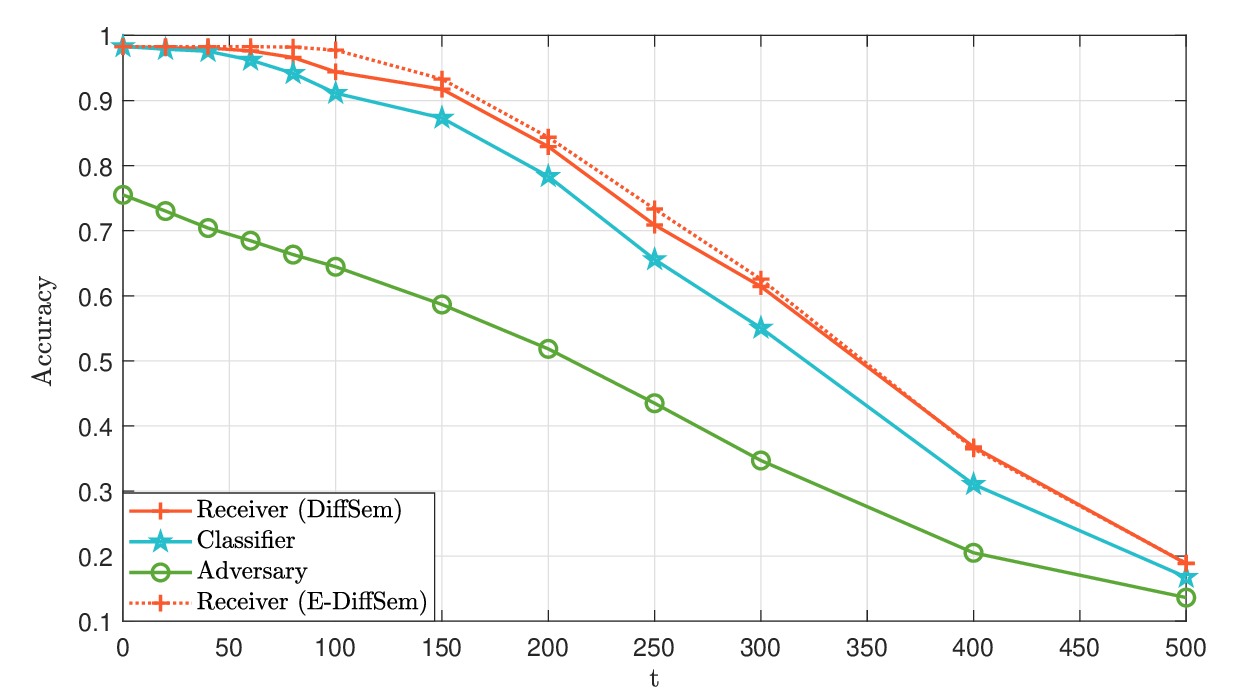}
    \fi
    \ifCFGcaseTwo
    \includegraphics[width=\linewidth]{retultfigs/cifar-1.eps}
    \fi
    \caption{System performances using CIFAR-10 dataset.}
    \label{fig:cifar-1}
\end{figure}

We also implement the system on the CIFAR-10 dataset, which has complex scenes that pose more challenges for semantic communication systems. Fig. \ref{fig:cifar-1} illustrates the receiver's classification accuracy and the attacker's success rate as functions of $t$. Given our assumption that the training of the $\mathcal{T}-\mathcal{R}$ pair takes precedence over the adversary's training, even at \( t = 0 \), the classification accuracy of the adversary's reconstructed images is not very high. This is because the $\mathcal{T}-\mathcal{R}$ pair's training prioritizes the completion of the label classification task, and the backpropagation process optimizes the $\mathcal{T}$'s neural network in a way that is not aligned with the adversary's image recovery requirements, which is acceptable since normal systems are not designed to meet attackers' needs.


The two red lines in Fig. \ref{fig:cifar-1} show that the system incorporating the diffusion module outperforms the classifier-only scenario. This indicates that imperfect labels can still provide useful coarse-grained constraints for the diffusion path. However, the additional performance improvement brought by E-DiffSem is less pronounced than that on the MNIST dataset. 
A more plausible explanation is that CIFAR-10 images have much richer intra-class variation and more complex local structure, so the preliminary labels estimated by the classifier are less effective as conditional guidance for the reverse diffusion process. At the same time, the unconditional DiffSem branch already recovers a substantial part of the task-relevant semantics, leaving less room for E-DiffSem to further improve upon it. In the future, we could explore integrating Transformer-based self-attention mechanisms or leveraging image semantic segmentation techniques to provide stronger guidance for the diffusion module, thereby further enhancing system performance.

As shown in Fig. \ref{fig:cifar_images}, our system actively degrades the semantic utility of adversarial reconstructions. For instance, at $t=150$ ($p=2.3$ dB), the adversary's recovered images (Fig. \ref{fig:cifar_images}(c)) retain only fragmented visual cues (e.g., partial textures), resulting in a classification accuracy of $58.7\%$, which is significantly below $87.3\%$ when bypassing diffusion, as well as $91.7\%$ and $93.3\%$ when using DiffSem and E-DiffSem. The gaps stem from our diffusion model's prioritized recovery of task-specific features (e.g., object categories) while discarding non-essential syntactic details. Such a design suggests that even if attackers partially reconstruct pixel-level content, they still cannot effectively infer semantic information.

\begin{figure}[!t]
    \centering
    \ifCFGcaseOne
    \includegraphics[width=0.65\linewidth]{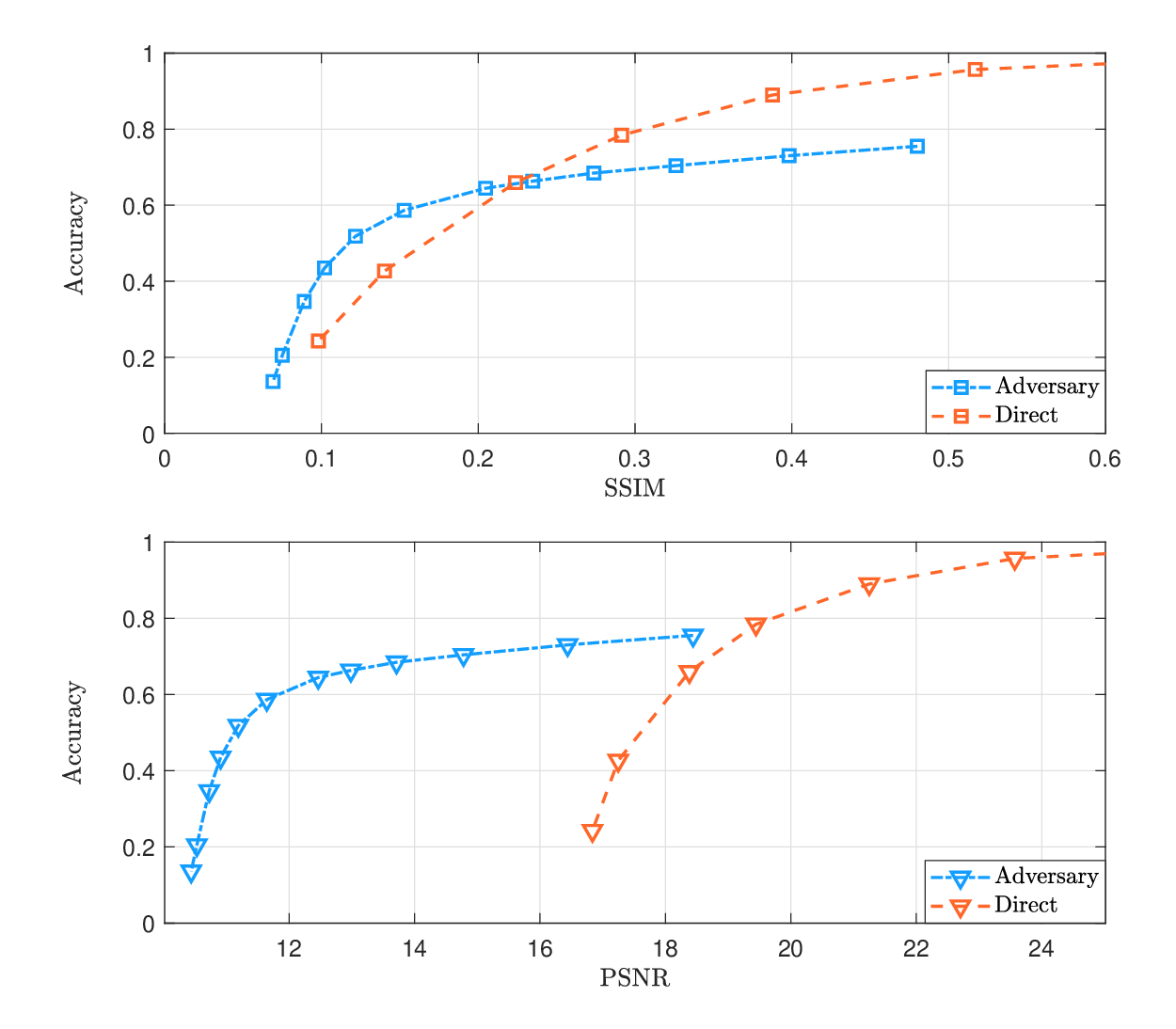}
    \fi
    \ifCFGcaseTwo
    \includegraphics[width=\linewidth]{retultfigs/cifar-4-20250917.eps}
    \fi
    \caption{Plot of accuracy versus SSIM and PSNR using CIFAR-10 dataset.}
    \label{fig:cifar-4}
\end{figure}

\ifCFGcaseTwo
\begin{figure}[!t]
\centering
\hfill
\begin{subfigure}{2.4cm}
  \includegraphics[height=2.4cm]{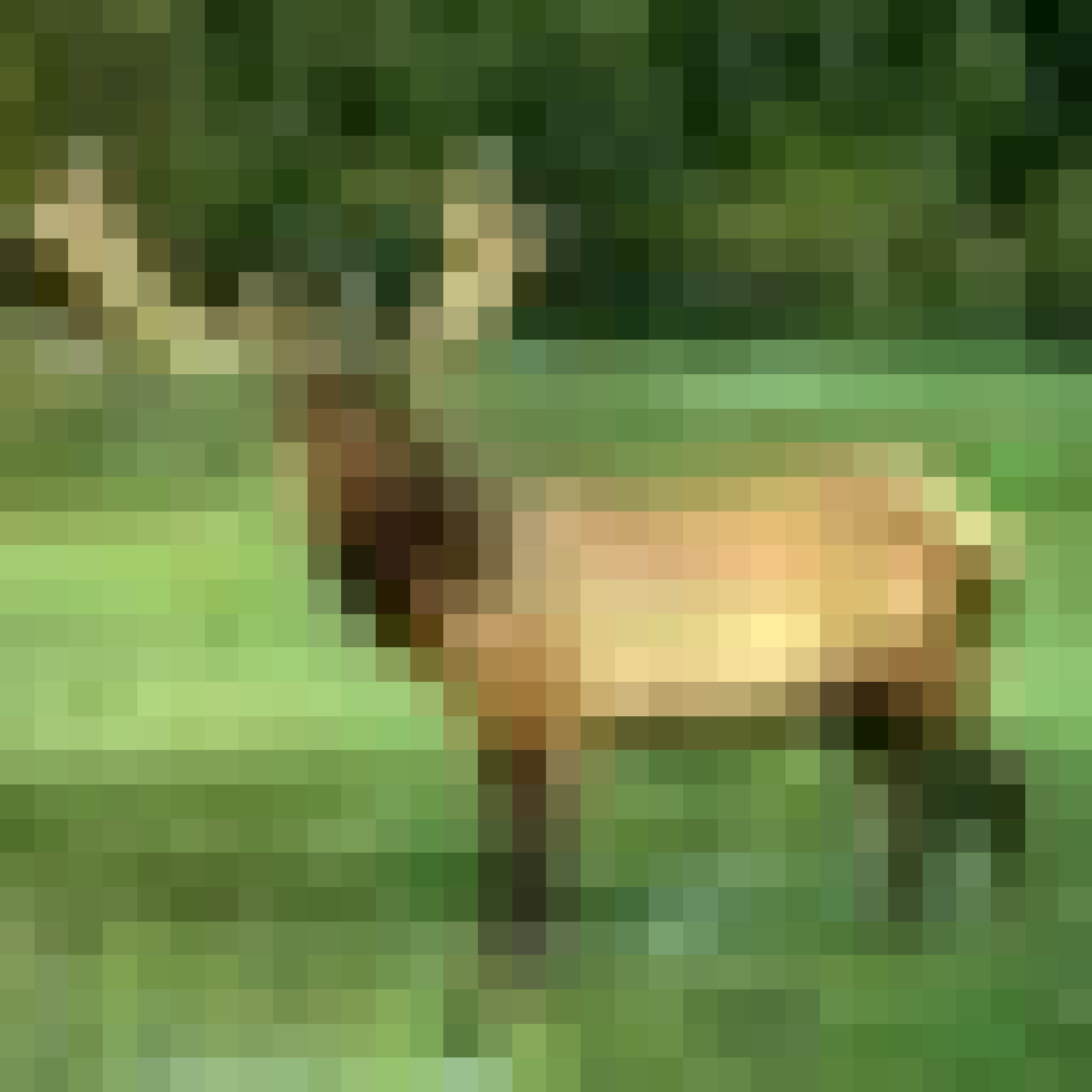}
  \caption{}
\end{subfigure}\hfill
\begin{subfigure}{2.4cm}
  \includegraphics[height=2.4cm]{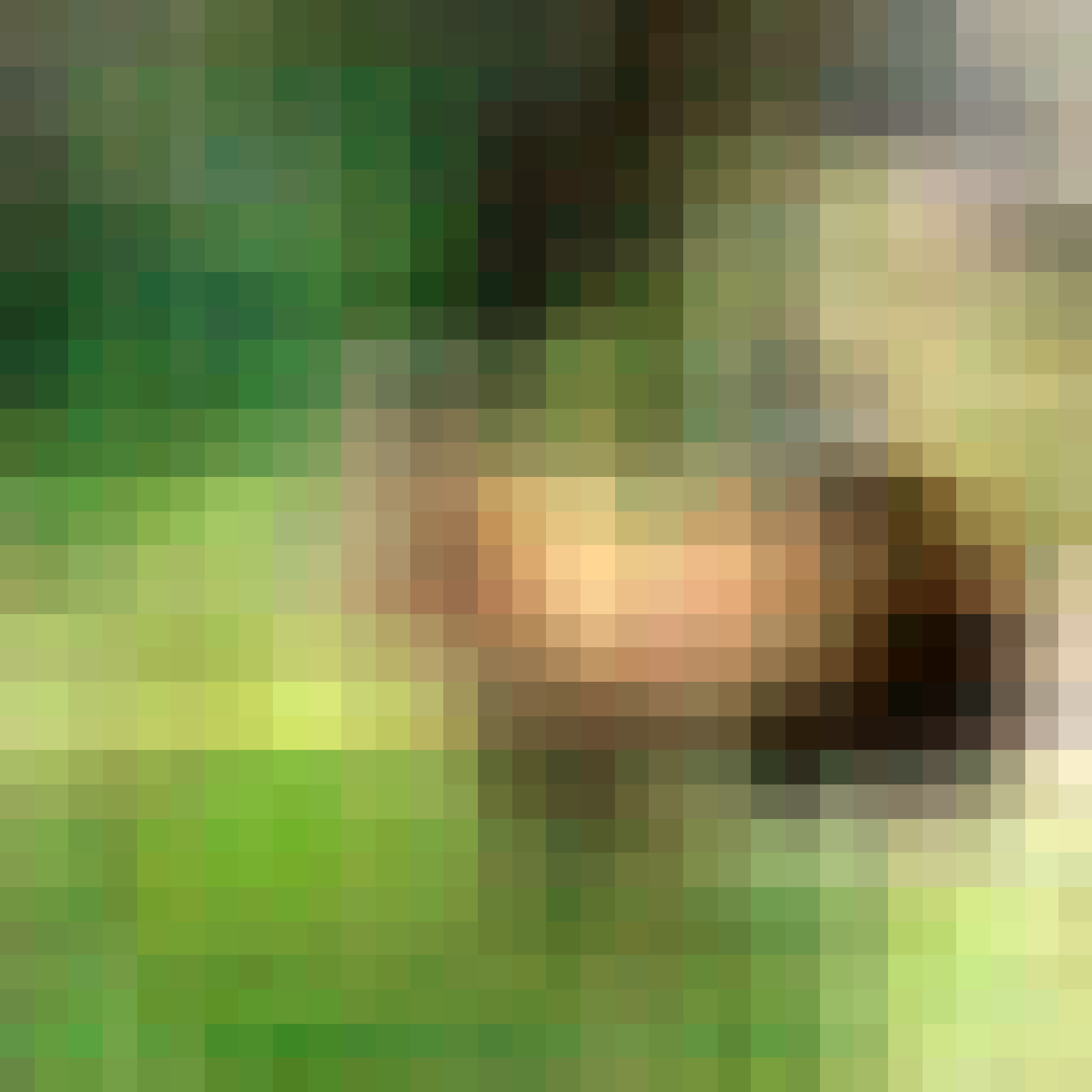}
  \caption{}
\end{subfigure}
\hfill\hfill

\vspace{2mm} 

\hfill
\begin{subfigure}{2.4cm}
  \includegraphics[height=2.4cm]{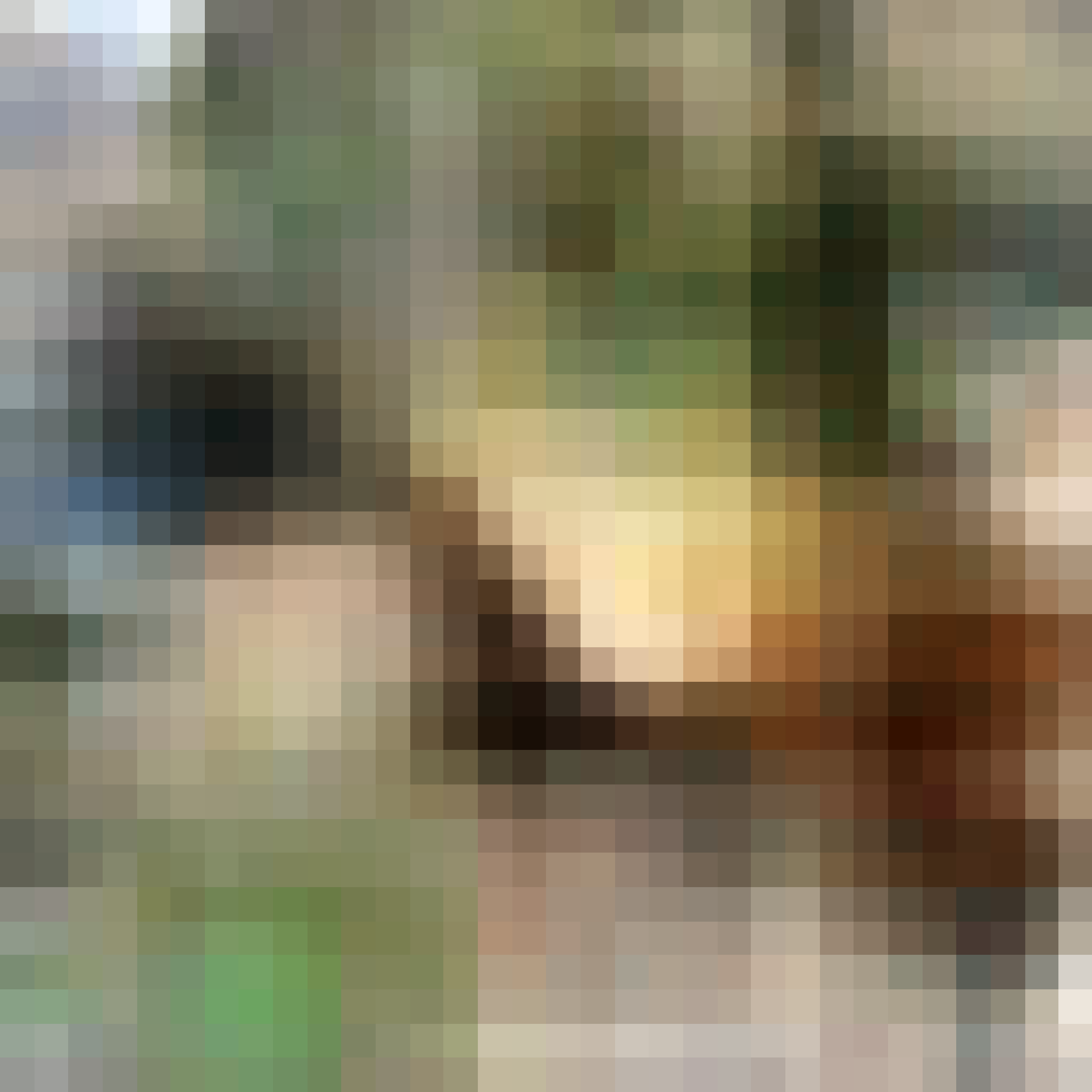}
  \caption{}
\end{subfigure}\hfill
\begin{subfigure}{2.4cm}
  \includegraphics[height=2.4cm]{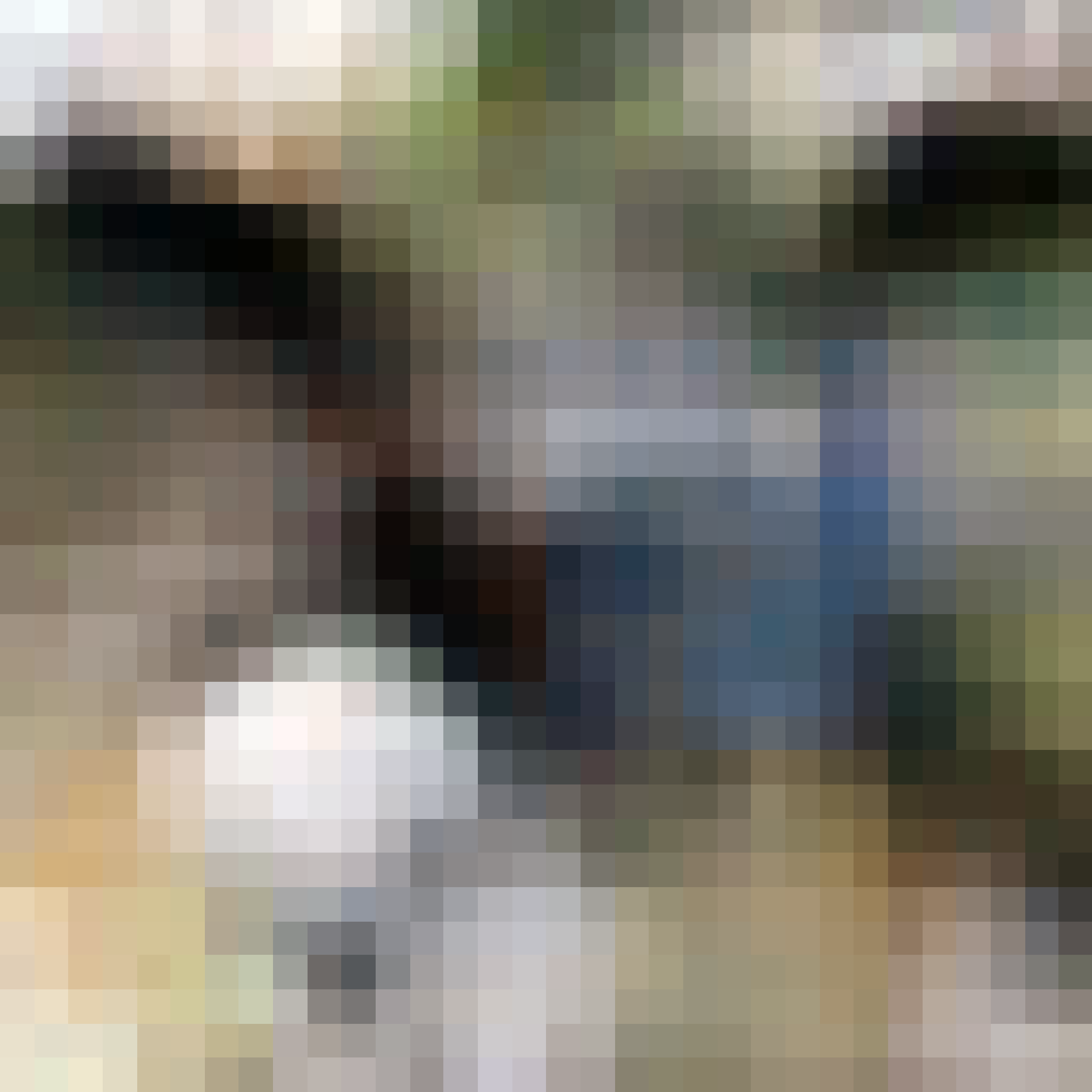}
  \caption{}
\end{subfigure}
\hfill\hfill
\caption{Adversary's output images with different $t$ under CIFAR-10 dataset. (a) Original image. (b)(c)(d) Adversary's recovery when $t=60$, $150$ and $300$.}
\label{fig:cifar_images}
\end{figure}
\fi
\ifCFGcaseOne
\begin{figure}[!t]
\centering
\hfill
\begin{subfigure}{2.4cm}
  \includegraphics[height=2.4cm]{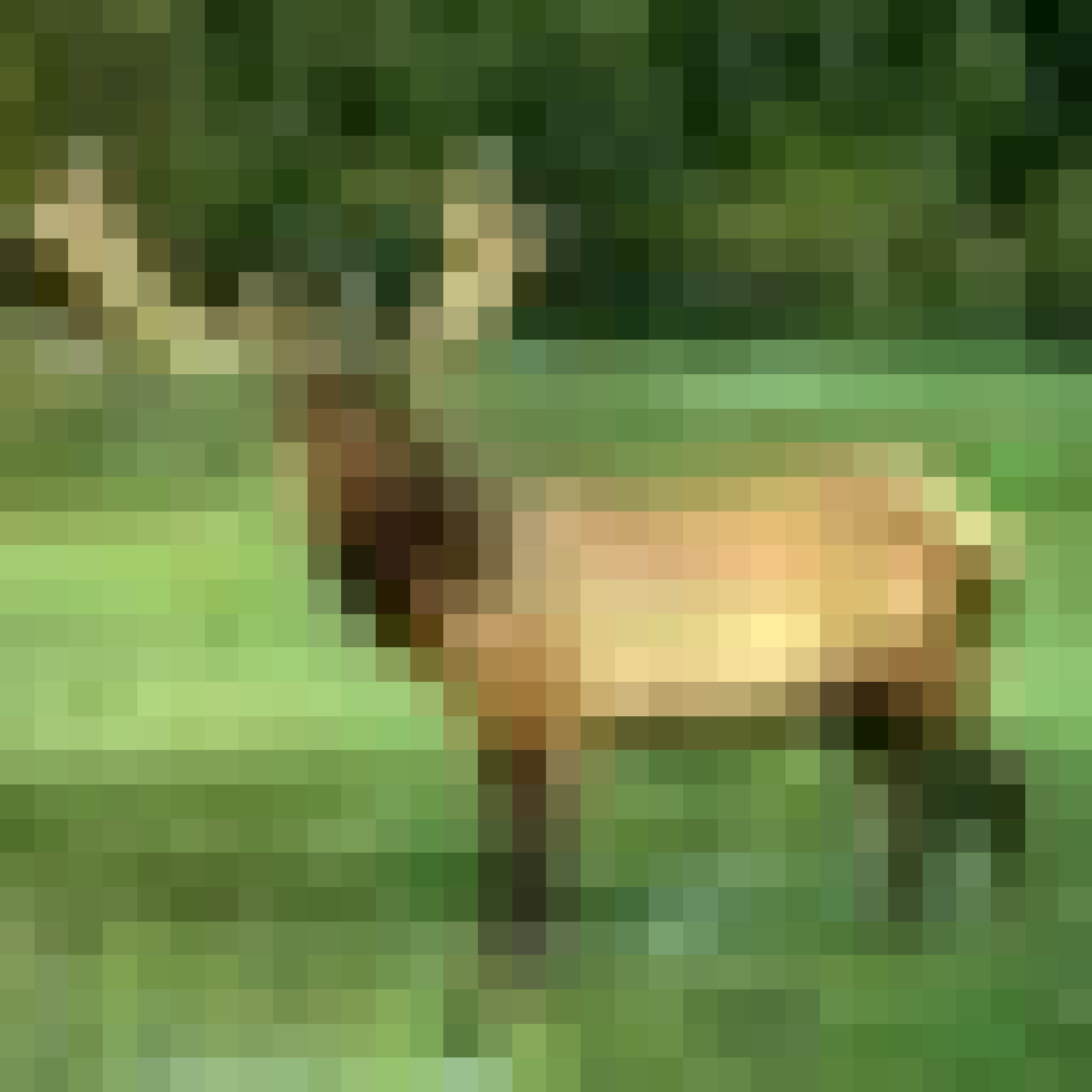}
  \caption{}
\end{subfigure}\hfill
\begin{subfigure}{2.4cm}
  \includegraphics[height=2.4cm]{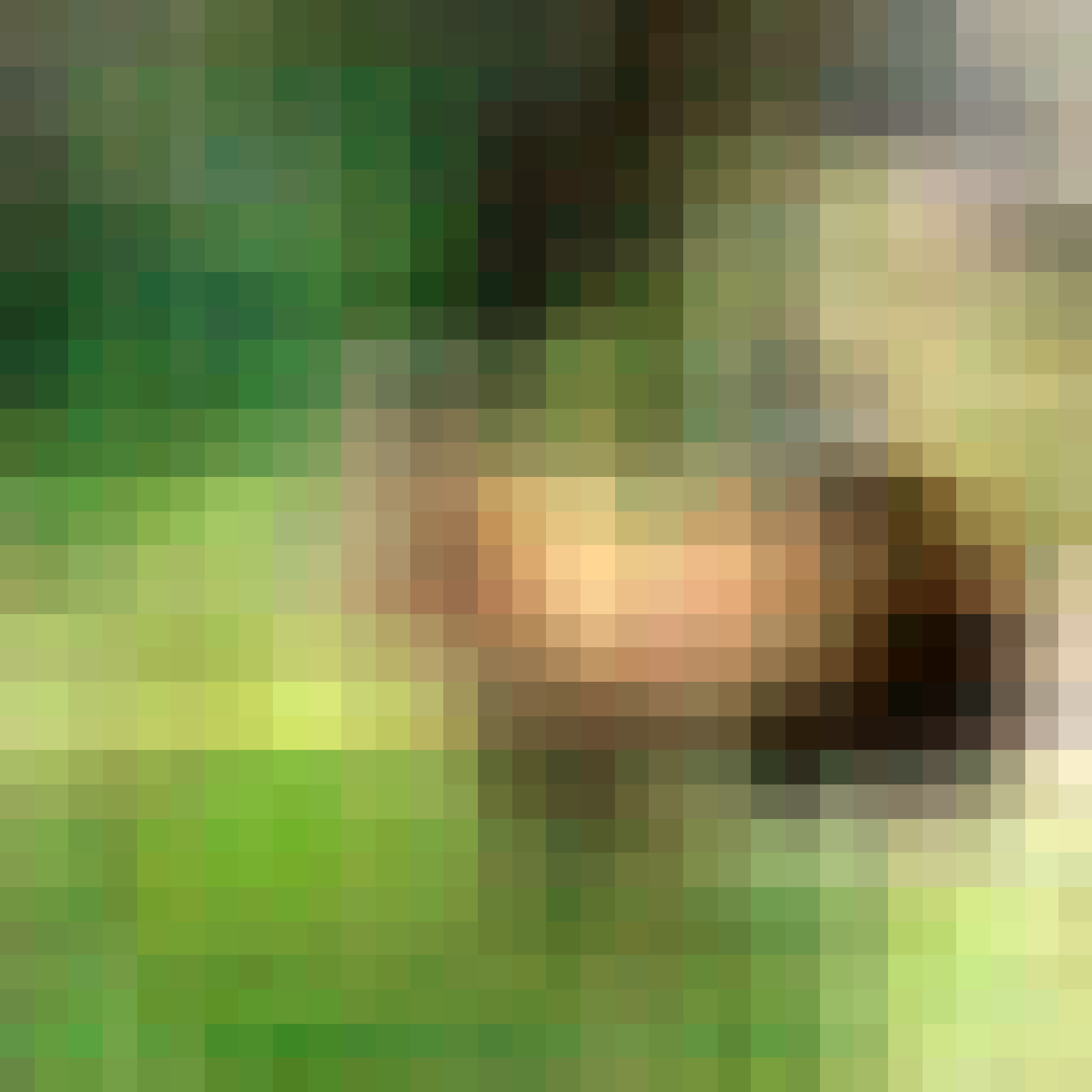}
  \caption{}
\end{subfigure}\hfill
\begin{subfigure}{2.4cm}
  \includegraphics[height=2.4cm]{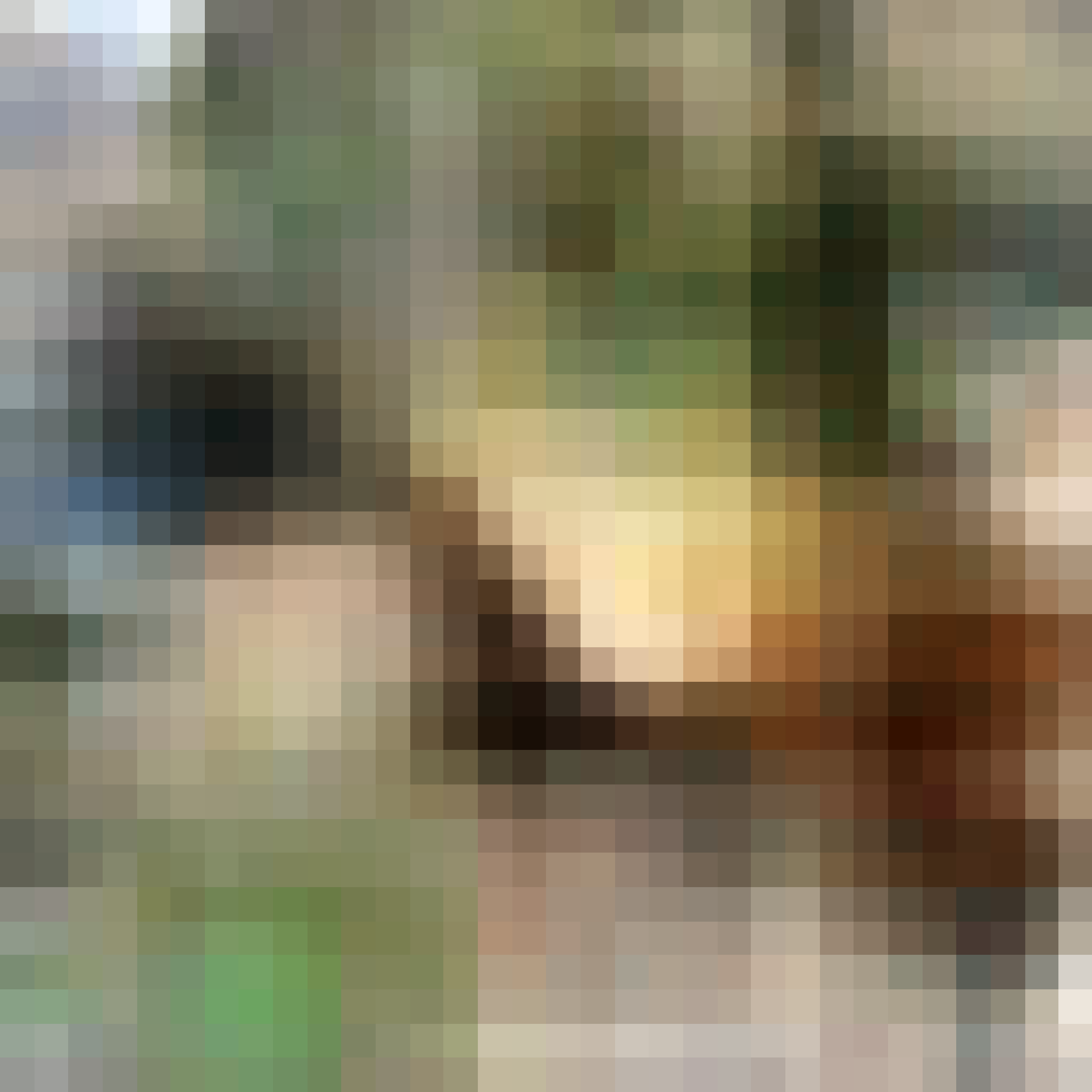}
  \caption{}
\end{subfigure}\hfill
\begin{subfigure}{2.4cm}
  \includegraphics[height=2.4cm]{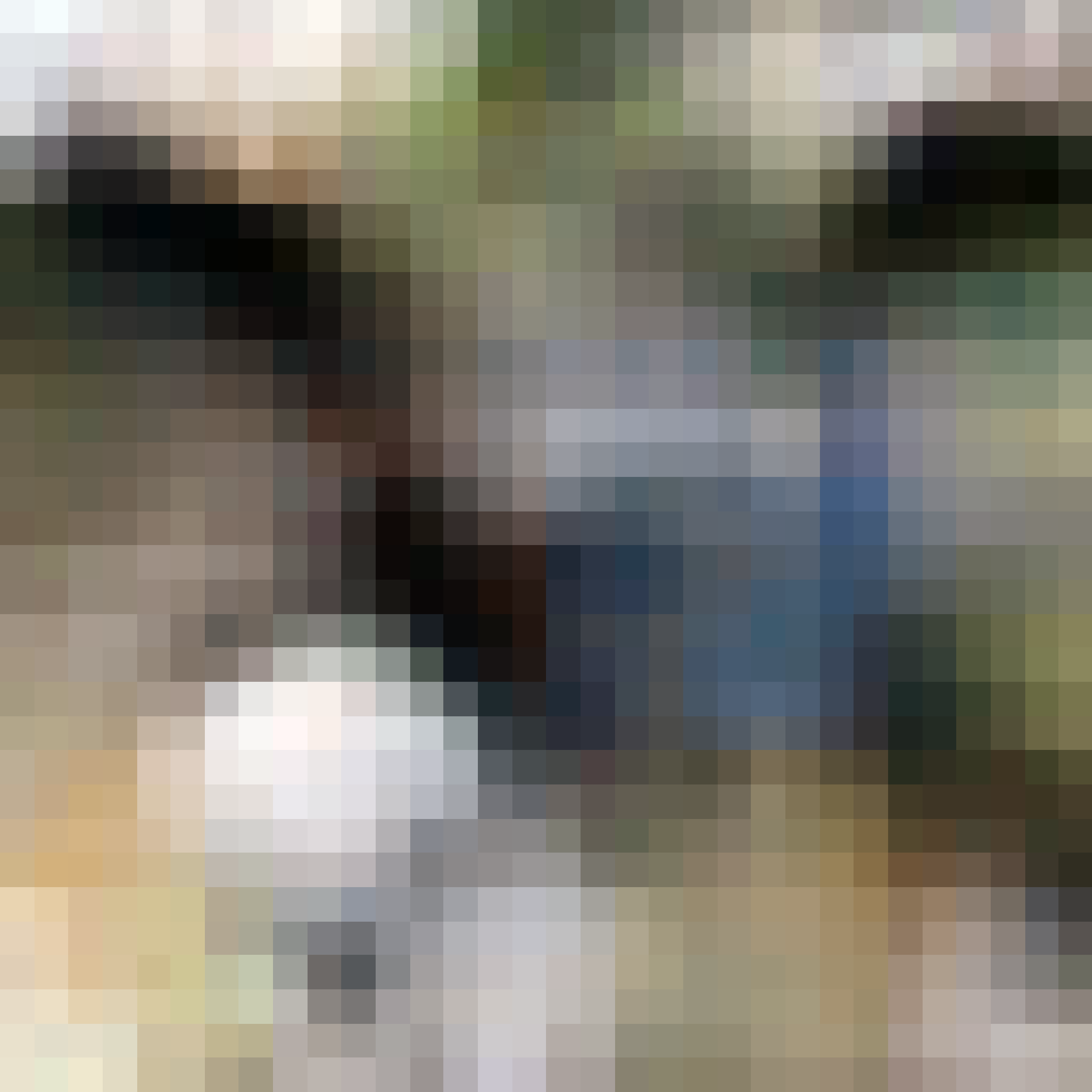}
  \caption{}
\end{subfigure}\hfill\hfill
\caption{Adversary's output images with different $t$ under CIFAR-10 dataset. (a) Original image. (b)(c)(d) Adversary's recovery when $t=60$, $150$ and $300$.}
\label{fig:cifar_images}
\end{figure}
\fi

Fig. \ref{fig:cifar-4} shows the accuracy versus SSIM and PSNR on the CIFAR-10 dataset. It highlights the weak correlation between traditional image quality metrics and semantic security. For instance, when SSIM is $0.12$, the attacker's classification accuracy still reaches $51.8\%$, whereas in the direct mode, with SSIM equal to $0.14$, the accuracy is only $42.6\%$. This difference arises from the attacker's ability to implicitly learn class-discriminative features, while variations in pixel texture are insufficient to prevent the attacker from acquiring these features. Other curves in this figure can further demonstrate that SSIM and PSNR are not suitable metrics for evaluating the amount of task-relevant information obtained by the attacker in task-oriented communication systems.

\begin{figure}[!t]
    \centering
    \ifCFGcaseOne
    \includegraphics[width=0.65\linewidth]{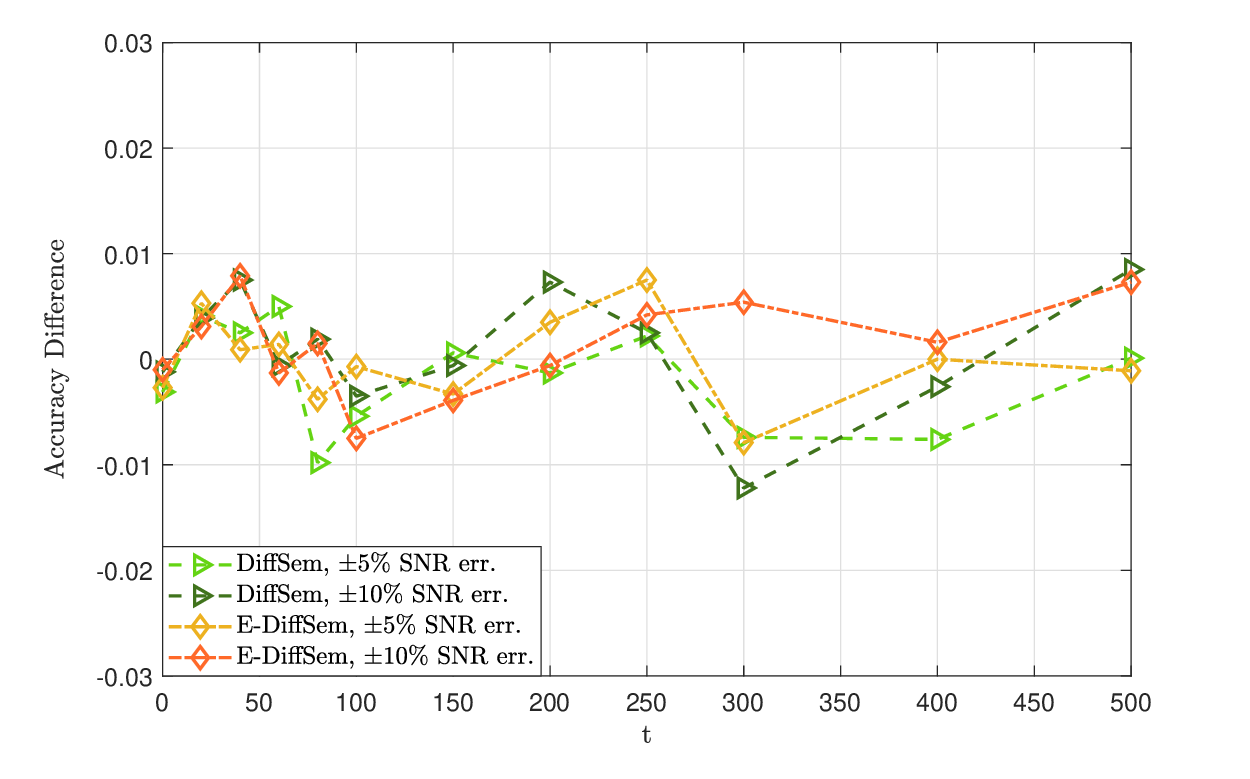}
    \fi
    \ifCFGcaseTwo
    \includegraphics[width=\linewidth]{retultfigs/cifar-snr-est.eps}
    \fi
    \caption{    
    Robustness of DiffSem to imperfect SNR estimation on CIFAR-10 dataset.    
    }
    \label{fig:cifar-snr-est}
\end{figure}

We further examine the effect of imperfect receiver-side SNR estimation using CIFAR-10 as an example. Specifically, instead of using the exact SNR $\gamma$, we use a multiplicatively perturbed estimate $\hat{\gamma}=\gamma(1+\kappa)$, where $\kappa\sim\mathcal{U}[-0.05,0.05]$ and $\kappa\sim\mathcal{U}[-0.1,0.1]$. Fig. \ref{fig:cifar-snr-est} plots the accuracy differences of DiffSem and E-DiffSem relative to the accurate-SNR case over different diffusion steps $t$. The four curves fluctuate around zero across the whole range of $t$, indicating that moderate estimation errors only cause small perturbations to the operating point.
\\\indent Quantitatively, when compared with the accurate-SNR case, the average absolute deviation is $4.09\times 10^{-3}$ under $\kappa\sim\mathcal{U}[-0.05,0.05]$ and $4.36\times 10^{-3}$ under $\kappa\sim\mathcal{U}[-0.1,0.1]$ for DiffSem, with the largest observed deviation being $0.0122$. For E-DiffSem, the corresponding average absolute deviations are $3.18\times 10^{-3}$ and $3.78\times 10^{-3}$, and the largest observed deviation is $0.0079$. Thus, the perturbations remain small across the tested error levels.

\begin{figure}[!t]
    \centering
    \ifCFGcaseOne
    \begin{subfigure}[t]{0.49\linewidth}
        \centering
        \includegraphics[width=\linewidth]{retultfigs/celeb-1-final.eps}
        \caption{Exact-match performance using CelebA dataset.}
        \label{fig:celeb_exact}
    \end{subfigure}
    \hfill
    \begin{subfigure}[t]{0.49\linewidth}
        \centering
        \includegraphics[width=\linewidth]{retultfigs/celeb-3-final.eps}
        \caption{Average attribute-wise performance using CelebA dataset.}
        \label{fig:celeb_avg}
    \end{subfigure}
    \fi

    \ifCFGcaseTwo
    \begin{subfigure}[t]{\linewidth}
        \centering
        \includegraphics[width=\linewidth]{retultfigs/celeb-1-final.eps}
        \caption{Exact-match performance using CelebA dataset.}
        \label{fig:celeb_exact}
    \end{subfigure}

    \vspace{0.5em}

    \begin{subfigure}[t]{\linewidth}
        \centering
        \includegraphics[width=\linewidth]{retultfigs/celeb-3-final.eps}
        \caption{Average attribute-wise performance using CelebA dataset.}
        \label{fig:celeb_avg}
    \end{subfigure}
    \fi
    \caption{CelebA performance under different self-noising levels $t$.}
    \label{fig:celeb_results}
\end{figure}

\subsection{Experiments on CelebA dataset}
\label{subsec:celeba_exp}

To complement with a more challenging visual benchmark, we further evaluate the proposed framework on CelebA, which is widely used in privacy and model inversion studies \cite{ye2025model}. Compared with the previous two datasets, CelebA is not only higher in resolution, but also lies on a much more structured facial manifold. This makes it a more suitable testbed for examining whether the proposed framework remains effective when the data prior is stronger and the semantic content is more concentrated. In our setting, each image is resized to $3\times128\times128$, encoded into a semantic latent of size $3\times32\times32$, and evaluated on a three-attribute multi-label task consisting of \textit{Male}, \textit{Smiling}, and \textit{Black\_Hair}. These three attributes correspond to distinct facial cues while maintaining relatively weak pairwise correlation, which makes them suitable for studying semantic preservation and semantic leakage under transmission noise.
\\\indent
The CelebA experiment follows the same transmission principle as in MNIST and CIFAR-10. The latent is power-normalized, transmitted over an AWGN channel with $\mathrm{SNR}=15$\,dB, and evaluated over several timesteps. Since the CelebA task is multi-label, we report both exact-match accuracy (Fig. \ref{fig:celeb_exact}) and average attribute-wise accuracy (Fig. \ref{fig:celeb_avg}). The overall ordering remains consistent under both accuracies, and is the same with that observed on MNIST and CIFAR-10, that is, the adversarial branch is the weaker than the classifier branch, and both diffusion-based receivers remain above direct latent classification over the whole noise range.
Therefore, even on a higher-resolution facial dataset, diffusion-based semantic recovery still improves robustness over direct latent classification and remains substantially stronger than inversion-based semantic leakage.
\\\indent
At the same time, CelebA exhibits a noticeably different curve shape from MNIST and CIFAR-10. In particular, the additional gain of \textit{E-DiffSem} over \textit{DiffSem} remains limited throughout the operating range. This is still consistent with the interpretation suggested by the CIFAR-10 results, that is, once the semantic structure is already largely recoverable, the extra benefit of imperfect conditional guidance becomes modest. On CelebA, this effect is even more natural because the data lie on a relatively concentrated face manifold, so unconditional latent denoising already benefits from a strong facial prior. Moreover, the target task here is milder than the class-label settings in MNIST and CIFAR-10: predicting three facial attributes requires only part of the facial semantics, and the $3\times32\times32$ latent already preserves these attributes reasonably well. Consequently, \textit{DiffSem} itself is comparatively strong, ``Classifier'' remains relatively competitive, and the role of \textit{E-DiffSem} is mainly to provide a modest refinement rather than a large additional gain.

As observed across the three datasets, the performance gap between diffusion-based semantic recovery and the other baselines does not evolve uniformly as the self-noising step $t$ increases. The asymptotic discussion in Section~\ref{subsec:mechanism} establishes that an intermediate regime may exist in which the equivalent channel can support reliable task recovery but not reliable lossless source recovery.
However, this separation does not determine the magnitude or evolution of the finite-system accuracy gap. The actual gap also depends on the source distribution, the prescribed task mapping, and the learned encoder, receiver, and inversion model. Consequently, the same self-noising level need not produce the same effective separation across different datasets and tasks. The different gap patterns observed in the three experiments are therefore not contradictory; rather, they suggest that the information separation between legitimate task decoding and inversion-based semantic leakage is inherently dataset- and task-dependent. A more precise characterization of this dependence remains an open problem.


\subsection{Complexity and Latency Discussion}

As a supplement to the main experimental results, we comment on the computational and communication overhead of the proposed system. 
\\\indent On CIFAR-10, the additional inference time remains on the order of $0.1$s under a batch size of 256, indicating that diffusion-based refinement does not impose a prohibitive burden on small-resolution data.
To make the discussion less sensitive to hardware differences, we focus on relative complexity trends from an end-to-end benchmark. Since classifier-only, DiffSem, and E-DiffSem share the same communication payload and differ mainly in the receiver-side refinement stage, they provide the most appropriate basis for complexity comparison. Under this benchmark, DiffSem and E-DiffSem incur about $7.9\%$ and $12.0\%$ higher end-to-end latency than classifier-only, respectively. Their parameter counts increase from $5.282$M for classifier-only to $7.826$M for DiffSem and $9.455$M for E-DiffSem. 
\\\indent This trend is consistent with the module scales observed across all three datasets. In particular, the encoder remains relatively lightweight, while the main additional model cost of DiffSem and E-DiffSem comes from the receiver-side diffusion U-Net. By contrast, the strong classifier can be larger on more complex datasets, but it is used only for evaluation and is not part of the online transmission pipeline. Therefore, the practical overhead of the proposed method is governed mainly by receiver-side refinement rather than by increased communication payload or attacker-side evaluation modules.
\\\indent Importantly, the communication payload remains unchanged across these methods. Hence, there is a computation--communication trade-off in our system design: the proposed diffusion-based receivers improve task performance without increasing communication cost, at the price of moderate receiver-side latency overhead. In the representative setting considered here, DiffSem also exhibits a slightly more favorable latency-efficiency trade-off than E-DiffSem.
\\\indent For higher-resolution data or stricter real-time requirements, this receiver-side overhead may become more critical. Future work will therefore consider more efficient diffusion implementations, including distillation, pruning, and lighter architectures.

\section{Conclusion}

In this paper, we introduce DiffSem, a diffusion-aided framework for task-oriented semantic communication, designed to balance task performance and privacy preservation under the considered model inversion threat model. We evaluate the success of model inversion using the task accuracy of an external strong classifier on adversarial reconstructions, rather than relying solely on pixel-level similarity metrics.
DiffSem exploits the inherent mismatch between the task-specific semantics required by the legitimate receiver and the broader source content that the attacker attempts to reconstruct.
Specifically, a transmitter-side self-noise mechanism adaptively regulates the radiated semantic information while compensating for channel noise, an unguided diffusion U-Net on the receiver enhances performance on downstream tasks, and an optional self-referenced label embedding further enhances receiver performance while maintaining the guidance-free constraint.
Experiments show that DiffSem enables the legitimate receiver to achieve higher accuracy than the baselines.
Nevertheless, the diffusion-based refinement also introduces additional receiver-side computation and latency, so further improving the resulting task performance, privacy, and complexity trade-off through more efficient diffusion designs remains an important direction for future work.

\appendices

\section{Derivation of Eq. (\ref{eq:alpha_t_primes})}
\label{apx:a}

Here we derive the Eq. (\ref{eq:alpha_t_primes}) in detail. 
Since we have
\begin{equation}
    \mathbb{E}\left(\|z'\|^2\right) = \bar{\alpha}_{T'}\|f_0\|^2 + \left(1-\bar{\alpha_{T'}}\right)D
\end{equation}
and
\begin{equation}
\begin{split}
    \mathbb{E}\left(\|\hat{z}''\|^2\right) &= \bar{\alpha}_{T'}\|f_0\|^2 + \left(1-\bar{\alpha}_{T''}\right)D + \sigma^2_{\text{ch}}D \\
    &= \bar{\alpha}_{T'}\|f_0\|^2 + \left(1-\bar{\alpha}_{T''}\right)D \\
    &~~~+ \frac{\bar{\alpha}_{T'}\mathbb{E}\left(\|f_0\|^2\right) + (1-\bar{\alpha}_{T''}) D}{\gamma}D,
\end{split}
\end{equation}
when $\mathbb{E}\left(\|z'\|^2\right) = \mathbb{E}\left(\|\hat{z}''\|^2\right)$, we can get
\begin{equation}
    1-\bar{\alpha_{T''}} + \frac{\bar{\alpha}_{T'}\mathbb{E}\left(\|f_0\|^2\right) + (1-\bar{\alpha}_{T''}) D}{\gamma} = 1-\bar{\alpha_{T'}}.
\end{equation}
Then
\begin{equation}
    \bar{\alpha}_{T''}\left(1+\frac{D}{\gamma}\right) = \bar{\alpha}_{T'}\left(1+\frac{\mathbb{E}(\|f_0\|^2)}{\gamma}\right) + \frac{D}{\gamma}.
\end{equation}
Arranging the formula above, we can get
\begin{equation}
    \bar{\alpha}_{T''} = \frac{\bar{\alpha}_{T'} \left(\gamma + \mathbb{E}\left(\|f_0\|^2\right)\right) + D}{\gamma + D}.
\end{equation}

\section{Robustness to Imperfect SNR Estimation}
\label{apx:b}

We briefly analyze the sensitivity of the mapping in Eq.~(\ref{eq:alpha_t_primes})
to imperfect SNR estimation. Let the estimated SNR be
\begin{equation}
\label{eq:snr_estimation_model}
\hat{\gamma} = \gamma(1 + \kappa), \qquad \kappa > -1,
\end{equation}
where $\kappa$ denotes the relative estimation error, and $\gamma$ follows
the same definition as in the main text and Appendix A. Define
\begin{equation}
\label{eq:Ef_B_def}
B \triangleq \bar{\alpha}_{T'} \mathbb{E}\|f_0\|^2 + (1 - \bar{\alpha}_{T'}) D.
\end{equation}

Replacing $\gamma$ in Eq.~(\ref{eq:alpha_t_primes}) by $\hat{\gamma}$ gives
\begin{equation}
\label{eq:alpha_T_rho}
\bar{\alpha}_{T_\kappa}
=
\frac{\bar{\alpha}_{T'}(\hat{\gamma} + \mathbb{E}\|f_0\|^2) + D}{\hat{\gamma} + D}.
\end{equation}

Under the true SNR $\gamma$, the resulting effective variance mismatch
relative to the ideal target variance $1 - \bar{\alpha}_{T'}$ is
\begin{equation}
\label{eq:Delta_rho_def}
\Delta_\kappa
=
\Bigl(\bar{\alpha}_{T'} \mathbb{E}\|f_0\|^2 + (1 - \bar{\alpha}_{T_\kappa}) D\Bigr)
\left(
\frac{1}{\gamma} - \frac{1}{\hat{\gamma}}
\right),
\end{equation}
where we used the same relation as in Appendix A, with $\hat{\gamma}$
replacing $\gamma$. Moreover, from Eq.~(\ref{eq:alpha_T_rho}),
\begin{equation}
\label{eq:alpha_T_rho_aux}
\bar{\alpha}_{T'} \mathbb{E}\|f_0\|^2 + (1 - \bar{\alpha}_{T_\kappa}) D
=
\frac{\hat{\gamma} B}{\hat{\gamma} + D}.
\end{equation}

Hence,
\begin{equation}
\label{eq:Delta_rho_closed}
\Delta_\kappa
=
\frac{B(\hat{\gamma} - \gamma)}{\gamma(\hat{\gamma} + D)}
=
\frac{B\kappa}{\gamma(1 + \kappa) + D}.
\end{equation}


Then the coefficient deviation is
\begin{equation}
\begin{split}
\label{eq:alpha_deviation}
\bar{\alpha}_{T_\kappa} - \bar{\alpha}_{T''}
&=-\frac{B(\hat{\gamma} - \gamma)}{(\hat{\gamma} + D)(\gamma + D)}\\
&=-\frac{B\gamma\kappa}{(\gamma(1 + \kappa) + D)(\gamma + D)}.
\end{split}
\end{equation}

If $|\kappa| \le \varepsilon < 1$, we have
\begin{equation}
\label{eq:rho_bounds}
\begin{split}
|\Delta_\kappa|
&\le
\frac{B\varepsilon}{\gamma(1 - \varepsilon) + D},\\
|\bar{\alpha}_{T_\kappa} - \bar{\alpha}_{T''}|
&\le
\frac{B\gamma\varepsilon}{(\gamma(1 - \varepsilon) + D)(\gamma + D)}.
\end{split}
\end{equation}

For small estimation errors,
\begin{equation}
\label{eq:rho_small_error}
\begin{split}
\Delta_\kappa
&=\frac{B}{\gamma + D}\kappa + O(\kappa^2),\\
\bar{\alpha}_{T_\kappa} - \bar{\alpha}_{T''}
&=-\frac{B\gamma}{(\gamma + D)^2}\kappa + O(\kappa^2).
\end{split}
\end{equation}

Therefore, under the adopted Gaussian-equivalent model, the induced
mismatch is first-order in the relative SNR estimation error $\kappa$.


\bibliographystyle{IEEEtran}
\normalem
\bibliography{References}

\end{document}